\pdfoutput=1

\documentclass[preprint2]{aastex}

\usepackage{hyperref,ifthen,psfig,graphicx}
\usepackage{epstopdf}
\usepackage{soul,color,amsmath}
\usepackage{ifthen,psfig,graphicx}
\usepackage{epstopdf,aas_macros}
\usepackage{soul,color,amsmath}
\usepackage{microtype}
\def\mathnew{\mathsurround=0pt}
\def\simov#1#2{\lower 2.5pt\vbox{\baselineskip0pt \lineskip-.5pt
\ialign{$\mathnew#1\hfil##\hfil$\crcr#2\crcr\sim\crcr}}}
\def\simless{\mathrel{\mathpalette\simov <}}
\def\simgreat{\mathrel{\mathpalette\simov >}}
\newcommand{\MeV}{Me\kern-0.11em V}
\newcommand{\keV}{ke\kern-0.11em V}

\newcommand{\fdenl}{erg~s$^{-1}$\,cm$^{-2}$\,\rm{\AA}$^{-1}$}

\newcommand{\kms}{\ensuremath{\rm km\;s}^{-1}}
\newcommand{\cgsbrid}{\ensuremath{\rm{erg\;cm}^{-2}\;\rm{s}^{-1}\rm{\AA}^{-1}\rm{arcsec}^{-2}}}
\newcommand{\cgsbri}{\ensuremath{\rm{erg\;cm}^{-2}\;\rm{s}^{-1}\rm{arcsec}^{-2}}}
\newcommand{\omegam}{$\Omega_{m,0}$\/}
\newcommand{\omegal}{$\Omega_{\Lambda,0}$\/}

\newcommand{\Mbh}{\ensuremath{M_{\bullet}}}
\newcommand{\Mdot}{\ensuremath{\dot{M}}}
\newcommand{\Msun}{\ensuremath{{\rm M}_{\odot}}}

\makeatletter
\newcommand\iont[2]{{#1$\;${\small\expandafter\@slowromancap\romannumeral #2@\relax}}}
\makeatother
\makeatletter
\newcommand{\raisemath}[1]{\mathpalette{\raisem@th{#1}}}
\newcommand{\raisem@th}[3]{\raisebox{#1}{$#2#3$}}
\makeatother

\slugcomment{Published as ApJ, 2017, ApJ, 844, 69}

\shorttitle{CHEERS Results from NGC 3393, II}
\shortauthors{Maksym et al.}

\begin{document}


\title{CHEERS Results from NGC 3393, II:\\ Investigating the Extended Narrow Line Region using Deep {\it Chandra} Observations and {\it Hubble} Narrow Line Imaging}


\author{W. Peter Maksym, Giuseppina Fabbiano, Martin Elvis, Margarita Karovska, Alessandro Paggi, John Raymond}
\affil{Harvard-Smithsonian Center for Astrophysics, 60 Garden St., Cambridge, MA 02138, USA}
\email{walter.maksym@cfa.harvard.edu; @StellarBones}

\author{Junfeng Wang}
\affil{Department of Astronomy, Physics Building, Xiamen University Xiamen, Fujian, 361005, China}

\and

\author{Thaisa Storchi-Bergmann}
\affil{Departamento de Astronomia, Universidade Federal do Rio Grande do Sul, IF, CP 15051, 91501-970 Porto Alegre, RS, Brazil}



\begin{abstract}

%@arxiver{f1c.jpg,f10.jpg}

The {\it CHandra Extended Emission Line Region Survey} (CHEERS) is an X-ray study of nearby active galactic nuclei (AGN) designed to take full advantage of {\it Chandra}'s unique angular resolution by spatially resolving feedback signatures and effects.  In the second paper of a series on CHEERS target NGC 3393, we examine deep high-resolution {\it Chandra} images and compare them with {\it Hubble} narrow line images of [\ion{O}{3}],  [\ion{S}{2}] and H$\alpha$, as well as previously-unpublished mid-ultraviolet (MUV) images.  The X-ray provide unprecedented evidence that the S-shaped arms which envelope the nuclear radio outflows extend only $\lesssim0.2\arcsec$ ($\lesssim50 pc$) across.  The high-resolution multiwavelength data suggest that the ENLR is a complex multi-phase structure in the circumnuclear ISM.  Its ionization structure is highly stratified with respect to outflow-driven bubbles in the bicone and varies dramatically on scales of $\sim10$ pc.  Multiple findings show likely contributions from shocks to the feedback in regions where radio outflows from the AGN most directly influence the ISM.  These findings include H$\alpha$ evidence for gas compression and extended MUV emission, and are in agreement with existing STIS kinematics.  Extended filamentary structure in the X-rays and optical suggests the presence of an undetected plasma component, whose existence could be tested with deeper radio observations.

\end{abstract}


\keywords{galaxies: active --- galaxies: individual (NGC 3393) --- galaxies: jets --- galaxies: Seyfert --- X-rays: galaxies}


\section{Introduction}

Active Galactic Nuclei (AGN) are well-known to play an important role in the evolution of galaxies, where a supermassive black hole (SMBH) central engine converts gravitational energy into radiative or kinetic energy. Kinematic outflows such as radio-emitting jets, or conically-shaped regions of highly ionized gas extending from the AGN itself have been associated with these nuclei.  The ionization cones emit brightly in narrow emission lines from species across the electromagnetic spectrum such as [\ion{O}{3}], or at higher photon energies such as in X-rays \citep[see, e.g.][for recent reviews]{HB14,Netzer15}.

This AGN Narrow Line Region (NLR) provides a major tool for studying the interaction of the AGN with its host galaxy on scales of hundreds of pc to $\sim1\;$kpc or more.  At larger scales ($\sim$few kpc) and in the absence of large-scale jets, we may see Extended Emission Line Regions (EELRs) where line excitation is dominated by a photoionizing spectrum emitted from the accretion disk surrounding the SMBH itself, and these EELRs may provide clues into the radiative history of the AGN on timescales of the light-crossing time of the galaxy \citep{Keel12,Keel15}.  However, the common presence of small-scale jets may complicate this picture, transferring accretion-powered kinematic energy into the surrounding gas and thereby stimulating X-ray or optical emission \citep{Wang11a,Wang11b,Wang11c,Paggi12,Keel15,Sartori16} on scales of the Extended Narrow Line Region (ENLR) or smaller.  Such small-scale outflows may be related to similar phenomena observed with regularity in stellar-mass black holes, whereby the accretion state switches between radiative and kinematic modes dependent in part upon the current Eddington fraction \citep[e.g.][]{FBG04}.

The processes forming the NLR and ENLR provide a particularly important window through which to understand AGN feedback, the process by which an AGN regulates its accretion rate from surrounding gas via radiative and kinematic processes \citep[see][for a review]{Fabian12}.  Via such feedback processes, the AGN can heat or eject its fuel supply, thereby reducing its accretion rate, and possibly starve itself completely.

The difficulty involved in modeling the ENLR implies considerable complexity.  In order to tease out the relative contributions of different feedback processes and the different states of ENLR gaseous media, we have undertaken a {\it CHandra survey of Extended Emission line Regions in nearby Seyfert galaxies} ({\it CHEERS}).  Using {\it Chandra} X-ray data of sufficient signal-to-noise to take advantage of {\it Chandra}'s sub-pixel resolving capability coupled with narrow-line {\it Hubble} images, we are able to distinguish X-ray and optical line-emitting regions on scales of $\sim50\;$pc, in order to distinguish between photo-ionized and shocked feedback regimes.

This work builds upon the study of NGC 4151 by \cite{Wang11a,Wang11b, Wang11c}, and that of Mkn 573 by \cite{Paggi12}.  In both cases, comparison of {\it Chandra}, {\it Hubble} and {\it VLA} images led to the identification of spatially resolved photoionized regions and collisional gas between its jet radio lobes and optical arcs.  Here, we investigate the galaxy \object{NGC 3393} in the second paper of a series on that object \citep{Maksym16}.

At $z=0.0125$, NGC 3393 is a nearby, bright \citep[$m_B=13.1$][]{deV91} Seyfert 2 galaxy.  Like Mkn 573, NGC 3393 has prominent S-shaped emission line arcs associated with a triple-lobed radio source within $\sim$kpc-scale ionization cones \citep{Cooke00}.  NGC 3393 is also Compton thick, as supported by observations from BeppoSAX \citep{Maiolino98}, {\it XMM-Newton} \citep{Guainazzi05}, the {\it Swift} Burst Alert Telescope \citep[][$N_H\sim4.5\times10^{24}\;\rm{cm}^{-2}$]{Burlon11}, and {\it NuSTAR} \citep{Koss15}.  

\cite{Cooke00} previously used {\it Hubble} pre-COSTAR narrow filter optical imaging of [\ion{O}{3}] and H$\alpha$+[\ion{N}{2}], as well as {\it Hubble} Faint Object Spectrograph (FOS) and ground-based spectroscopy, and {\it VLA} radio data, to identify the S-shaped arcs in NGC 3393 and to study the galaxy's extended narrow line emission.  They determined the predominant method of ENLR excitation was likely to be photoionization, but could not rule out a role for shocks.

Subsequent resolved X-ray studies using {\it Chandra} \citep{Bianchi06,Levenson06} determined that the soft X-rays associated with the AGN are extended on scales of $\sim2\;$kpc and show strong morphological correlation with the extended  [\ion{O}{3}] features.  \cite{Bianchi06} suggest this correlation points to origins of a single photoionized medium giving rise to the [\ion{O}{3}]  and X-rays.  This correlation is supported by \cite{Koss15}, who find that deeper {\it CHEERS} observations and zeroth-order {\it Chandra} imaging support prior associations between [\ion{O}{3}], X-rays, and radio emission.  

In \cite{Maksym16}, we used continuum-subtracted  {\it HST} narrow-line observations to demonstrate that the ENLR of NGC 3393 was predominantly Seyfert-like with a LINER cocoon surrounding the nuclear bicone-and-jets structure.  Here, we produce a more extensive analysis of the {\it HST} narrow-line observations, complemented by X-ray images which take advantage of {\it Chandra's} $\sim0.2\arcsec$ mirror resolution.  We directly compare the spatial distribution of [\ion{O}{3}], [\ion{S}{2}] and H$\alpha$ using the {\it CHEERS} post-COSTAR data, and expand upon previous work by \cite{Cooke00} and \cite{Koss15} by using [\ion{O}{3}]/H$\alpha$ and [\ion{S}{2}]/H$\alpha$ ratio maps to distinguish regions of relative ion species dominance.  These line ratio maps are of interest because [\ion{O}{3}]/H$\alpha$ typically indicates photoionization, while [\ion{S}{2}]/H$\alpha$ is a common indicator of shocks or enhanced density, and hence kinematic feedback.

The {\it CHEERS} WFC3 data are deeper than the narrow line images used by \cite{Cooke00} and do not require deconvolution, allowing more detailed investigation into the narrow line morphology.  \cite{Koss15} note a general correspondence between \ion{O}{3} and X-ray emission, as well as the expected role of the radio jets in shaping the ENLR, but here we examine the ENLR X-ray morphology in greater detail, examining the physical origin of ENLR substructure.  Such inquiry is enabled by our use of EMC2 Bayesian deconvolution \citep[][see Fig. \ref{fig:xray}]{esch04,karovska05,karovska07} with respect to the X-ray images, as well as our analysis of the \ion{S}{2} images, which were not used by \cite{Cooke00} or \cite{Koss15}.

In \S\ref{sec-obs}, we summarize the origin and processing of our {\it HST}, {\it Chandra} and {\it VLA} observations of NGC 3393.  In \S\ref{sec-im-an}, we describe the methods we use to obtain deconvolved X-ray images and narrow line image maps from processed data.  In \S\ref{sec-results}, we examine and compare the relative morphologies of the resulting narrow-line, X-ray and radio images, particularly with respect to the inner $\sim2\;$kpc of the galaxy.  In \S\ref{sec-disc}, we discuss the physical implications of the multi-wavelength morphology in the NLR and ENLR and examine the physical implications for the origins of this emission.  In \S\ref{sec-con}, we summarize our results and the implications for future research.

Throughout this paper, we adopt concordant cosmological parameters\footnote{Distances are calculated according to http://www.astro.ucla.edu/~wright/CosmoCalc.html} of
$H_0=70\ $km$^{-1}$ sec$^{-1}$ Mpc$^{-1}$, \omegam=0.3 and \omegal=0.7. All coordinates are J2000.  In all figures, celestial North is up and color scales are logarithmic (unless otherwise noted).  For distance evaluation we use the \cite{Theureau98} determination of redshift $z=0.0125$ from observations of the 21-cm neutral hydrogen emission line, such that NGC 3393 is at distance $D=53\;$Mpc with linear scale $257\;\rm{pc\,arcsec}^{-1}$.
%, and calculate distances using \cite{Wright06}. All coordinates are J2000.  

\section{Observations and Data Reduction}\label{sec-obs}

\subsection{ACIS/Chandra Data}

\begin{table*}
\centering
\caption{Chandra Observation Properties}
\label{table:cxo-obstable}
\vspace{0.1in}
\begin{tabular}{cccccc}
\tableline
\tableline
Obsid	&	Obs Date	&	Exposure (ks)	 & Grating & Net Counts$^{\rm a}$ 	& Net Counts$^{\rm a}$ \\
		&				&			& 		& 0.3$-$2 keV 			&	2$-$10 keV	\\
%		&							&\multicolumn{3}{c}{($\times10^{-14}$\ecmss)}	\\
\tableline
04868	&	2004 Feb 28	& 29.33	&	NONE	& $1959\pm45$ & $110\pm17$	\\
12290	&	2011 Mar 12	& 69.16	&	NONE	&  $3688\pm62$& $183\pm26$	\\
\tableline
Imaging	&	... 			& 98.49	&	NONE	&  $5647\pm77$& $293\pm31$	\\
\tableline 
13967	&	2012 Feb 29	&  176.78	&	HETG	&  $594\pm26$& $540\pm31$	\\
14403	&	2012 Mar 06	& 77.72	&	HETG	&  $264\pm18$& $214\pm20$	\\
14404	&	2012 Apr 02	& 56.68	&	HETG	&  $180\pm15$& $164\pm17$	\\
13968	&	2012 Apr 08	& 28.06	&	HETG	&  $88\pm10$ & $113\pm13$	\\
\tableline
Gratings	&	... 			& 339.24	&	HETG	&  $1126\pm36$& $1031\pm43$	\\
\tableline
Merged	&				&437.73	&	MERGE	& $6773\pm85$ & $1324\pm53$ \\

%{\tt zpowerlw} &	$\Gamma$	& $2.88\pm0.28$	& $5.2\pm1.1$	& $5.45/5$\\
%{\tt zbbody} &	$kT_{bb}$ (keV)	& $0.11\pm0.02$	& $2.4\pm0.6$	& $6.40/5$\\
 %{\tt diskbb} &	$kT_{bb}$	(keV)	& $0.17\pm0.04$	& $2.9\pm0.7$	& $5.60/5$\\

\tableline

\end{tabular}
%% Any table notes must follow the \end{tabular} command.\\
\tablenotetext{a}{Counts extracted from a circular region of radius 15\arcsec.}
\end{table*}

NGC 3393 was observed by the {\it Chandra} Advanced CCD Imaging Spectrometer (ACIS) six times (See Table \ref{table:cxo-obstable}); of these observations, four used the High Energy Transmission Grating (HETG).  The two imaging observations were centered on the back-illuminated S3 chip of the ACIS-S array: once on 2004 February 28 for 29 ks (obsid 4868) and once on 2011 March 12 for 69 ks (obsid 12290).  Obsid 12290 was performed as part of the CHEERS survey.   The source was also observed four times with the High Energy Transmission Grating (HETG), on 2012 February 29 (obsid 13967), 2012 March 06 (obsid 14403), 2012 April 02 (obsid 14404), and 2012 April 08 (obsid 13968).  The observations performed using the grating produce a zeroth-order image which is functionally similar to imaging observations, and we include these observations in some of our analysis.  We retrieved the data from the {\it Chandra} Data Archive\footnote{http://cda.harvard.edu/chaser} and reduced them using the {\it Chandra} Interactive Analysis of Observations package \citep[CIAO;][]{ciao} version 4.7, and version 4.6.8 of the {\it Chandra} Calibration Data Base (CALDB).   In order to take better advantage of the angular resolution of the {\it Chandra} High Resolution Mirror Assembly (HRMA), which has $\rm{FWHM}\sim0.2\arcsec$ but is undersampled by the native $\sim0.492\arcsec$ ACIS pixels, we reprocessed the data using the ACIS Energy-Dependent Subpixel Event Repositioning algorithm \citep[EDSER;][]{edser}.  EDSER removes the artificially-introduced position blurring of the ACIS pipeline and instead takes advantage of event charge patterns and aspect dithering to improve event position estimates, and has been used reliably in many cases \citep[e.g.][]{Harris04,Wang11a,Paggi12}.

We used the CIAO tool {\tt wavdetect} to identify point sources in the field-of-view, which we exclude from analysis of extended emission.  Examining only source-free regions, we excluded periods of high background by using the {\tt deflare} CIAO tool to identify field count rates in excess of $3\sigma$ above the quiescent level (though in practice these `flaring periods' were negligible).  To determine the significance of pileup, we generated a pileup map using the {\tt pileup\_map} tool.  Pileup is minor, reaching $\sim3\%-4\%$ within $\sim1\arcsec$ of the nucleus.  We corrected astrometry relative to the longest imaging observation (obsid 12290) by using {\tt wcs\_match} to compare source lists across observations, then adjusted the aspect and event file astrometry using CIAO tools {\tt reproject\_aspect} and {\tt wcs\_update}.  

In order to identify and examine fainter features, we created merged event files using {\it merge\_obs}.  The Point Spread Function (PSF) for zeroth-order images using ACIS-HETG images is not as well understood at subpixel scales as for non-grating images.  We therefore generate separate merged event files and images for pure imaging and zeroth-order imaging observations and treat the pure imaging observations as our primary dataset.  We note, however, that we observe good agreement between pure imaging and zeroth-order imaging observations on scales of $\sim1\arcsec$, as allowed by the limits of photon statistics. 

\subsection{Optical/UV/IR Data}

\begin{table*}
\centering
\caption{Hubble Observation Properties}
\label{table:hst-obstable}
\begin{tabular}{cccccc}
\tableline
\tableline

Dataset	&	Obs Date	&	Exposure (s)	 & Instrument & Filter & Note	\\
%		&							&\multicolumn{3}{c}{($\times10^{-14}$\ecmss)}	\\
\tableline
5730			&	1994 Nov 22	& 260	&	WFPC2	&	F218W	& MUV	\\
IBIG06010	&	2011 Nov 11	& 147	&	WFC3/IR	& F110W & $1.1\mu\rm{m}$	\\
IBIG06020	&	2011 Nov 11	& 422	&	WFC3/IR	&  F160W& $1.6\mu\rm{m}$	\\
IBIG06030	&	2011 Nov 11	& 1230	&	WFC3/UVIS	&  F336W & NUV 	\\
IBIG06040	&	2011 Nov 11	&  444	&	WFC3/UVIS	&  F438W & B-band	\\
IBIG06050	&	2011 Nov 11	& 2040	&	WFC3/UVIS	&  F814W & I-band	\\
\tableline
IBLY01011	&	2011 May 16	& 566	&	WFC3/UVIS	&  FQ508N & [\ion{O}{3}]	\\
IBLY01021	&	2011 May 17	& 466	&	WFC3/UVIS	&  F665N & H$\alpha$+[\ion{N}{2}]	\\
IBLY01GWQ	&	2011 May 17	& 208	&	WFC3/UVIS	&  F547M & line continuum	\\
IBLY01GXQ	&	2011 May 17	& 208	&	WFC3/UVIS	&  F621M & line continuum	\\
IBLY01GYQ	&	2011 May 17	& 314	&	WFC3/UVIS	&  F673N & [\ion{S}{2}]	\\

%{\tt zpowerlw} &	$\Gamma$	& $2.88\pm0.28$	& $5.2\pm1.1$	& $5.45/5$\\
%{\tt zbbody} &	$kT_{bb}$ (keV)	& $0.11\pm0.02$	& $2.4\pm0.6$	& $6.40/5$\\
 %{\tt diskbb} &	$kT_{bb}$	(keV)	& $0.17\pm0.04$	& $2.9\pm0.7$	& $5.60/5$\\

\tableline

\end{tabular}
%% Any table notes must follow the \end{tabular} command.\\
\tablenotetext{}{IBIG exposures are from program 12185 (PI: Greene).  IBLY exposures are from CHEERS, program 12365 (PI: Wang). Program 5730 is previously unpublished data observed by PI: Baldwin.}
\end{table*}

To compare the spatial distribution of X-ray emission from NGC 3393 with optical emission, we also obtained {\it HST} images from the Hubble Legacy Archive\footnote{http://hla.stsci.edu/hlaview.html}.  NGC 3393 has been extensively observed by {\it HST}, but in particular we examined CHEERS data (program 12365, PI: Wang) obtained using WFC3 in the UVIS mode.  CHEERS observations were taken on 2011 May 16 and 17 in filters FQ508N, F665N, F547M, F621M, and F673N for 566s, 466s, 208s, 208s and 314s respectively (see Table \ref{table:hst-obstable}).

The quad filter FQ508N covers a narrow $\sim42$\AA\ range around [\ion{O}{3}]$\lambda5007$\AA\ at the redshift of NGC 3393 and is cleanly separated from other prominent emission lines (notably [\ion{O}{3}]$\lambda4959$\AA).  F665N spans $\sim42$\AA\ around redshifted H$\alpha$,$\lambda6563$, and includes [\ion{N}{2}]$\lambda\lambda6548,6584$.  F673N covers $\sim42$\AA\ around the redshifted [\ion{S}{2}]$\lambda\lambda$6716,6731 doublet.  Measurements for the [\ion{O}{3}]  continuum are taken from the F547M band, while the continuum of H$\alpha$+[\ion{N}{2}] and [\ion{S}{2}] is inferred from F621M, both of which we expect to be relatively free of prominent emission lines.
We also use the 2011 November 11 observations taken for {\it HST} program 12185 (PI: Greene) in F336W (NUV), F438W (B-band), F814W (I-band), F110W ($1.1\mu$m) and F160W ($1.6\mu$m), and older but previously unpublished {\it HST} WFPC2 images taken on 1994 November 22 using F218W (PI: Baldwin).

We processed the {\it HST} images according to the same techniques described in \cite{Maksym16}, using the standard {\it HST} data processing package AstroDrizzle \citep{astrodrizzle} for Pyraf installed through the version 1.5.1 release of Ureka\footnote{http://ssb.stsci.edu/ureka/}.  As before, we were unable to correct the images for charge transfer efficiency (CTE).

 {\it Spitzer} 3.5$\,\mu$m IRAC data were observed under Program \#10098, PI: Stern, and are taken from the {\it Spitzer} Heritage Archive\footnote{http://sha.ipac.caltech.edu/applications/Spitzer/SHA/}.

\subsection{Radio Data}

\begin{table}
%\centering
\caption{VLA Observation Properties}
\label{table:vla-obstable}
\begin{tabular}{ccc}
\tableline
\tableline

Obs Date	&	Frequency (GHz)	 & Resolution (\arcsec) 	\\
%		&							&\multicolumn{3}{c}{($\times10^{-14}$\ecmss)}	\\
\tableline
1991 Oct 11 &		1.51 &	2.77\\
1991 Oct 11 &		4.89 &	0.83\\
1991 Oct 11 &		8.46 &	0.50\\
1992 Nov 29 &	1.45 &	1.79\\
1992 Nov 29 &	4.89 &	0.50\\
1992 Nov 29 &	8.46 &	0.29\\
1993 Feb 4 &		4.89 &	1.04\\
\tableline

\end{tabular}
%% Any table notes must follow the \end{tabular} command.\\
%\tablenotetext{}{IBIG exposures are from program 12185 (PI: Greene).  IBLY exposures are from CHEERS, program 12365 (PI: Wang). }
\end{table}

To analyze radio emission from NGC 3393, we obtained {\it VLA} images from the NRAO Science Data Archive\footnote{https://archive.nrao.edu/archive/archiveimage.html}.
NGC 3393 has been observed by the VLA at 1.51 GHz, 4.89 GHz and 8.46 GHz on 1991 October 11, at 1.45 GHz, 4.89 GHz and 8.46 GHz on 1992 November 29, and at 4.89 GHz on 1993 February 4.  The angular resolution of these images ranges between $0.29\arcsec$ for 8.46 GHz on 1992 November 29 and $2.77\arcsec$ for 1.51 GHz on 1991 October 11.  These observations are summarized in Table \ref{table:vla-obstable}.

We examine the 1.51 GHz image and see that the central nuclear structure is marginally resolved, with no evidence for extended structure at $>59\mu\rm{Jy\;beam}^{-1}$ which might otherwise be ``resolved out" at higher resolutions.  Other observations show compact northeastern and southwestern radio lobes at $\simless1.5\arcsec$ from the nucleus.  We will henceforth exclusively use the 8.46 GHz images with $0.29\arcsec$ resolution, since we are primarily interested in structural morphology of the circumnuclear region, and all other images lack additional structural information due to lack of resolution or elongation of the beam PSF.

\section{Image Analysis}\label{sec-im-an}

\subsection{Chandra PSF Modeling and Image Deconvolution}

\begin{figure*} 
\noindent
\includegraphics[width=0.32\textwidth]{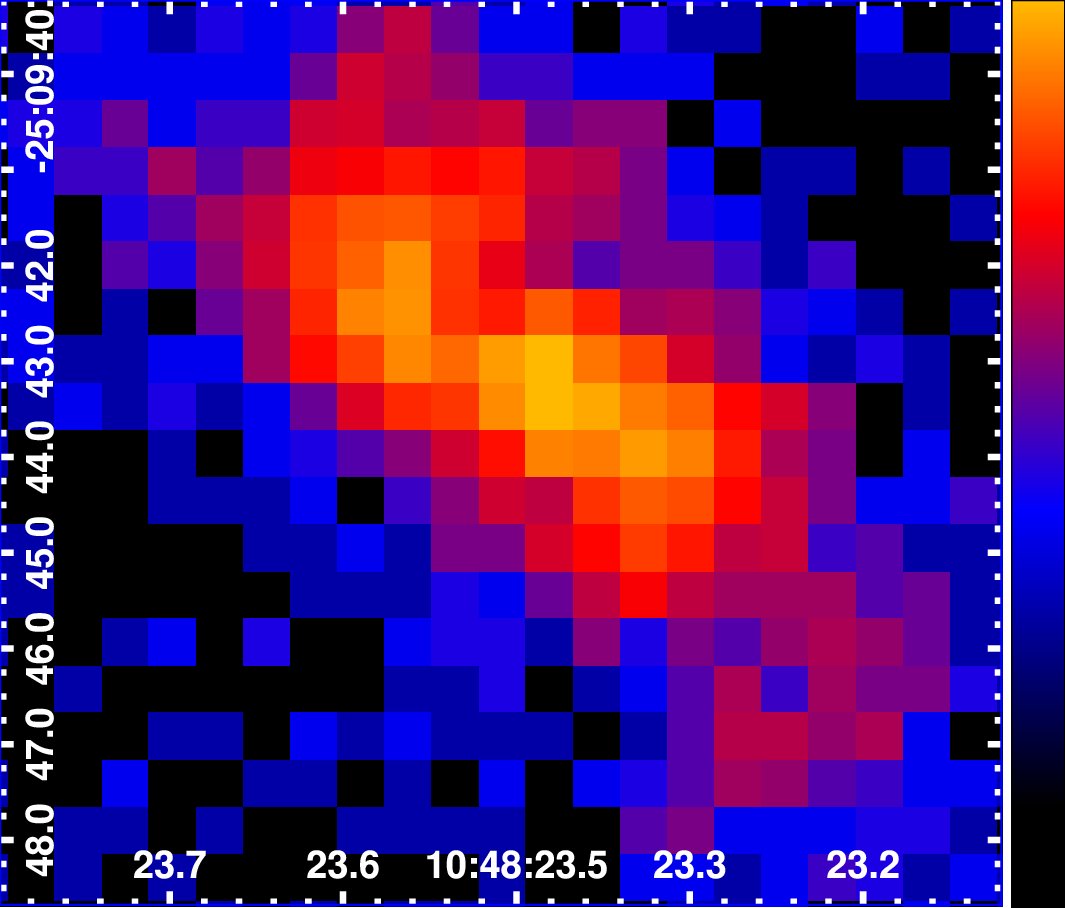}\hspace{0.01\textwidth}%
\includegraphics[width=0.32\textwidth]{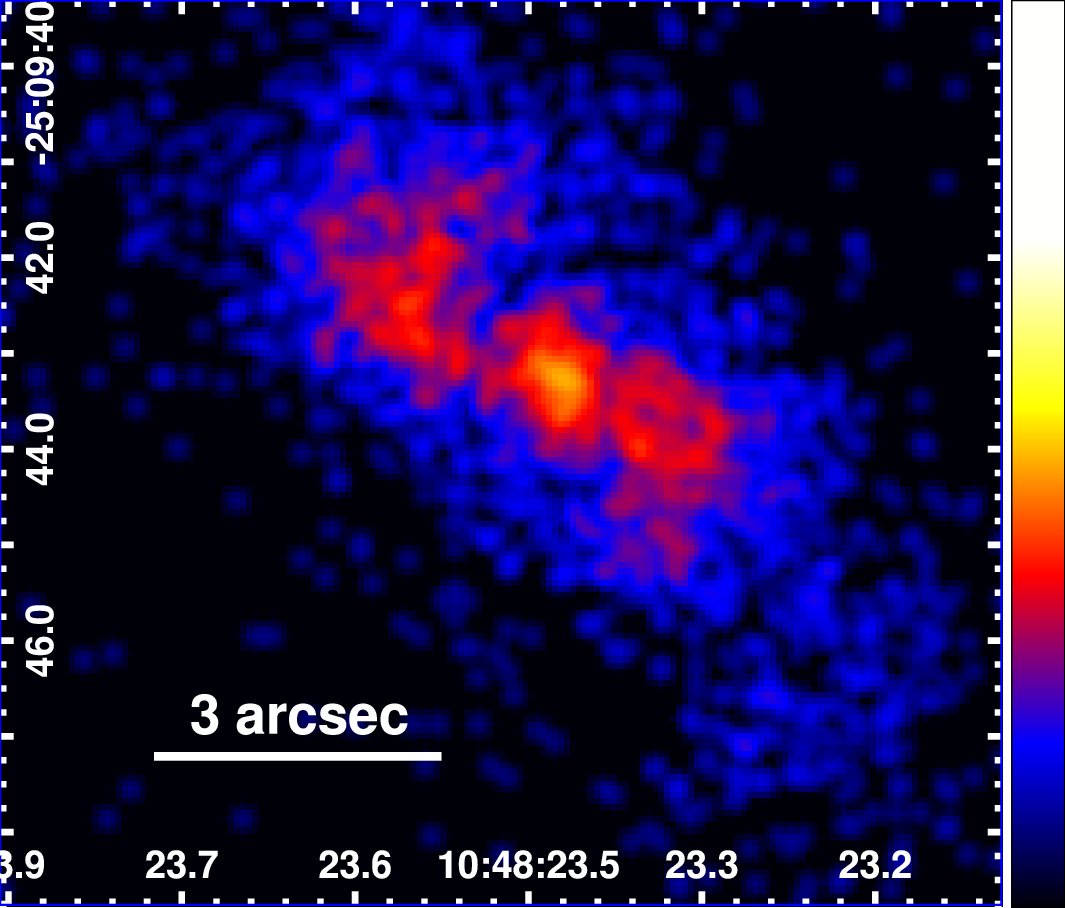}\hspace{0.01\textwidth}%
\includegraphics[width=0.32\textwidth]{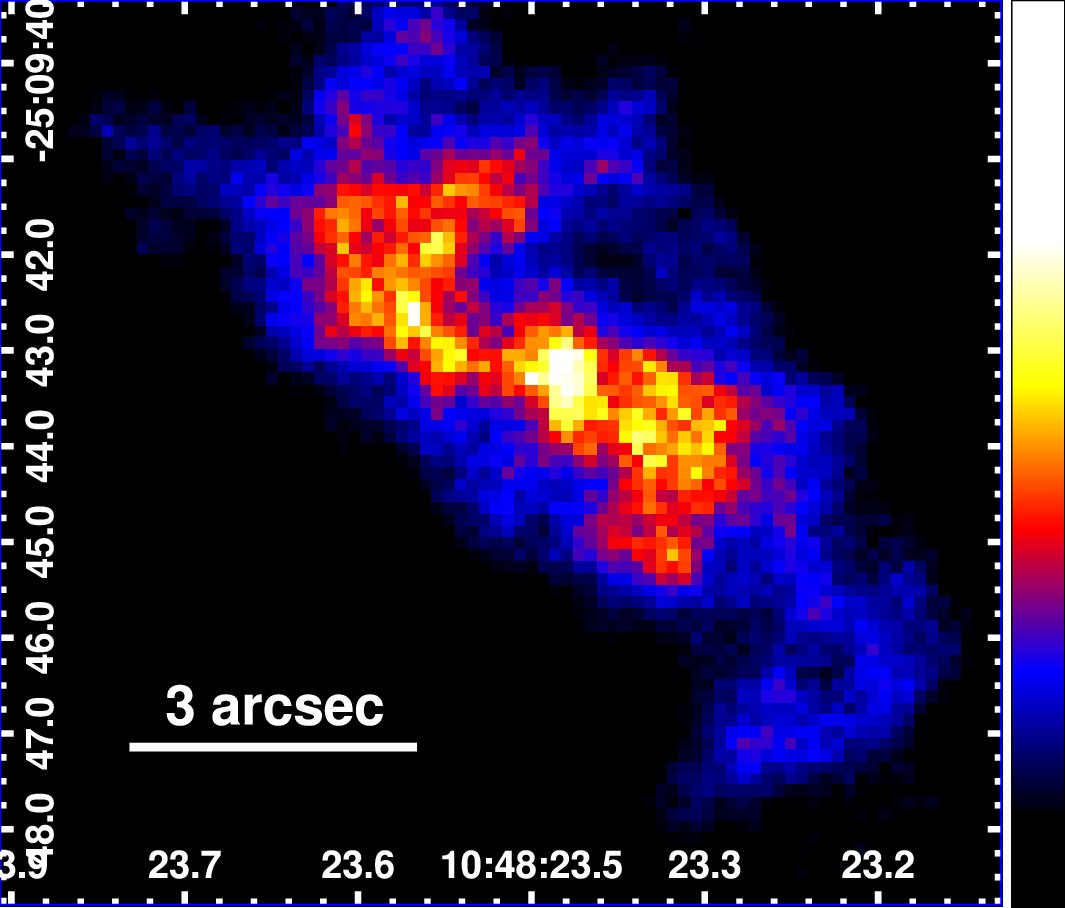}\par
\caption{X-ray images ($0.3-8.0\,$keV) of  NGC 3393 binned to native {\it Chandra} pixel size (left), binned to 1/8 native pixel size [$0.062\arcsec$] and smoothed with a Gaussian of FWHM 3 image pixels [$0.186\arcsec$] (middle), and deconvolved using EMC2 from an image with 1/4 native pixel bins [$0.124\arcsec$] (right).  Although 1/8 pixel scale oversamples the {\it Chandra} PSF, the 1/4 pixel scale is the practical limit for deconvolution given the counting statistics of the data.
} 
\label{fig:xray}
\end{figure*}

Following standard {\it Chandra} science threads\footnote{http://cxc.harvard.edu/ciao/threads/prep\_chart/}, we simulated the point spread function (PSF) of the imaging observations (obsids 4868 and 12290) using the {\it Chandra} Ray Tracer \citep[ChaRT\footnote{http://cxc.harvard.edu/chart/} v2;][]{chart}, taking into account source spectrum, off-axis angle, and aspect solution.  We then used MARX \citep{marx} to generate a PSF image from the raytraced simulation.   

%\subsection{Chandra Image Deconvolution}
\label{decon}

We deconvolved the X-ray image using the Expectation through Markov Chain Monte Carlo \citep[EMC2;][]{esch04,karovska05,karovska07} algorithm.  We used the longest  imaging observation (obsid 12290) in order to avoid systematic effects due to combining different PSFs in deconvolution.  Grating zeroth order images were not used for deconvolution because the PSF is not reliable on relevant scales in this mode.  We achieved the best-significance results from EMC2 using an input image with subpixel spatial binning at 1/4 of the native pixel size ($\sim0.124\arcsec$) using a broad band ($0.3-8.0\;$keV) for obsid 12290, excluding only high and low energies with typically poor ACIS signal-to-noise levels.  For comparison, we display the broad $0.3-8.0\;$keV image of NGC 3393 in Fig. \ref{fig:xray} binned at native ACIS pixel size, as well as a smoothed image binned to 1/8 native pixel size, and the broad image deconvolved at 1/4 native pixel size using EMC2.  Although 1/8 pixel scale oversamples the {\it Chandra} PSF, the 1/4 pixel scale is the practical limit for deconvolution given the counting statistics of the data.

\subsection{Hubble Narrow Line Mapping}

\begin{figure*} 
\noindent
\includegraphics[width=0.32\textwidth]{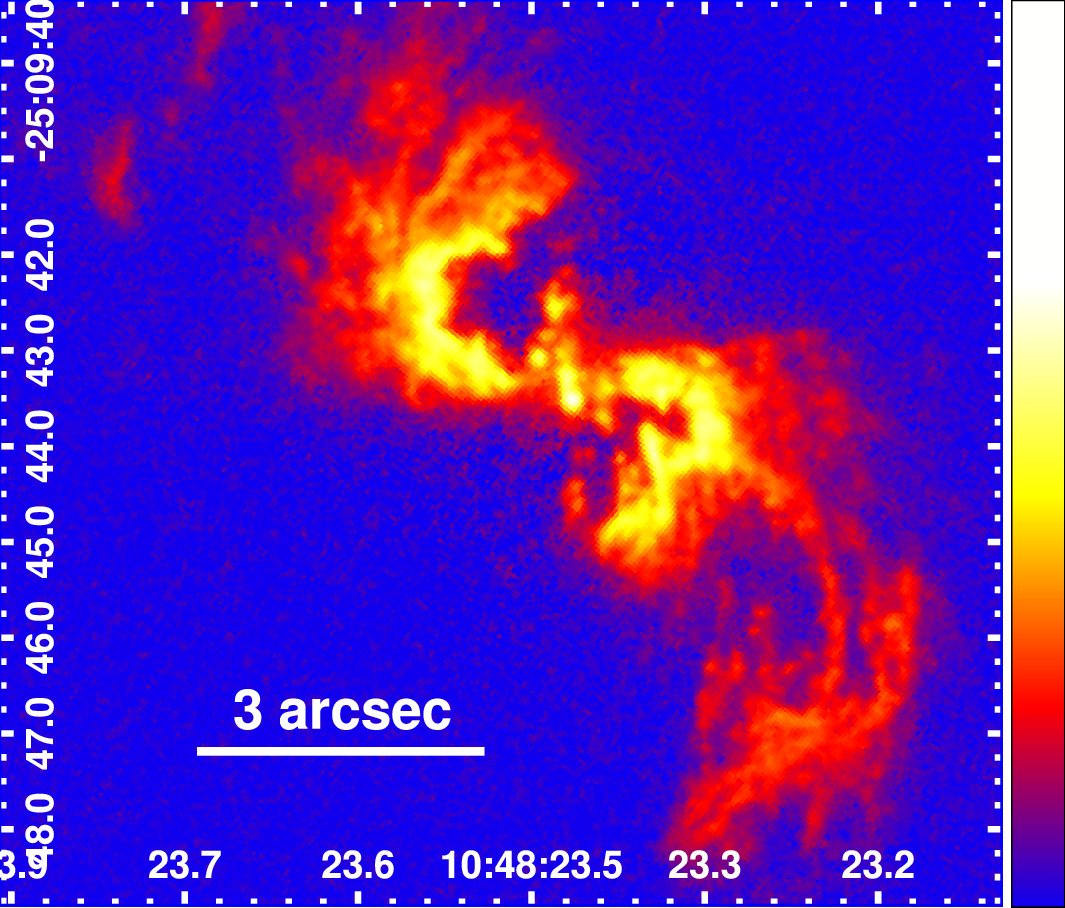}\hspace{0.01\textwidth}%
\includegraphics[width=0.32\textwidth]{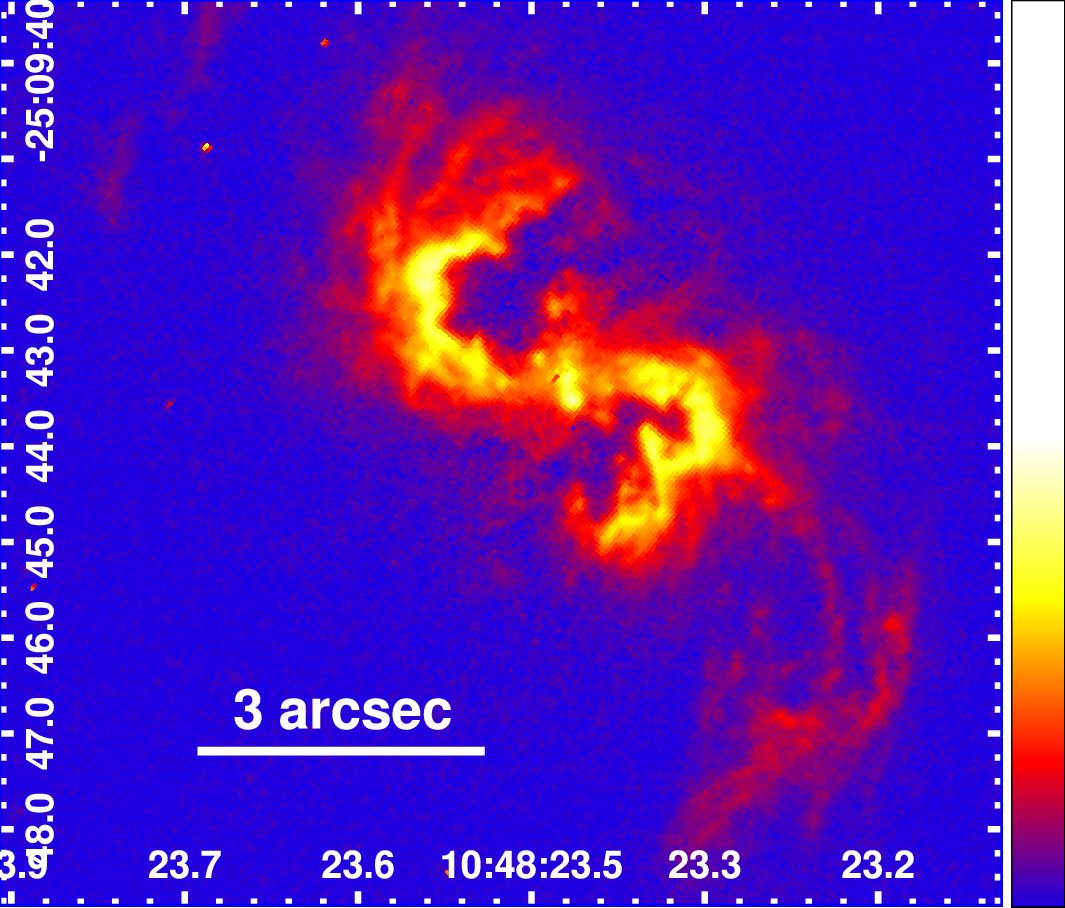}\hspace{0.01\textwidth}%
\includegraphics[width=0.32\textwidth]{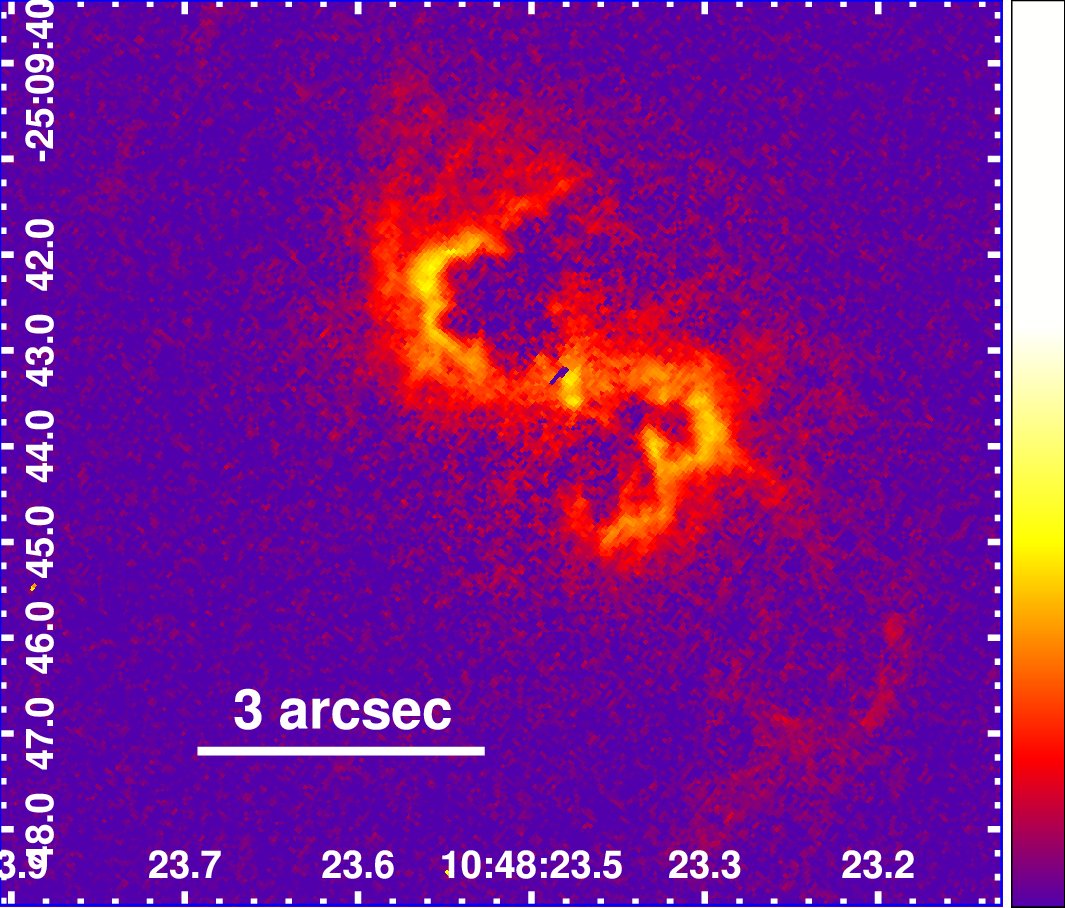}\par
\caption{Continuum-subtracted surface brightness images of  [\ion{O}{3}] (left), H$\alpha$ (middle) and [\ion{S}{2}] (right).  In H$\alpha$ and [\ion{S}{2}], there is a flaw near the nucleus a few pixels in size due to poor cosmic ray subtraction in the continuum band.
} 
\label{fig:line-sb}
\end{figure*}

\begin{figure*} 
\noindent
\includegraphics[width=0.45\textwidth]{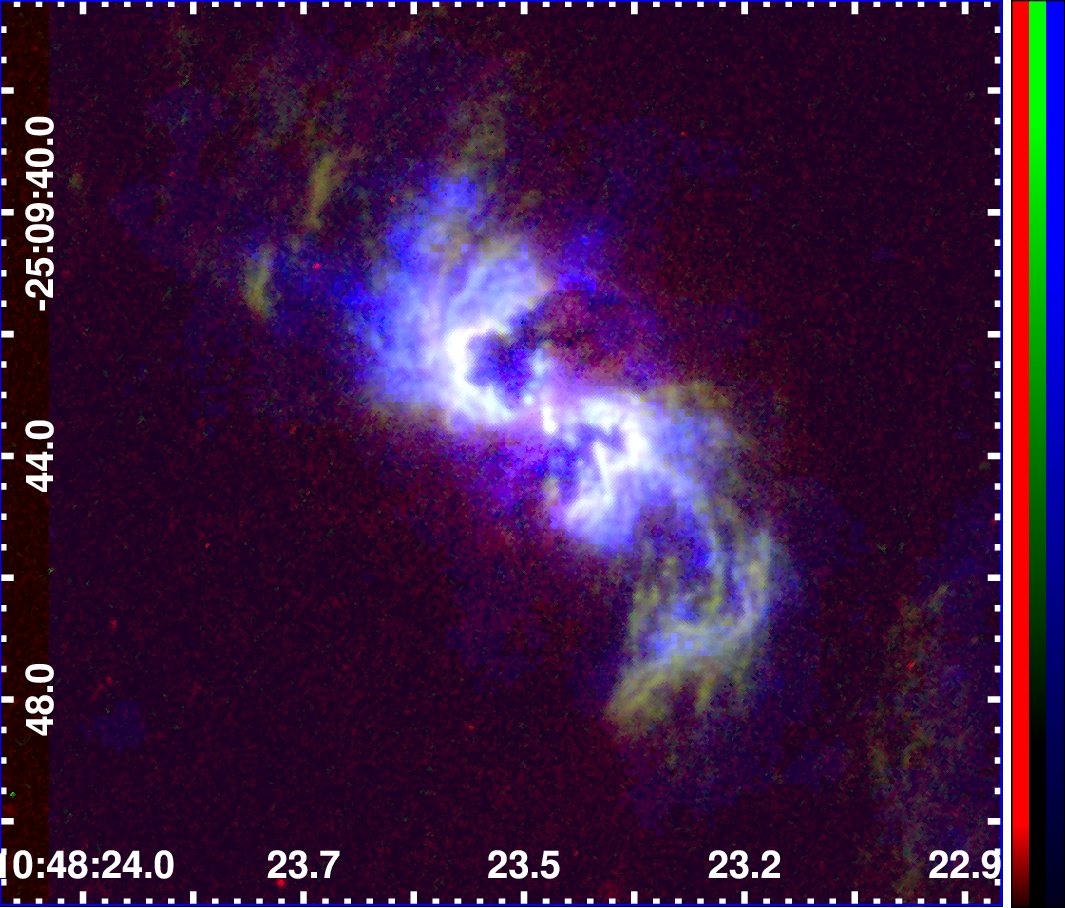}\hspace{0.1\textwidth}%
\includegraphics[width=0.45\textwidth]{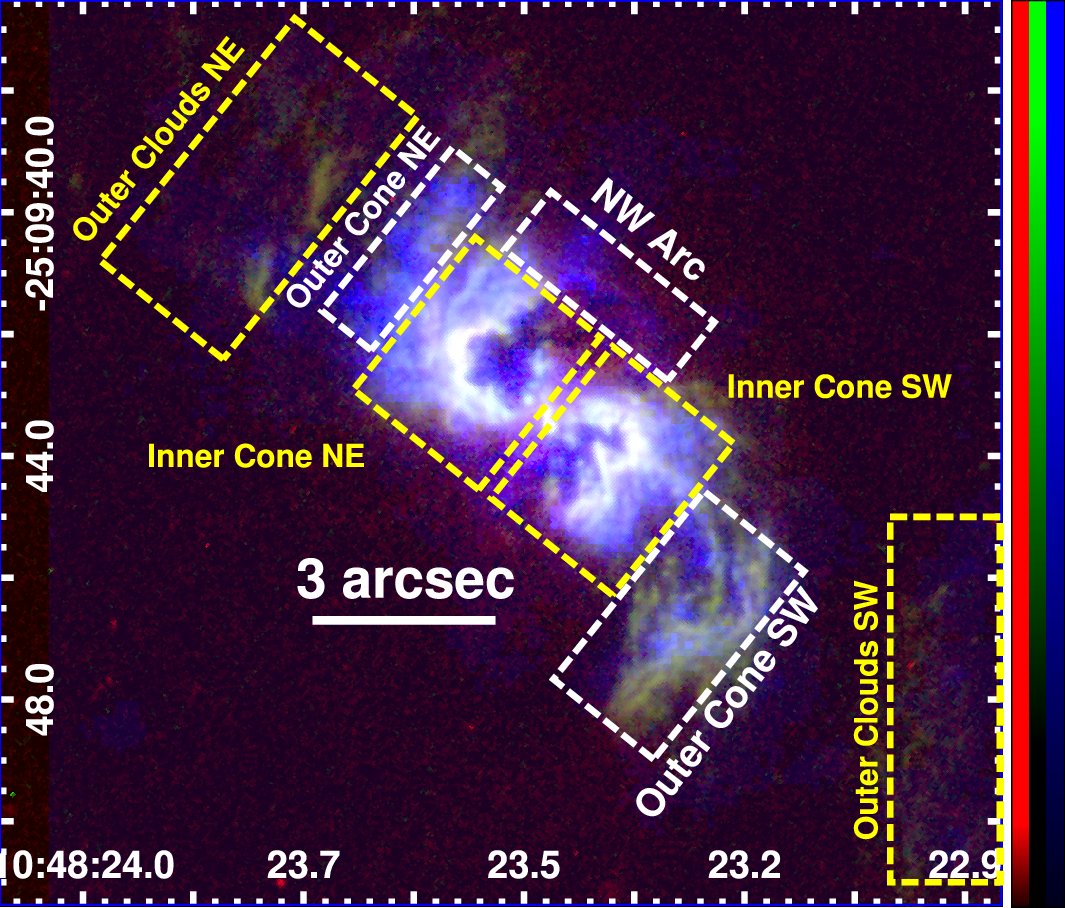}\par
\caption{Left: 3-color image of continuum-subtracted line emission from H$\alpha$ (red) and [\ion{O}{3}] (green), as well as EMC2-deconvolved 0.3-8\;keV X-ray emisson (blue). Right: same as left, but with boxes (dashed lines) to indicate regions discussed in \S\ref{hao3xray-text}.  Note for the outer clouds SW, the X-ray emission is parallel to [\ion{O}{3}] but  $\sim1\arcsec$ NW of the [\ion{O}{3}].} 
\label{fig:line-colors}
\end{figure*}

\begin{figure*} 
\noindent
\includegraphics[width=0.29\textwidth]{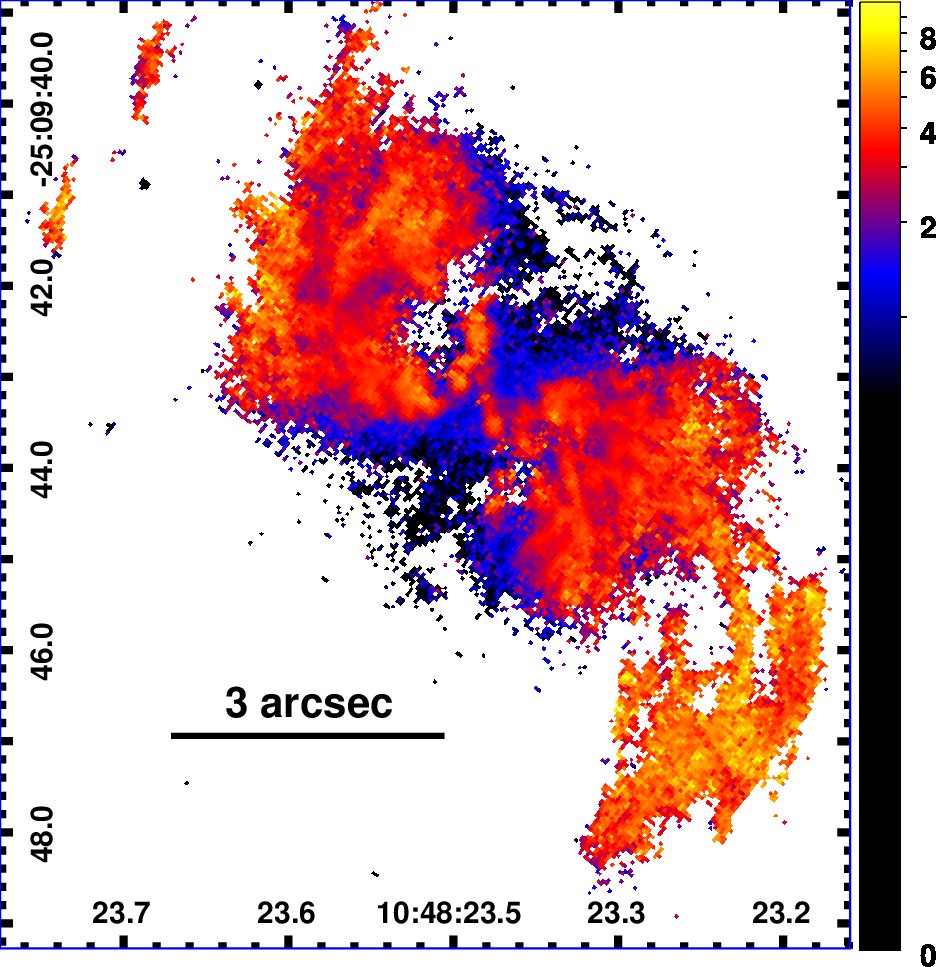}\hspace{0.05\textwidth}%
\includegraphics[width=0.31\textwidth]{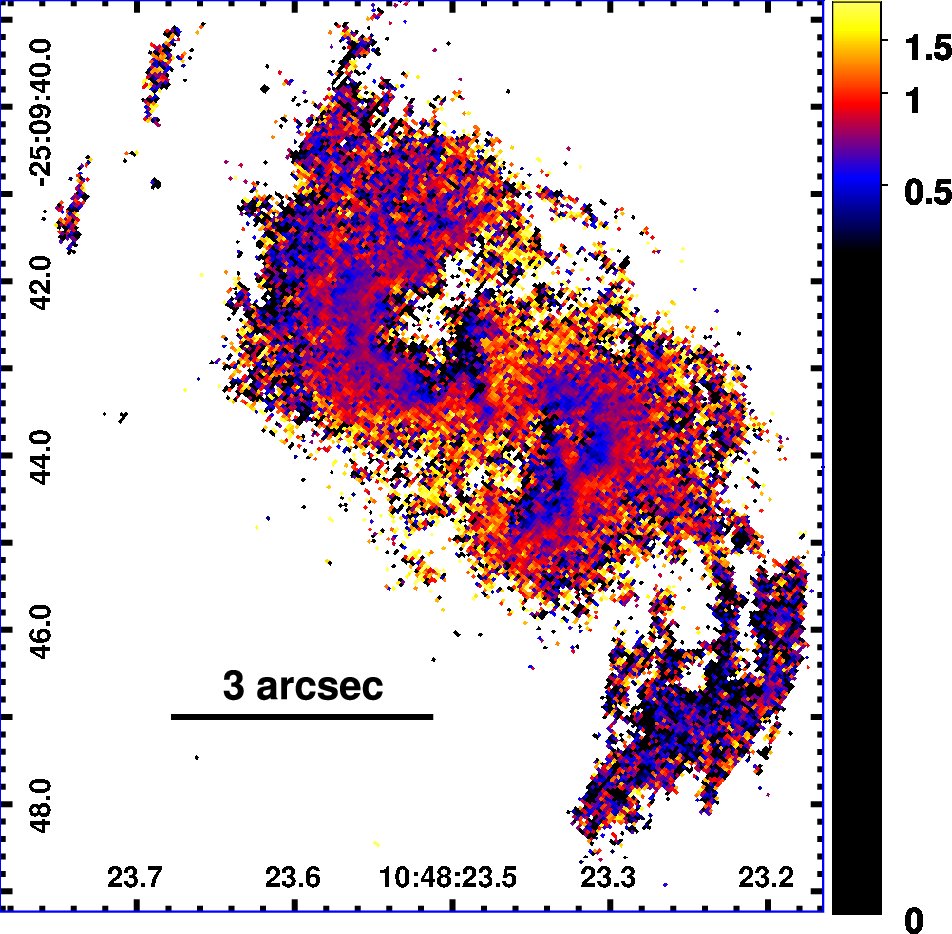}\hspace{0.05\textwidth}%
\includegraphics[width=0.3\textwidth]{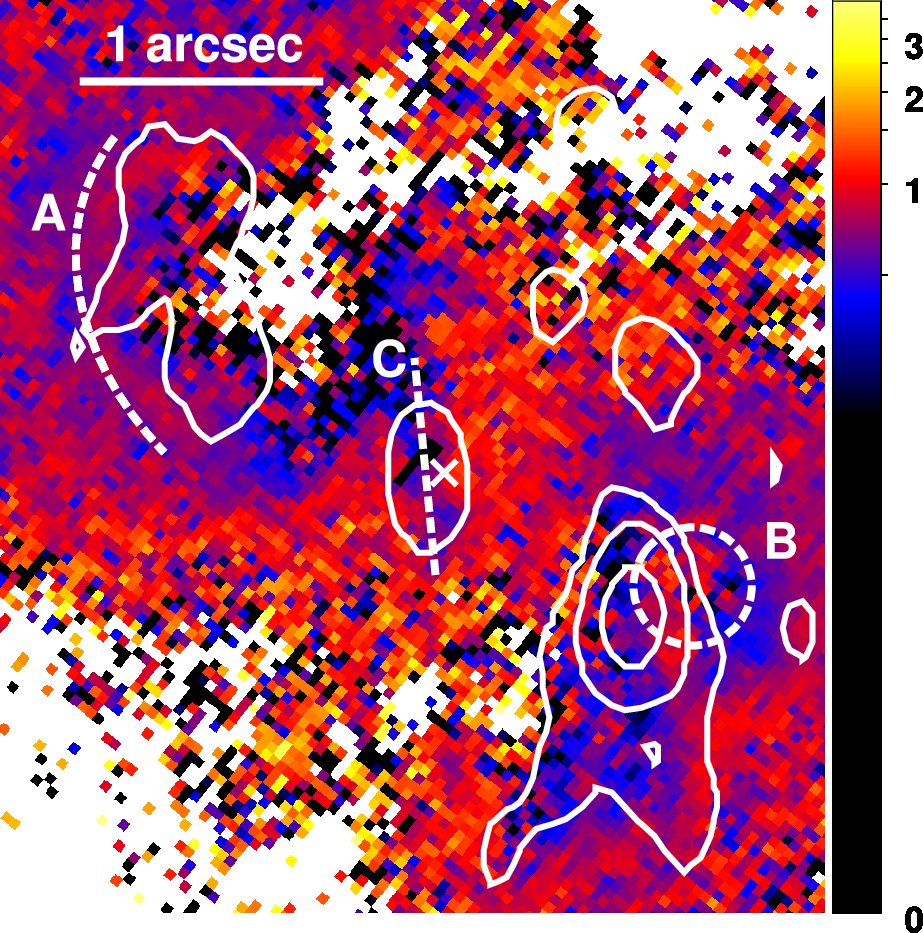}\par
\caption{Line ratio maps with black/blue indicating low ratios and yellow indicating high ratios.  Left: Line ratio map of  [\ion{O}{3}]/H$\alpha$ (excitation).  Middle: Line ratio map of [\ion{S}{2}]/H$\alpha$ (shocks and low photoionization). Right: Enlargement of the inner $\sim3\arcsec\times3\arcsec\times$ of the [\ion{S}{2}]/H$\alpha$ line map.  Solid white contours represent logarithmically-spaced emission at 8.46 GHz, indicating 0.10, 0.56, and 3.1 $\rm{mJy\;beam}^{-1}$ (for beam size $\sim0.29\arcsec$ and map noise $\sim27\mu\rm{Jy}$). 
For all images, white coloration indicates regions which have been masked due to low H$\alpha$ values ($<3\times10^{-17}\,$\cgsbrid).  The white `X' indicates the centroid of $4-8\;$keV emission, which is expected to come from the central source of the AGN.  Dashed lines indicate regions of interest, described in \S\ref{s2ha-text}: Arc `A' corresponds to a ridge of elevated [\ion{S}{2}]/H$\alpha$ at the interface between the NE radio lobe and NE line emission arm.  Circle `B' corresponds to a blob of elevated [\ion{S}{2}]/H$\alpha$ adjacent to the SW radio lobe.  Line `C' corresponds to a nearly north-south region of depressed [\ion{S}{2}]/H$\alpha$ at the nucleus.} 
\label{fig:ratio}
\end{figure*}

We produced continuum-subtracted emission line maps of [\ion{O}{3}]$\lambda5007$\AA, H$\alpha$,$\lambda6563$ and [\ion{S}{2}] $\lambda\lambda$6716,6731 from the HST narrow line filter observations (Table \ref{table:hst-obstable}) using the same methods described in \cite{Maksym16}.  The final continuum-subtracted line surface brightness images are displayed in Fig. \ref{fig:line-sb}.  In Fig. \ref{fig:line-colors}, we display a 3-color map of H$\alpha$, [\ion{O}{3}], and the deconvolved $0.3-8.0\;$keV emission.  In order to map the relative strengths of [\ion{O}{3}]$\lambda5007$, H$\alpha$\;$\lambda6563$, and [\ion{S}{2}]$\lambda\lambda$6716,6731, we produce ratio maps of [\ion{O}{3}]/H$\alpha$ and [\ion{S}{2}]/H$\alpha$ using {\tt dmimgcalc}.  These line ratio maps are shown in Fig. \ref{fig:ratio} and discussed in \S\ref{omorph}.  Typical $1\sigma$ uncertainties in the low flux limit are $(4,13,14)\times10^{-18}\,$\cgsbrid\ for (H$\alpha$,[\ion{O}{3}],[\ion{S}{2}]).

\section{Results}\label{sec-results}

\subsection{Optical/UV/IR Morphology}\label{omorph}

\begin{figure*} 
\noindent
\includegraphics[width=\textwidth]{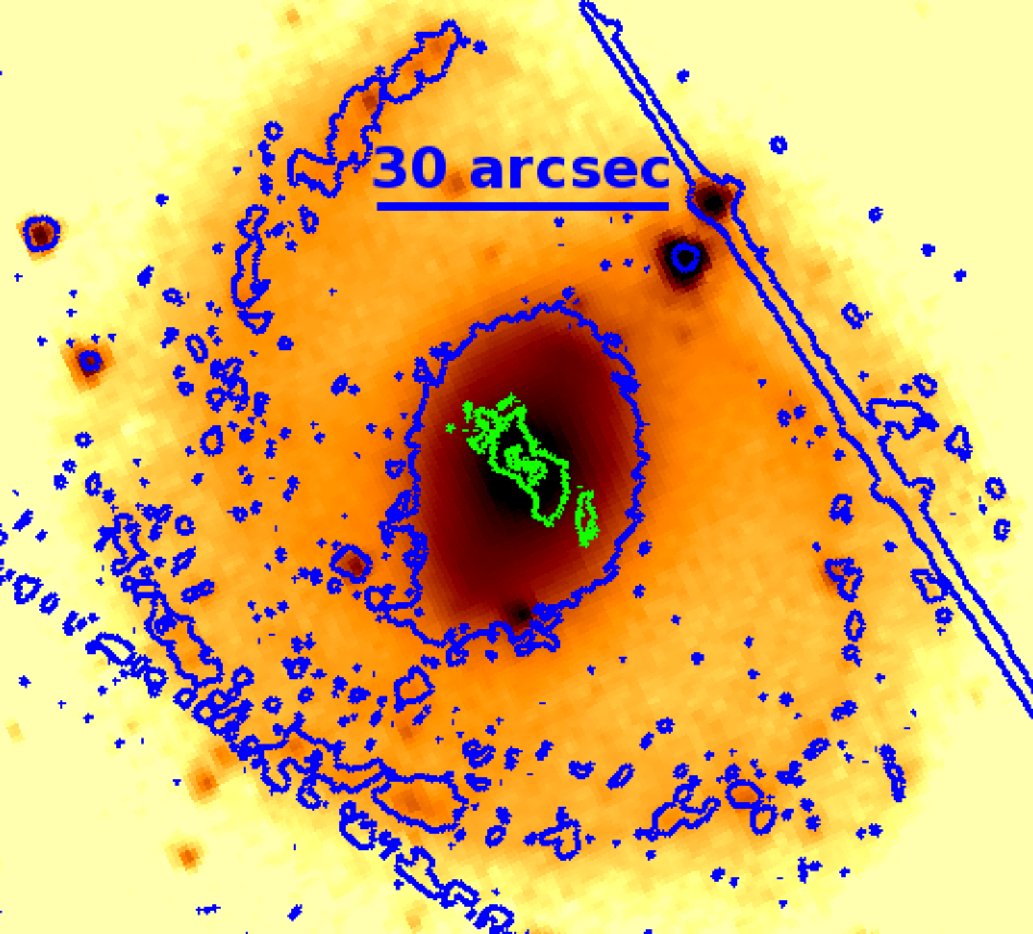}\par
\caption{Spitzer $3.6\,\mu$m image of NGC 3393 (Program \#10098, PI: Stern) with contours to indicate areas of bright F336W NUV emission ($\lambda3355\pm26\AA$; blue) and [\ion{O}{3}]\,$\lambda5007\,$\AA (green).  Extended linear NUV features composed of blue contours are instrumental artifacts.} 
\label{fig:bigpic}
\end{figure*}

For the purpose of clarity, we sub-divide the central ENLR into multiple spatial regions, shown in Fig. \ref{fig:line-colors}.  In particular, we refer to the ionization cone as having northeastern (NE) and southwestern (SW) directions, each of which is sub-divided into inner and outer portions.  The NE  and SW ionization cones are also associated with narrow-line outer clouds at $\simgreat5\arcsec$ in either direction.  
\label{hao3xray-text}

In order to investigate the physical origins of these structures and their relations to ionizing radiation from the AGN and kinematic feedback from the jets, we also compare the line ratio maps for  [\ion{O}{3}]/H$\alpha$ and [\ion{S}{2}]/H$\alpha$ of the H$\alpha$-bright regions in Fig. \ref{fig:ratio}.  

In Fig. \ref{fig:bigpic}, we show a large scale map and disk for context in the galaxy structure, using Spitzer 3.6\,$\mu$m (IR; starlight), [\ion{O}{3}], and F336W (NUV; young stars unobscured by dust).  With different scaling, [\ion{O}{3}]  from the outer clouds SW appears associated with the outer edge of the central F336W and F438W stellar structure, indicating a possible association with dust or a gap.

\begin{figure*} 
\noindent
\includegraphics[width=0.32\textwidth]{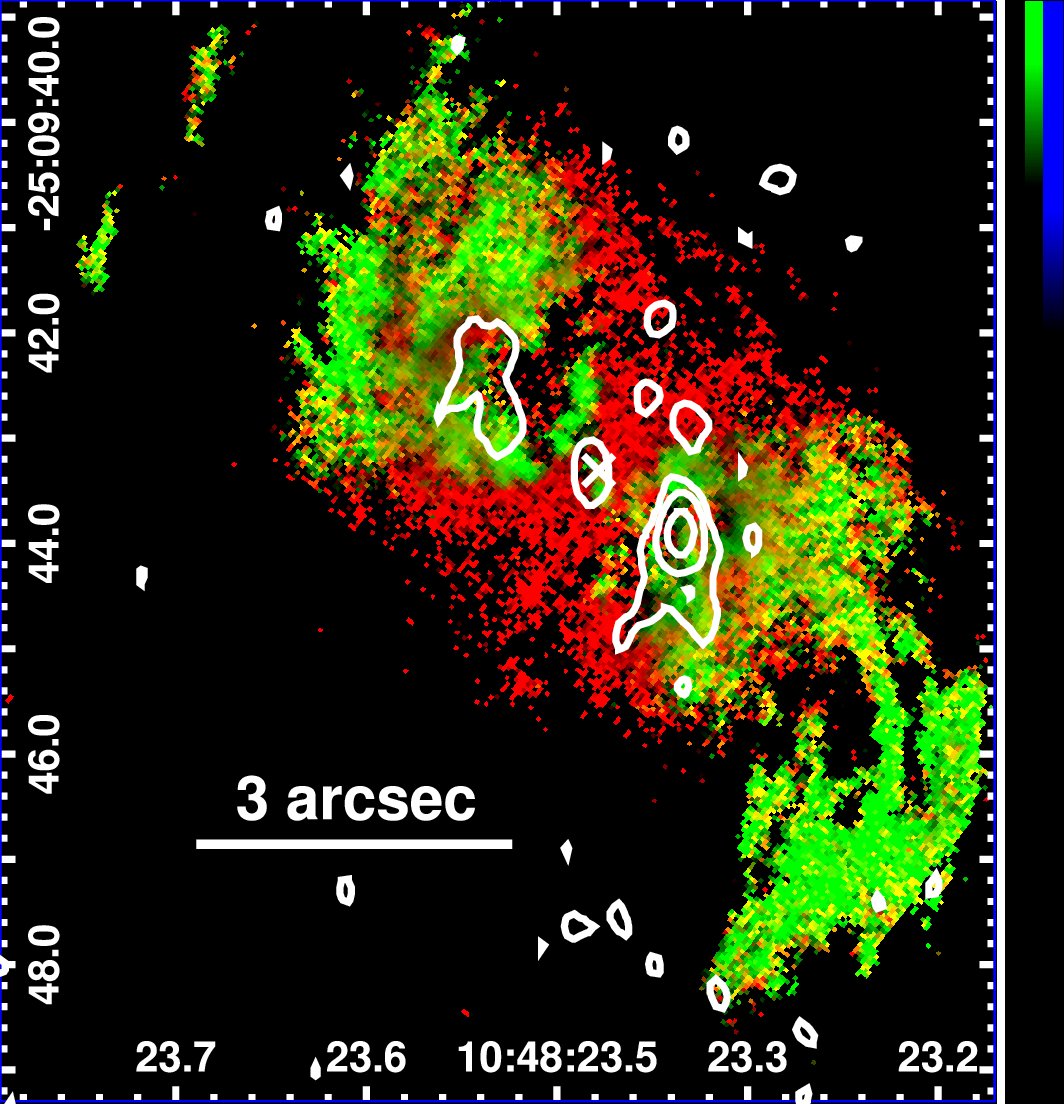}\hspace{0.01\textwidth}%
\includegraphics[width=0.32\textwidth]{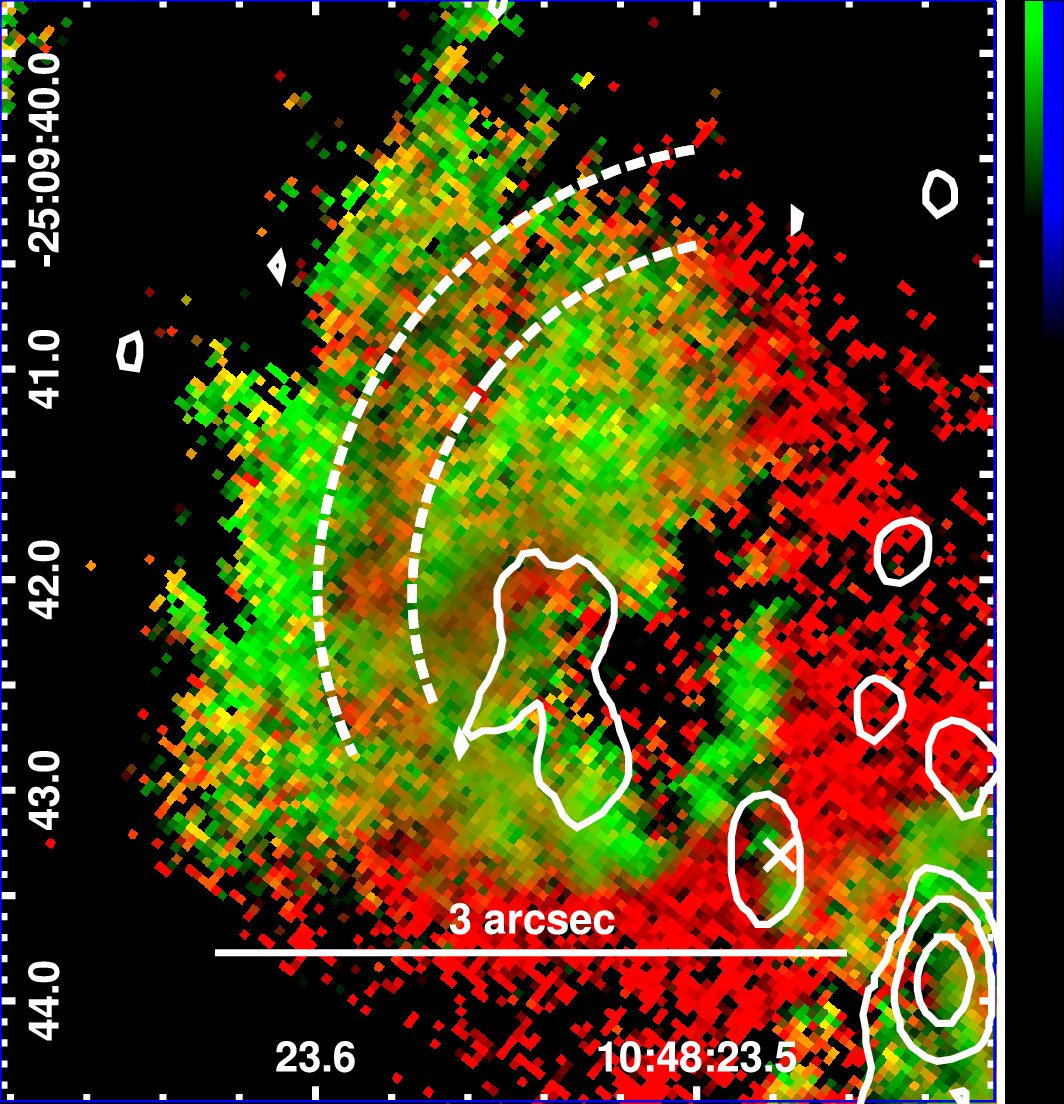}\hspace{0.01\textwidth}%
\includegraphics[width=0.32\textwidth]{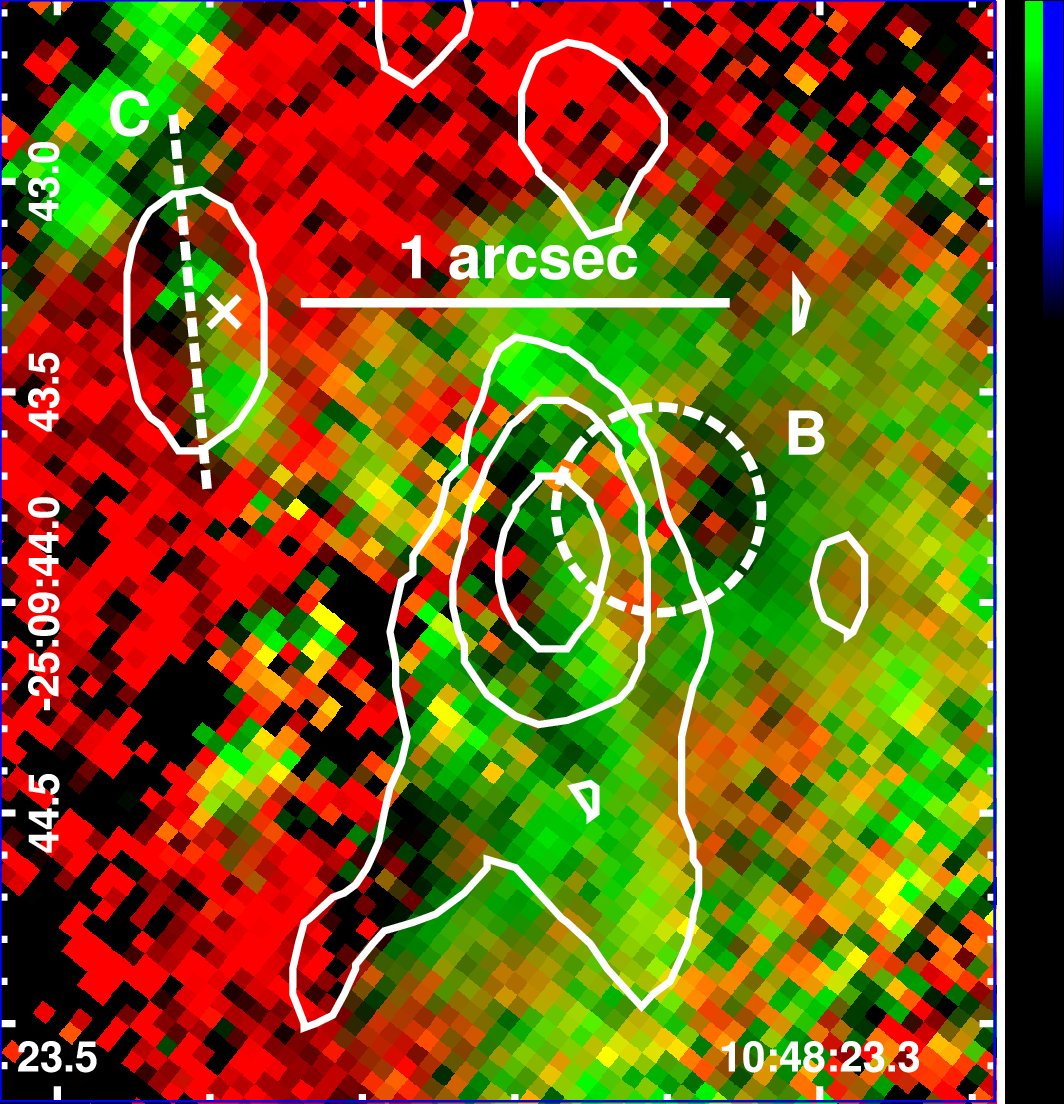}\par
\caption{Detailed [\ion{S}{2}]/H$\alpha$ (red) and [\ion{O}{3}]/H$\alpha$ (green).  {\bf Left:} Radio contours are overlaid, as in Fig. \ref{fig:ratio} (right).  Radio emission from the jets traces cavities in both [\ion{S}{2}]/H$\alpha$ (red) and [\ion{O}{3}]/H$\alpha$ (green), suggesting a role in the excitation of these features. 
{\bf Center:} magnification of the NE cone, again with radio contours.  The edge of the [\ion{S}{2}]/H$\alpha$ inner cavity shows enhanced [\ion{O}{3}]/H$\alpha$.  Dashed curves mark a gap ($\sim0.3\arcsec$ across; $\sim80\,$pc) between [\ion{O}{3}]/H$\alpha$ features on the outer edge of the NE cone.  The [\ion{O}{3}]/H$\alpha$ gap contains a ridge of enhanced [\ion{S}{2}]/H$\alpha$.
{\bf Right:} SW cone, again in [\ion{S}{2}]/H$\alpha$ and [\ion{O}{3}]/H$\alpha$.  Radio contours and labels are overlaid, as in Fig. \ref{fig:ratio} (right).  The SW radio lobe 
is adjacent to the edge of the [\ion{S}{2}]/H$\alpha$ inner cavity, separated from a knot of strong [\ion{S}{2}]/H$\alpha$ (B) by a high-excitation [\ion{O}{3}]/H$\alpha$ interface.  The high-excitation interface is traced by the eastern edge of the SW radio lobe.  The western edge of the SW radio lobe traces the other side of a cavity within the [\ion{S}{2}]/H$\alpha$ cocoon.
 } 
\label{fig:shock-ion}
\end{figure*}

From larger ($\simgreat1\arcsec$) to smaller ($\simless0.1\arcsec$) scales, the [\ion{O}{3}]/H$\alpha$ and [\ion{S}{2}]/H$\alpha$ maps are strikingly complementary.   This morphology becomes clearer when we overlay the line ratio maps directly in Fig. \ref{fig:shock-ion}.   Regions of enhanced  [\ion{S}{2}]/H$\alpha$ typically occupy regions of weak  [\ion{O}{3}]/H$\alpha$, and vice-versa.

The high [\ion{S}{2}]/H$\alpha$ region has an hourglass-shaped cocoon morphology which encloses the regions of enhanced [\ion{O}{3}]/H$\alpha$.  
Instead, high [\ion{O}{3}]/H$\alpha$, which measures excitation, is largely confined to the ionization cones.  Regions with strong [\ion{S}{2}]/H$\alpha$ and weak [\ion{O}{3}]/H$\alpha$ include the central cross-cone region, the regions immediately outside the inner cones, and a second, larger ridge along the NE arm.

\begin{figure} 
\centering
\includegraphics[width=0.45\textwidth]{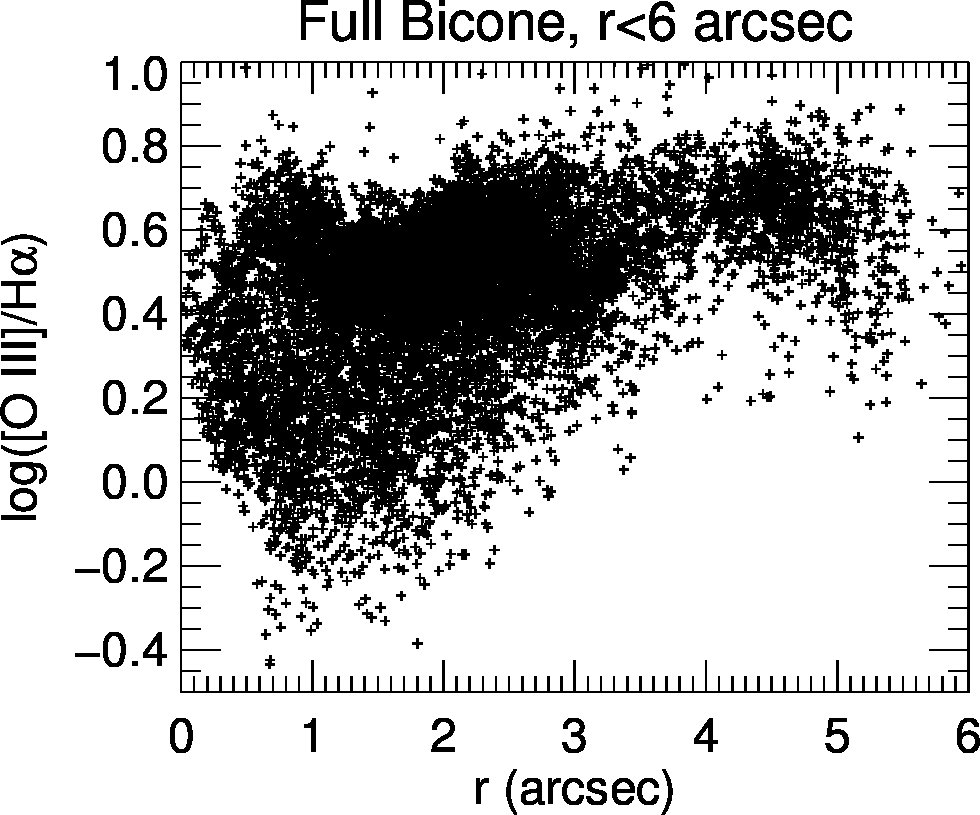}
\caption{Plot of [\ion{O}{3}]/H$\alpha$ vs. radius in arcseconds for the regions surrounding the nucleus of NGC 3393. Each WFC3 pixel detected above $3\sigma$ in [\ion{O}{3}], [\ion{S}{2}] and H$\alpha$ is represented as a cross.  Extending beyond $r\sim3\arcsec$, the ``outer cones" have the brightest average [\ion{O}{3}]/H$\alpha$.
} 
\label{fig:o3ha-radprof}
\end{figure}

Although several prominent knots of enhanced  [\ion{O}{3}]/H$\alpha$ are near the nucleus, average [\ion{O}{3}]/H$\alpha$ becomes stronger at larger radii (Fig. \ref{fig:o3ha-radprof}), particularly in the outer cones (defined in Fig. \ref{fig:line-colors}).  Regions with strong [\ion{O}{3}]/H$\alpha$ and weak [\ion{S}{2}]/H$\alpha$ include the outer ionization cones and outer clouds, as well as a high-ionization north-south region (marked C in Fig. \ref{fig:ratio}) at the nucleus.  

The edges of the inner cones which transition to the outer cones tend to be strong in both [\ion{O}{3}]/H$\alpha$ and [\ion{S}{2}]/H$\alpha$.

%In order to quantify the asymmetry in the S-shaped arms in the narrow line maps relative to the [\ion{S}{2}]/H$\alpha$ hourglass shape, we measure the ridge of  [\ion{S}{2}] along the edge of the ionization cones.  

Both [\ion{S}{2}]/H$\alpha$  and  [\ion{O}{3}]/H$\alpha$ exhibit low-strength cavities inside the inner cones.  Filamentary structures of strong [\ion{O}{3}]/H$\alpha$ dominate the edges of these cavities (Fig. \ref{fig:shock-ion}).  Regions with strong [\ion{O}{3}]/H$\alpha$ and strong [\ion{S}{2}]/H$\alpha$ include the transitions between the inner and outer cones.  

The S-shape in NGC 3393 is defined by the fact that each bubble appears bounded by narrow line emission on its far and counter-clockwise sides, but has a gap on its clockwise side where the emission is faint (taking the galactic nucleus as the origin in radial coordinates).  On the bright ridge of the SW arm in a strip measured 136$\degr$ east of north ($\sim1.3\arcsec$ in length, spanning $\sim50\degr$ in azimuth), we find typical [\ion{S}{2}] surface brightness, $\Sigma\sim(2-11)\times10^{-16}\;$\cgsbrid\ within 1.7\arcsec\ of the nucleus.  At 137$\degr$ west of north from the nucleus, we find $\Sigma\lesssim2.5\times10^{-16}\;$\cgsbrid, with no detection over a $\sim0.4\arcsec$ gap.   For a segment oriented $43\degr\pm10$ east of north connecting the NE arm to the SE end of the gap, we find $\Sigma\lesssim4.0\times10^{-17}\;$\cgsbrid, with no detection over a $\sim0.6\arcsec$ gap.  For the same angle along the bright ridge of the outer cone NE, we find $\Sigma\sim(2-11)\times10^{-16}\;$\cgsbrid, as in the SW arm.  Examining nearby circular regions with $r=2\arcsec$, we determine an upper limit of $\Sigma\sim2.0\times10^{-17}\;$\cgsbrid\ at $1\sigma$ significance.  The counter-clockwise sides of both bubbles therefore emit [\ion{S}{2}] $\simgreat4.4$ times brighter than the clockwise sides.  The clockwise edge of the cocoon  has fainter [\ion{S}{2}], however, with typical $\Sigma\lesssim2\times10^{-16}\;$\cgsbrid\ in the areas of largest [\ion{S}{2}]/H$\alpha$ ($1.5\pm0.5$) immediately outside the ionization cones.

\begin{figure} 
\centering
\noindent
\includegraphics[width=0.45\textwidth]{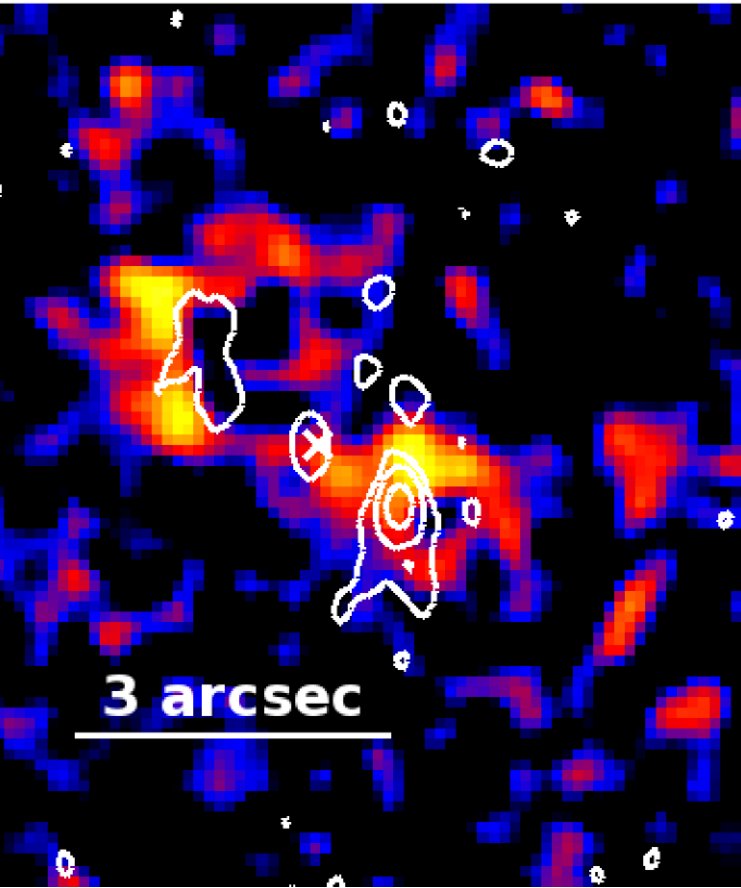}\\
\includegraphics[width=0.45\textwidth]{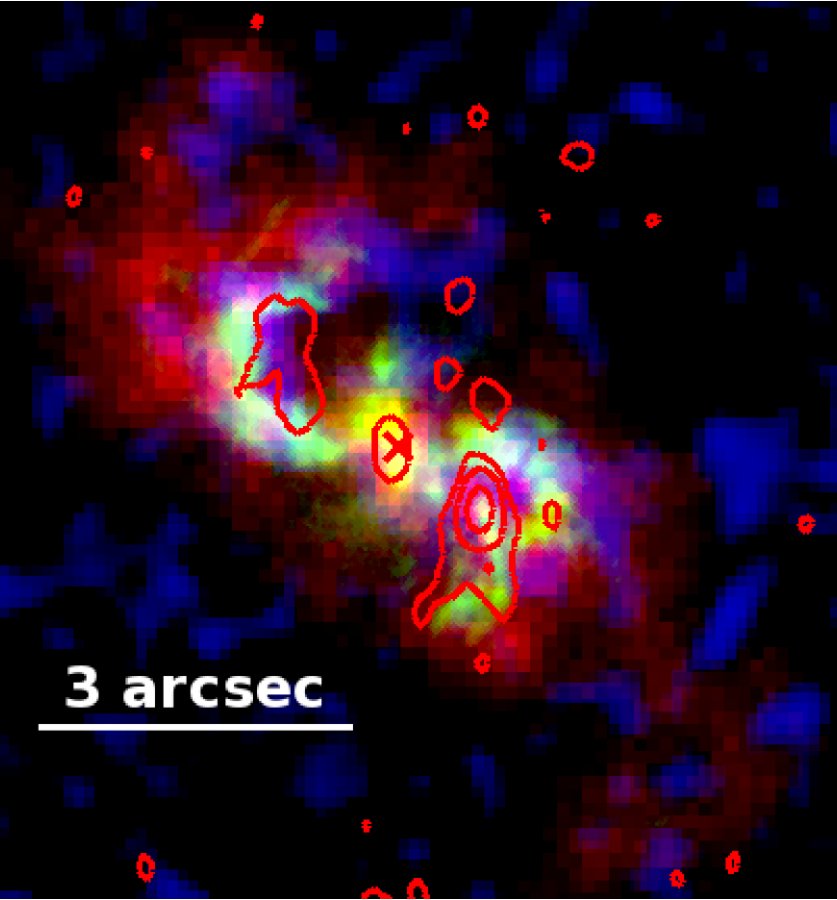}\par
\caption{Top: Mid-UV {\it HST} Image of the NGC 3393 center taken with the F218W filter for WFPC2.  Contours indicate 8.46 GHz VLA radio data for reference.  Bottom: 3-color image of the same region, with F218W (blue), NUV (F336W on WFC3, green), and EMC2-deconvolved X-rays as in Fig. \ref{fig:xray} (red).  Contours as above, but red for better contrast near bright regions.} 
\label{fig:UV}
\end{figure}

\begin{figure} 
\centering
\noindent
\includegraphics[width=0.45\textwidth]{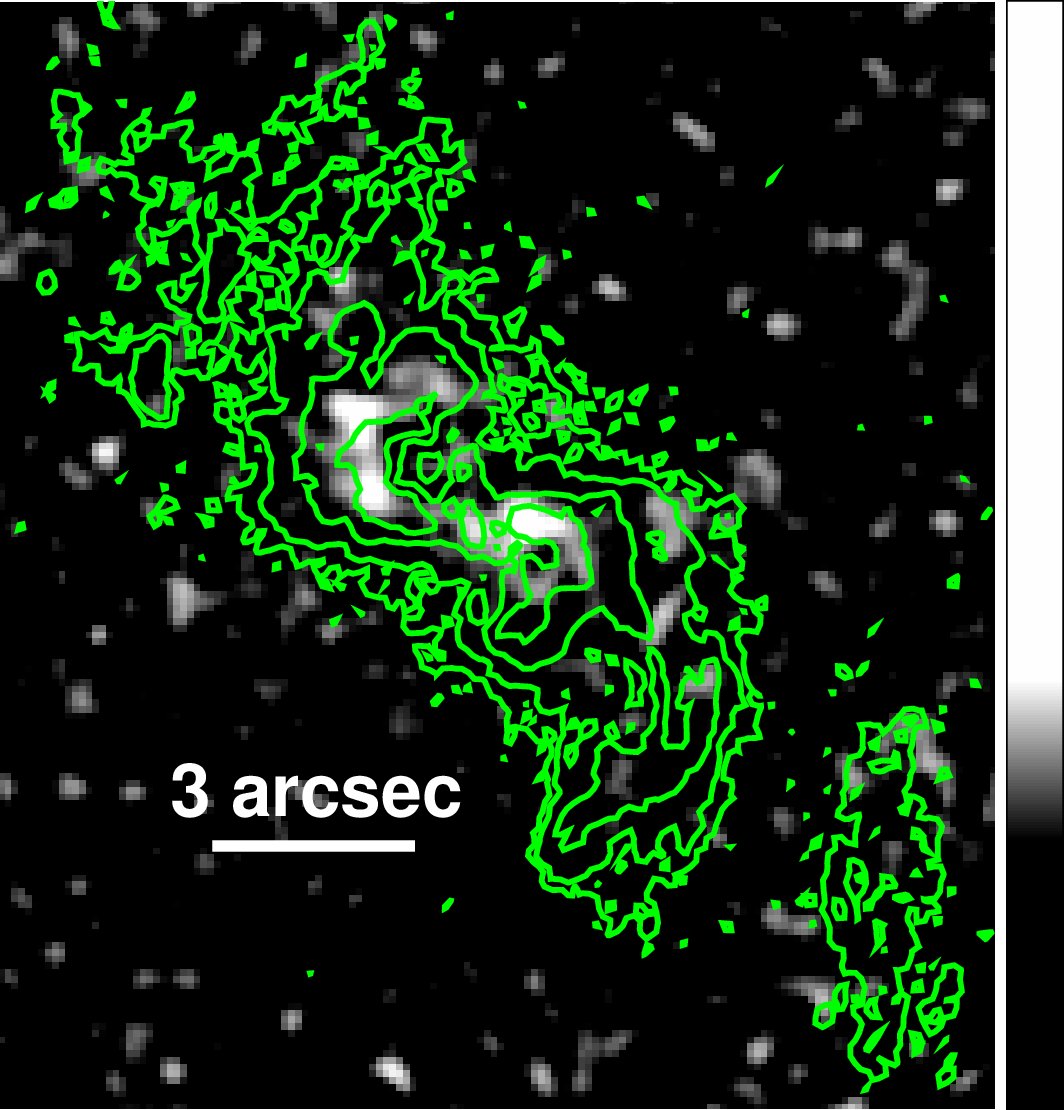}\\
\includegraphics[width=0.45\textwidth]{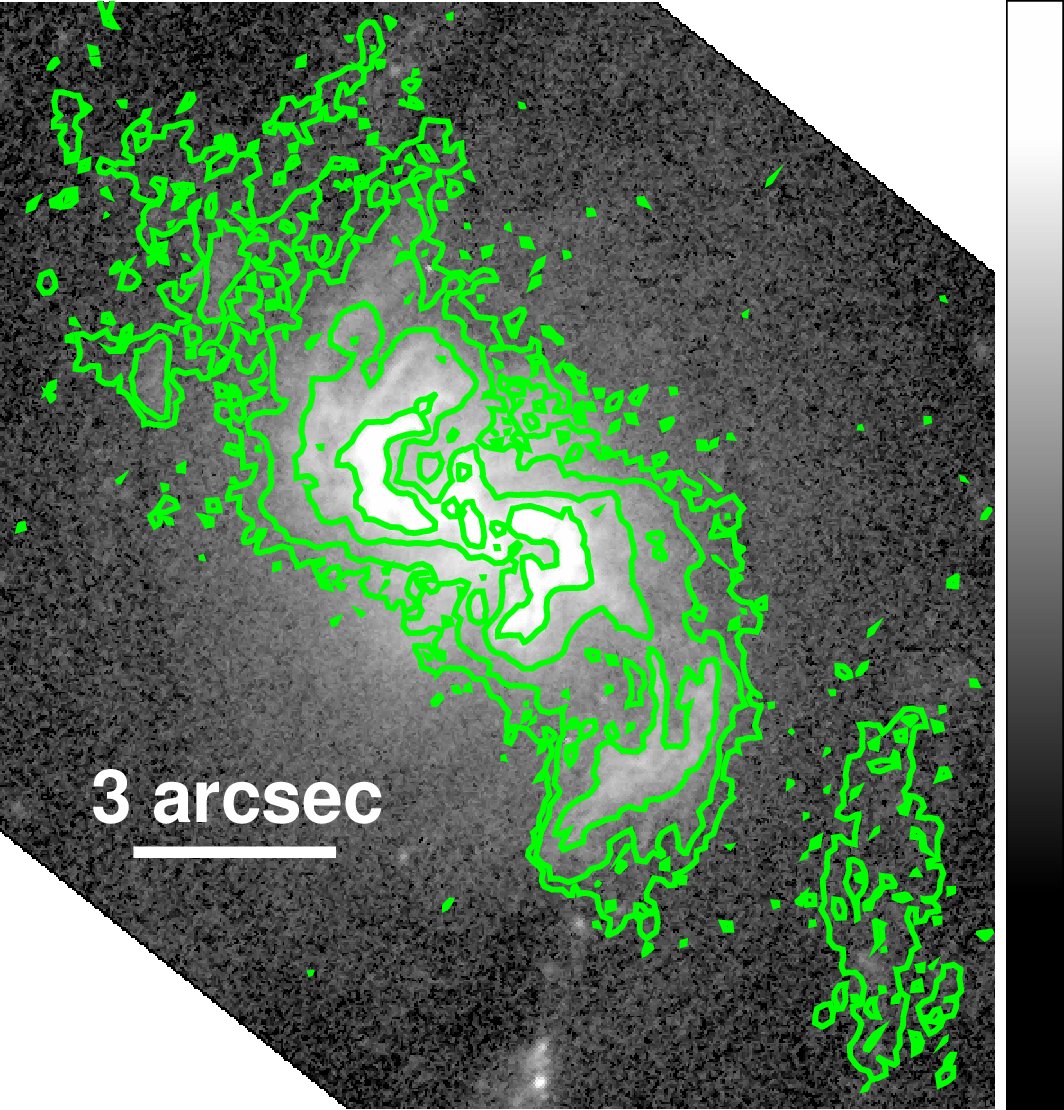}\par
\caption{Top: Smoothed F218W {\it HST} image from Fig. \ref{fig:UV}.  Contours taken from {\it HST} [\ion{O}{3}] data show the correlation between MUV and narrow line emission, particularly in the S-shaped arms.  Due to lack of bright F218W objects in the field, we have not corrected the F218W astrometry for an apparent $\sim0.2\arcsec$ offset from our reference dataset.  Bottom: F336W {\it HST} image of the same, with contours taken from {\it HST} [\ion{O}{3}] data for comparison.  The bright UV sources near the center of the southern edge of the frame are outside the ionization cones and could correspond to knots of recent star formation associated with a dusty spiral arm.} 
\label{fig:UV-O3}
\end{figure}

In Fig. \ref{fig:UV}, we show the F218W and F336W UV images.  Extracting F218W flux from circles of $r\sim1.4\arcsec$, we identify mid-UV (MUV) emission from both the NE and SW arms.  Each arm is detected as a structure with integrated flux at $\simgreat12\sigma$ significance above the observed background, and roughly coincides with emission seen in NUV and H$\alpha$.  This confirms the \cite{Koss15} finding that F336W emission is correlated with narrow line emission, and we also observe that the extended F218W emission is similarly correlated with narrow line emission (see Fig. \ref{fig:UV-O3}).  Some of the MUV emission in both the NE and SW may be within the bubble rather than directly overlapping the optical arm emission.  There is some evidence ($\simgreat3.4\sigma$) of an extended UV filament parallel with the NE arm which continues clockwise beyond the edge of the ionization cones and falls within a relatively low-H$\alpha$ region.  

{\it HST} F218W images clearly show that (as inferred by \citealt{Cooke00} from a $1.2\sigma$ detection using the {\it HST} FOC) the MUV emission observed with {\it IUE} is entirely extended.  From the F218W image we measure $F_{\lambda}\rm{(2100\,\AA)}=5.3\times10^{-16}$\,\fdenl\ in the NE arm and $4.5\times10^{-16}$\,\fdenl\ in the SW arm, consistent with the upper limits set by \cite{Cooke00}.  We do not detect a central point source.

\subsection{X-ray Morphology}\label{xmorph}

Taking regions of the EMC2-deconvoved image (Fig. \ref{fig:xray}) which encompass a $r=0.2\arcsec$ circular region about the nuclear peak, and contours of $100\;\rm{count\;arcsec}^{-2}$ and $3.3\;\rm{count\;arcsec}^{-2}$, we find that only $11\%$ of the photons come from the inner $\sim0.2\arcsec$.  By comparison, $53\%$ of the photons come from the next-brightest $6.6\;\rm{arcsec}^{2}$, and the remaining $36\%$ of this region come from an area spanning $52.3\;\rm{arcsec}^{2}$.

\begin{figure} 
\centering
\includegraphics[width=0.45\textwidth]{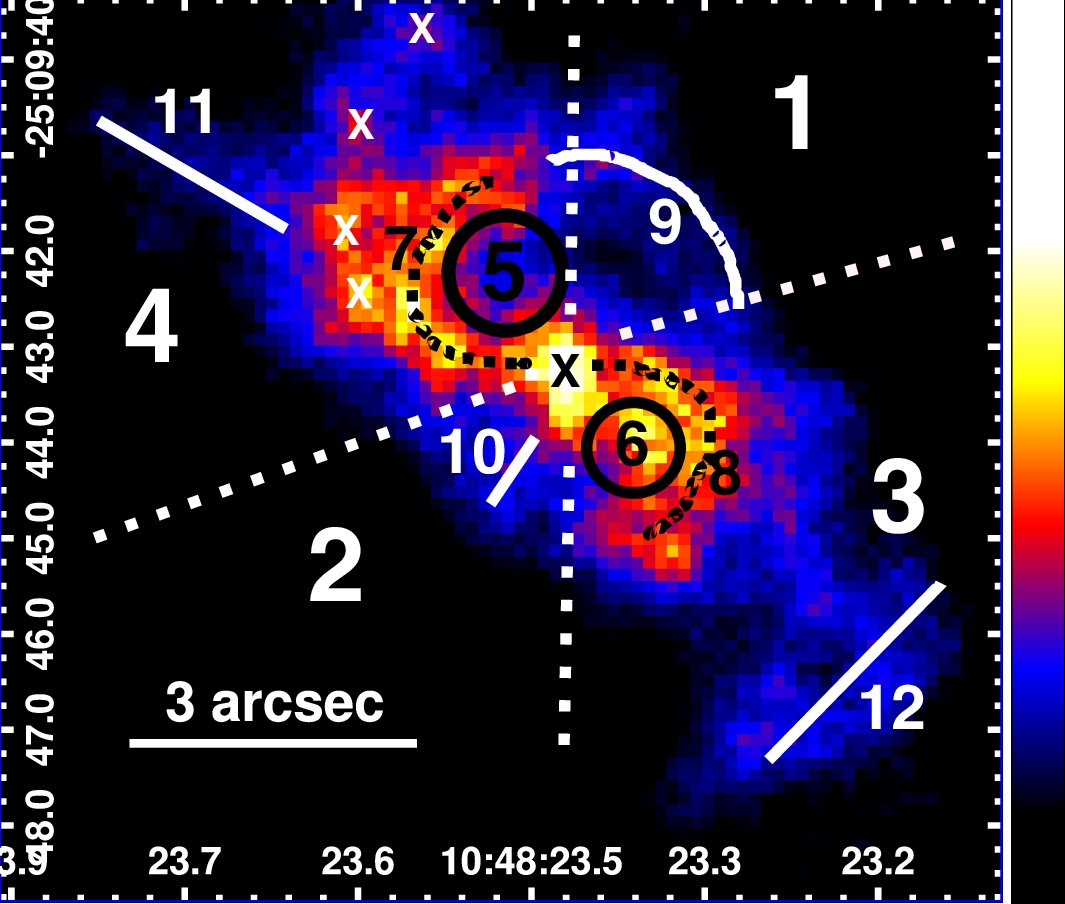}
\caption{Deconvolved X-ray image from Fig. \ref{fig:xray} (right), overlaid with labels to describe features referenced in the text:  1) NW cross-cone, bounded by dashed white lines.  2) SE cross-cone, also bounded by dashed white lines.  3) SW cone, bounded by dashed white lines. 4) NE cone, bounded by dashed white lines. 5) NE cavity, indicated by a black circle 6) SW cavity, indicated by a black circle. 7) NE arm, indicated by a dashed black curve. 8) SW arm, indicated by a dashed black curve.  9) NW arc, indicated by a solid white curve. 10) SE spur, indicated by a solid white line. 11) NE linear feature extending from the nucleus, indicated by a solid white line.  12) Structure parallel to the SW arm.  The nucleus is indicated with a black `X'.  Blobs in the NE cone are indicated with a white `X'.}
\label{fig:xlabel}
\end{figure}

The X-ray emission is dominated by the S-shaped arms (Fig. \ref{fig:xlabel}), which extend to $r\simgreat2.5\arcsec$ from the nucleus.  These arms run approximately east-west at the nucleus, then curve through nearly $\sim180\degr$ and extend almost to the north-south axis.  The brightest X-ray emission arises from a ridge $\lesssim0.4\arcsec$ thick at several points, and $\sim8\arcsec$ in length, measured along the ridge line.  The SW arm encircles a bright knot of X-ray emission with diameter $\sim0.6\arcsec$, containing $\sim4\%$ of the deconvolved flux.

At $r\sim3\arcsec-5\arcsec$, the NE cone shows multiple sub-arcsecond-scale blobs, and a linear structure which extends from the nucleus at $\sim60\degr$ east of north.  The SW cone, also at $r\sim3\arcsec-5\arcsec$, shows a $\sim3.5\arcsec$ structure running parallel to the S-shaped arm.  At $\sim6\arcsec-11\arcsec$ SW of the nucleus (in the outer clouds SW, not shown in Fig. \ref{fig:xlabel}; see Fig.\ref{fig:line-colors}), we see a faint, elongated $\sim6\arcsec$ structure that runs approximately north-south and is apparently separate from the main structure of the SW cone.

At $r\sim2\arcsec$, the northwestern cross-cone shows an arc roughly concentric with the nucleus (NW arc; Fig.\ref{fig:xlabel}).  In the southeastern cross-cone, we see tenuous evidence of a linear feature (spur; Fig. \ref{fig:xlabel}) extending approximately southwest and south from the nuclear region.

The inner $\sim1\arcsec$ of the nucleus is dominated by an elongated north-south feature $\sim0.8\arcsec$ in length which contains the nuclear peak itself.
 
\subsection{Comparison of Optical and X-ray Emission}

As is shown in Figs. \ref{fig:xray}, \ref{fig:line-sb}, and \ref{fig:line-colors}, there is good correlation on sub-arcsecond scales between X-rays, H$\alpha$ and [\ion{O}{3}], as has previously been established by \cite{Cooke00}, \cite{Bianchi06}, and \cite{Koss15}.  We see the same correlation in [\ion{S}{2}] as well, dominated by the flux in the S-shaped arms as in other bands.  

\begin{figure*} 
\noindent
\includegraphics[width=\textwidth]{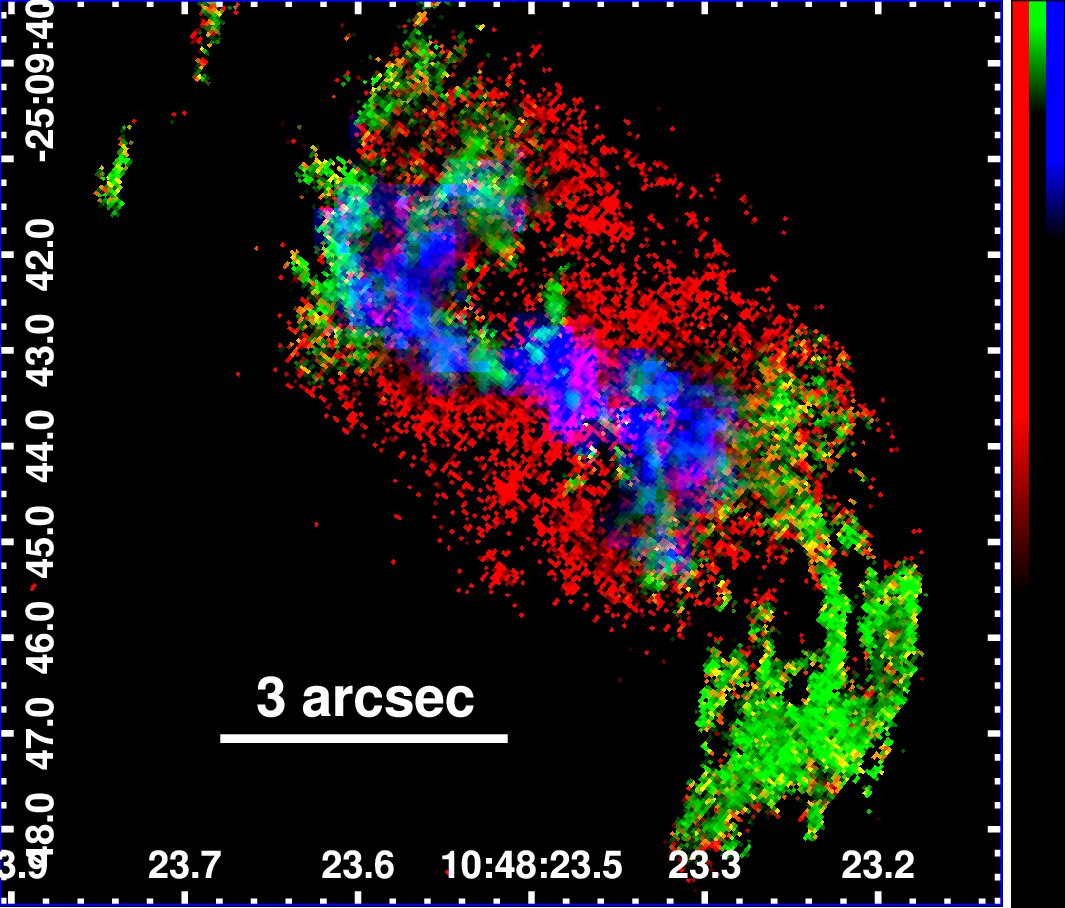}\hspace{0.05\textwidth}\par
\caption{Three-color image of line ratio maps for [\ion{S}{2}]/H$\alpha$ (red) and [\ion{O}{3}]/H$\alpha$ (green) as in Fig. \ref{fig:ratio}, and EMC2-deconvolved 0.3-8\;keV X-ray emisson (blue) as in Fig. \ref{fig:xray}.  Note that both low-value and masked low-H$\alpha$ regions are black in this color scheme.  Color contrast is chosen to emphasize the strongest regions in each component, as the X-rays in particular have significant extended components with surface brightness an order of magnitude fainter than displayed here.  Low-photoionization regions, possibly indicative of shocks, form a cocoon about the ionization cones.  The inner ionization cone and nuclear ridge C (Fig. \ref{fig:ratio}) are dominated by X-ray emission.  High-excitation regions are present in knots within the inner cones, but dominate the outer cones.} 
\label{fig:shock-ion-xray}
\end{figure*}

\begin{figure*} 
\noindent
\includegraphics[width=0.32\textwidth]{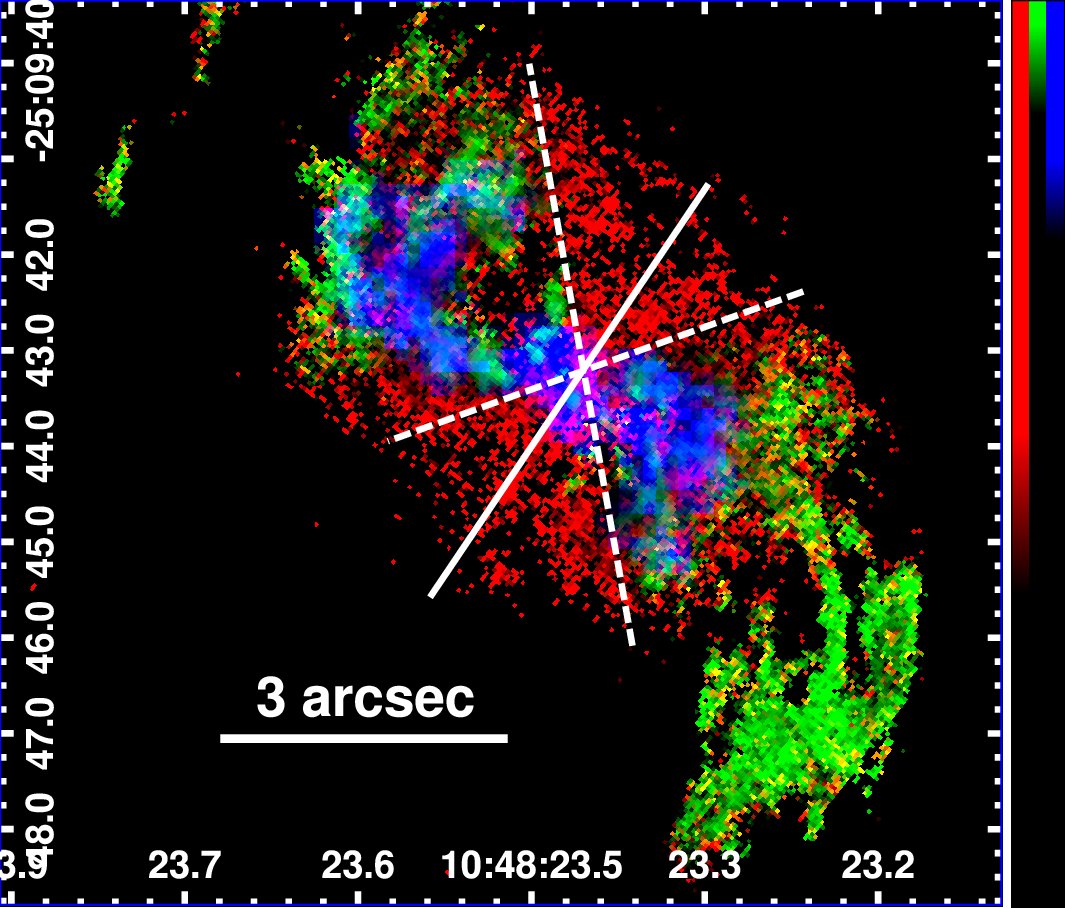}\hspace{0.01\textwidth}%
\includegraphics[width=0.32\textwidth]{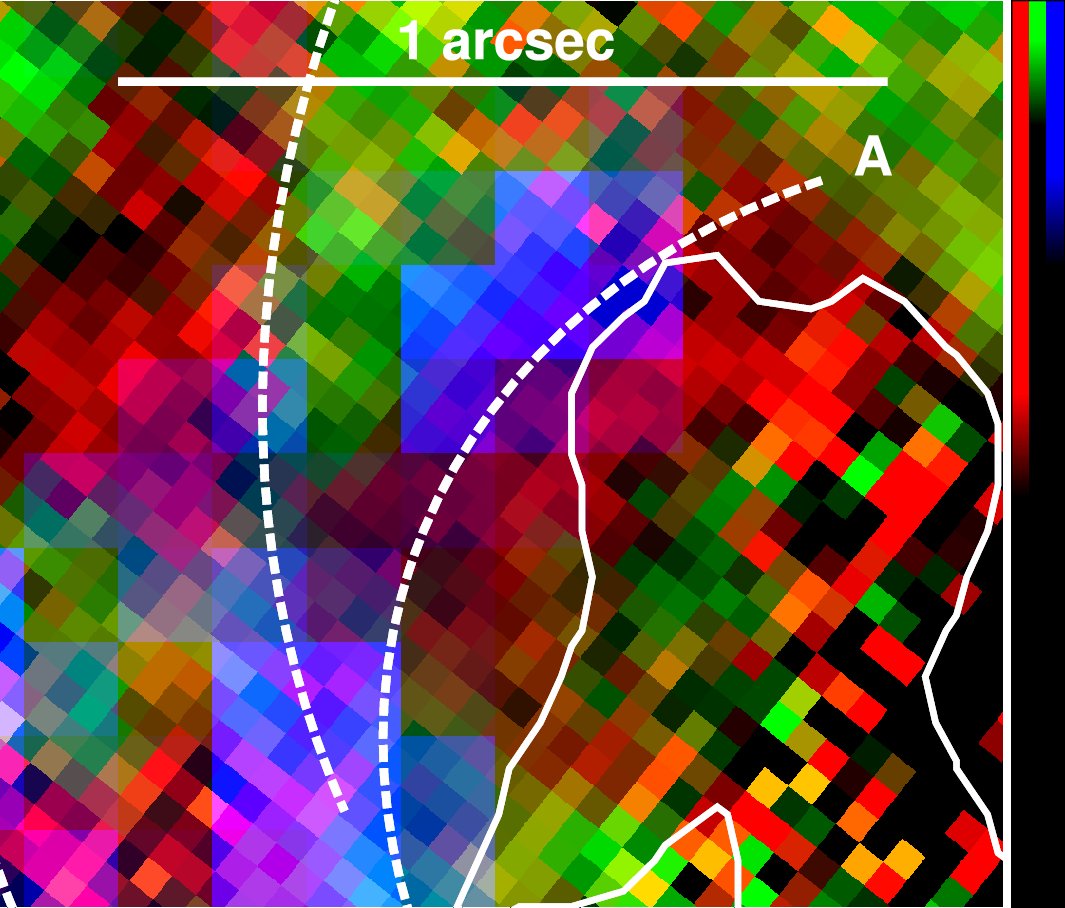}\hspace{0.01\textwidth}%
\includegraphics[width=0.32\textwidth]{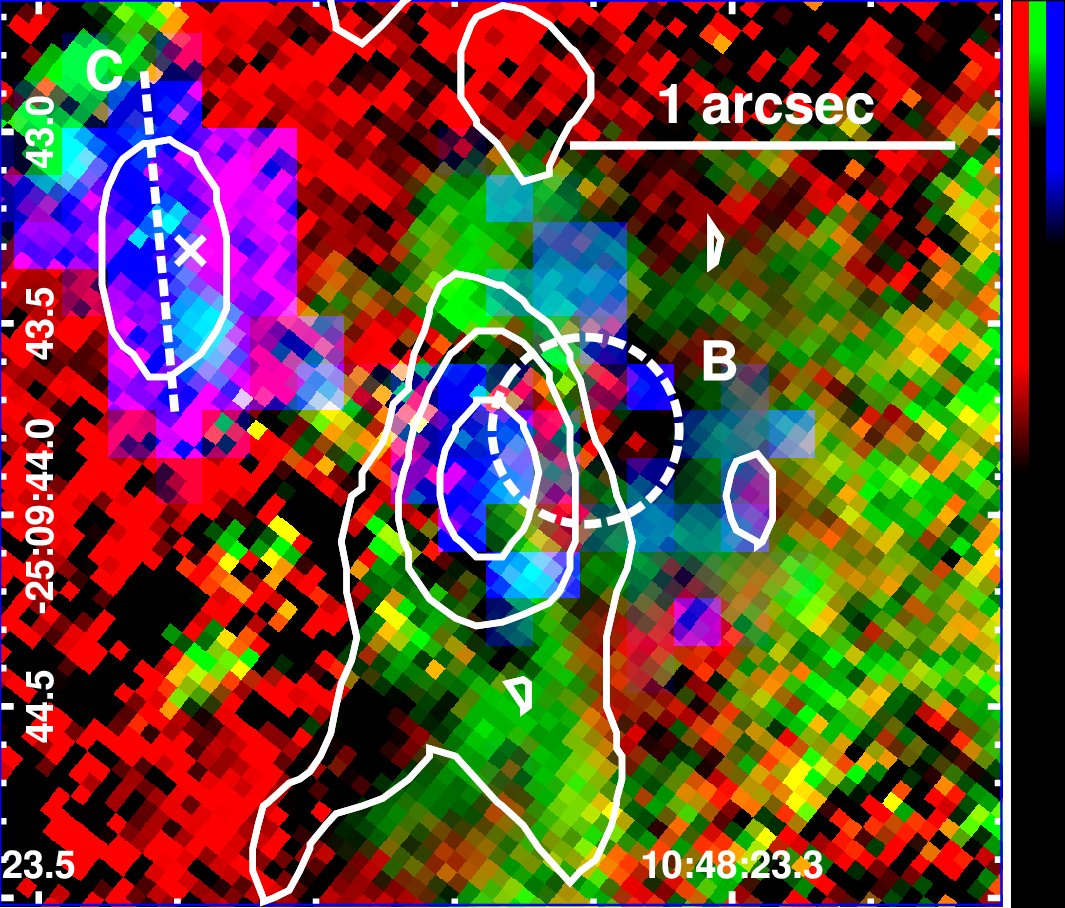}\par
\caption{{\bf Left:} same as Fig. \ref{fig:shock-ion-xray}, but with white lines to indicate the angle of the maser disk \citep[][ solid]{Kondratko08}, and previously-identified ionization cone edges \citep[][ dashed]{Cooke00}.  {\bf Center:} magnification of the NE inner cone, with colors as Fig. \ref{fig:shock-ion-xray}.  The ridge (A) of enhanced [\ion{S}{2}]/H$\alpha$ from Fig. \ref{fig:ratio} is indicated with a dashed white curve, as before.  Here, enhanced  [\ion{S}{2}]/H$\alpha$ falls to the right of curve A, whereas the X-ray ridge of the NE arm is left of curve A.  Moving leftward away from the nucleus and from the radio lobe, a green ridge shows enanced [\ion{O}{3}]/H$\alpha$, also parallel to A, suggesting a layered structure parallel to the radio lobe surface.  The dashed white curve on the left indicates a boundary between enhanced  [\ion{O}{3}]/H$\alpha$ and [\ion{S}{2}]/H$\alpha$, as described in Fig. \ref{fig:shock-ion}.  {\bf Right:} magnification of nucleus and SW inner cone, again with colors as Fig. \ref{fig:shock-ion-xray}. X-ray emission and  [\ion{O}{3}]/H$\alpha$ excitation dominate the nuclear ridge C (Fig. \ref{fig:ratio}) where [\ion{S}{2}]/H$\alpha$ is conspicuously absent.  Protrusions from the SW radio lobe trace the edge of the [\ion{S}{2}]/H$\alpha$ cavity (west) and a north-south [\ion{O}{3}]/H$\alpha$ excitation filament extending from the peak of radio emission, suggesting a role for the jet or kinematic outflow in the formation of these transitions.  We provide additional morphological description in two-color comparisons in Figs.  \ref{fig:shock-ion}, \ref{fig:two-comp}, \ref{fig:ion-xray},
 and \ref{fig:shock-xray}. }
\label{fig:shock-ion-xray-panel}
\end{figure*}

\begin{figure*} 
\noindent
\includegraphics[width=0.315\textwidth]{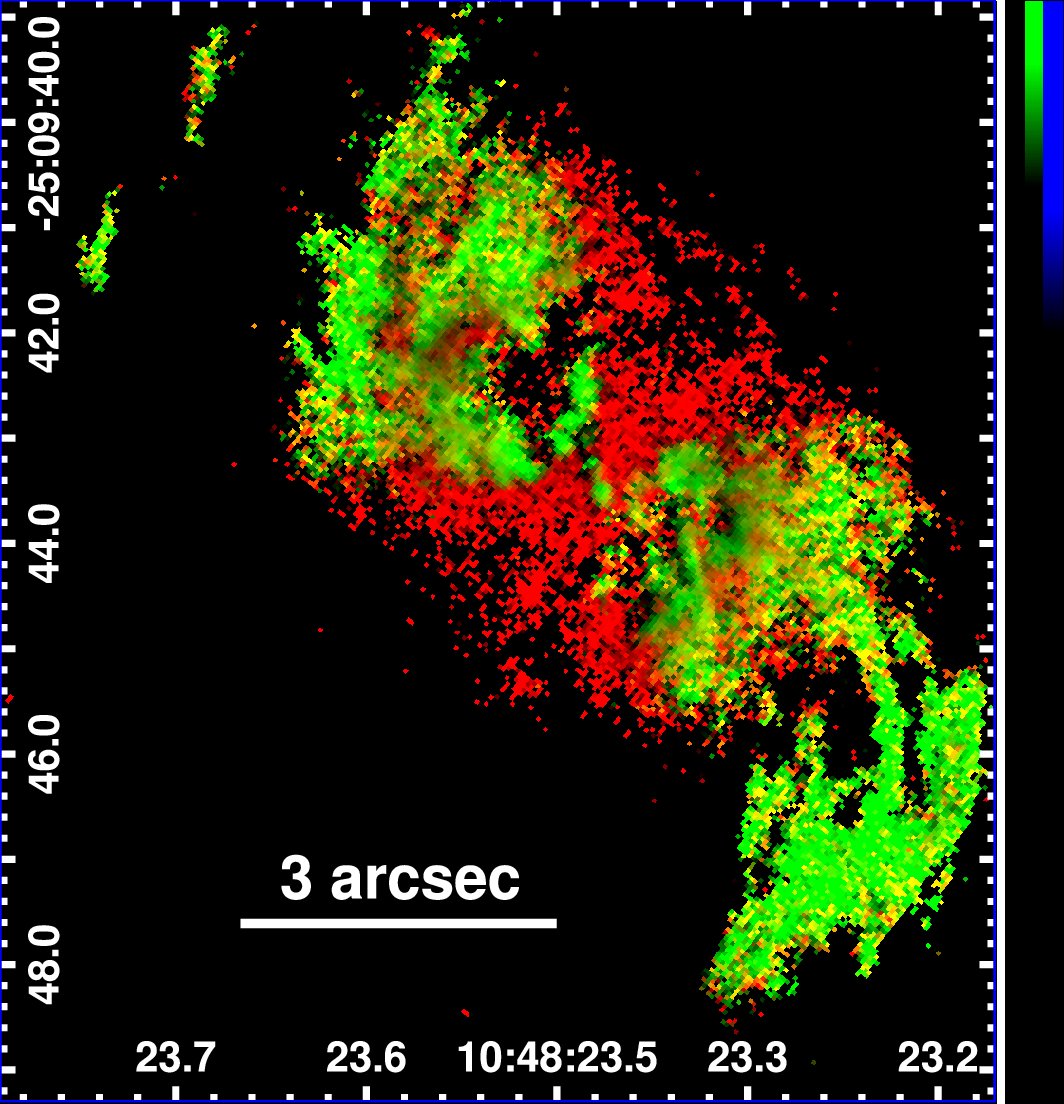}\hspace{0.01\textwidth}%
\includegraphics[width=0.315\textwidth]{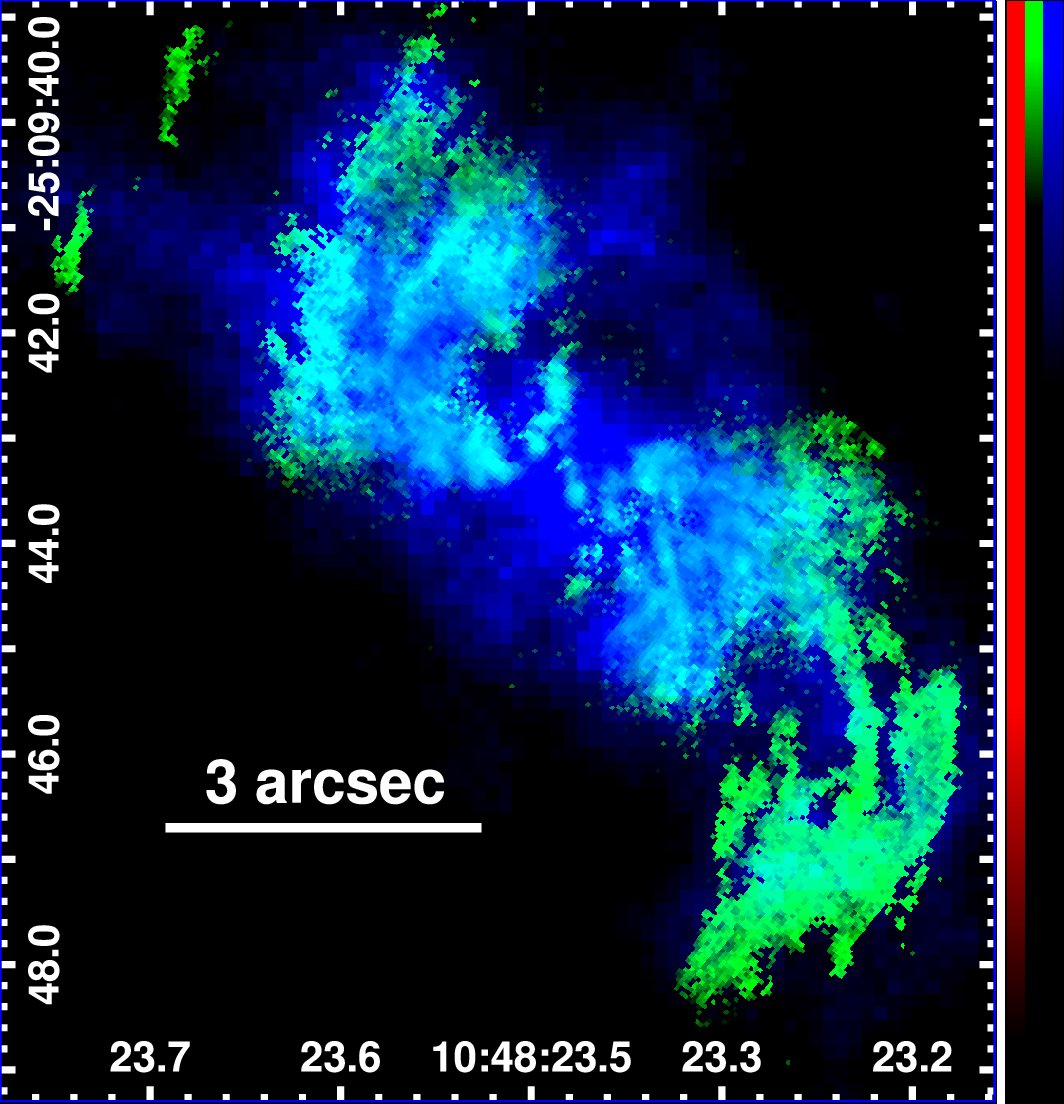}\hspace{0.01\textwidth}%
\includegraphics[width=0.325\textwidth]{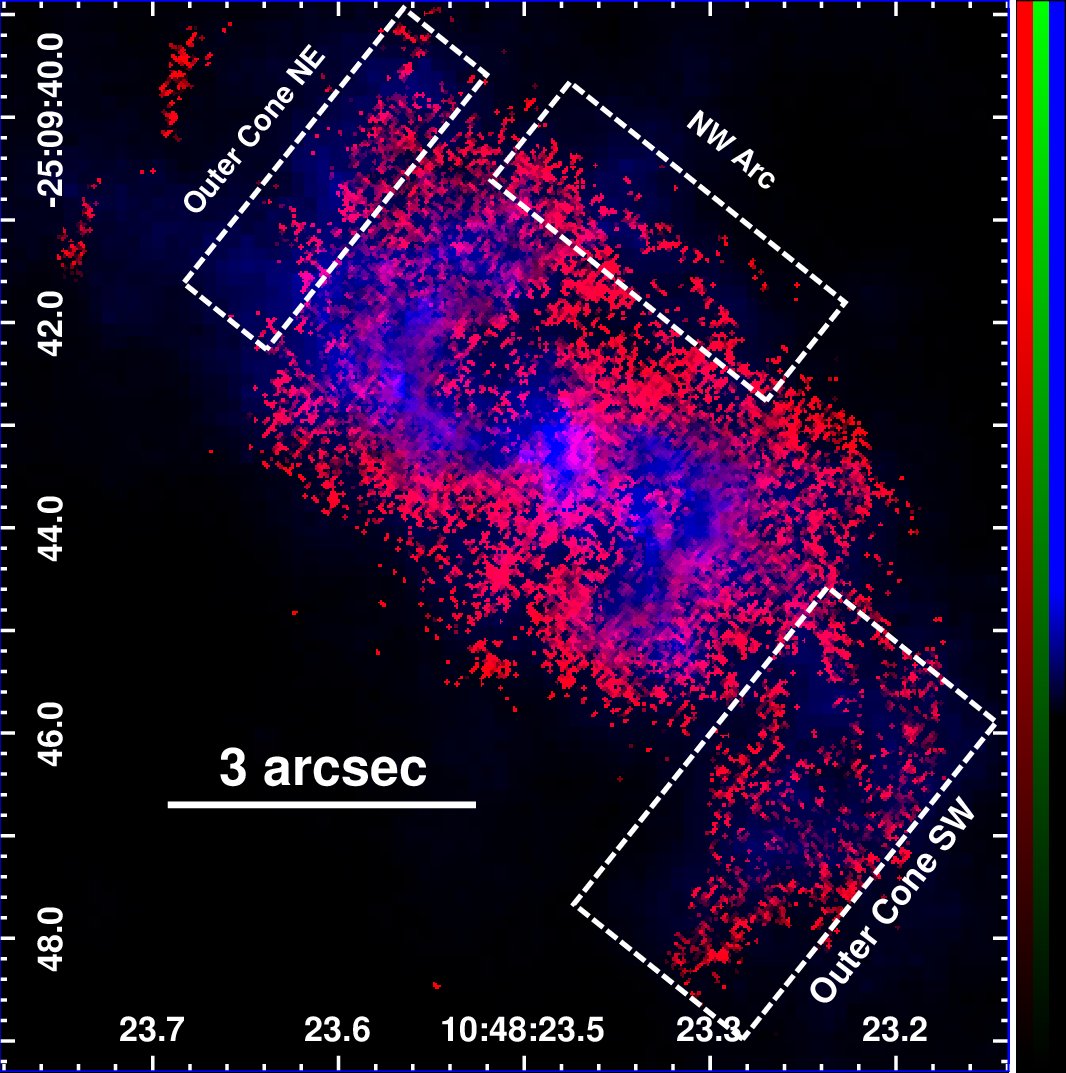}\par
\caption{The ionization cone from Fig. \ref{fig:shock-ion-xray}, reduced to two-color images for clarity.  {\bf Left:}  [\ion{S}{2}]/H$\alpha$ (red) and [\ion{O}{3}]/H$\alpha$ (green).  The [\ion{O}{3}]/H$\alpha$ traces the inside of the ionization cone, whereas [\ion{S}{2}]/H$\alpha$ traces an hourglass-shaped outer cocoon which surrounds high-excitation [\ion{O}{3}]/H$\alpha$, particularly in the near cross-cone region, but also co-spatial with [\ion{O}{3}]/H$\alpha$ along the outer edges of the inner cones.  [\ion{O}{3}]/H$\alpha$ dominates the outer cone SW completely, with only a trace of [\ion{S}{2}]/H$\alpha$ along the outer SW edge.
{\bf Center:}  0.3-8\;keV X-ray emission (blue) with [\ion{O}{3}]/H$\alpha$ (green).  X-rays in the inner clouds are largely co-spatial with [\ion{O}{3}]/H$\alpha$.  Major exceptions include the outer clouds NW, as well as enhanced X-ray emission in the NW arc (see also right and Fig. \ref{fig:line-sb}).  The NW arc may be part of a larger spheroidal shell structure which includes extended X-ray emission which corresponds to lower-excitation optical emission within $\sim2\arcsec$ of the nucleus.  Some of this X-ray emission is co-spatial with a possible extended low-excitation optical SW-NE spur from the nucleus (roughly in the plane of the maser disk; see right and Fig. \ref{fig:shock-ion-xray-panel}, left).  {\bf Right:}  [\ion{S}{2}]/H$\alpha$ (red) with 0.3-8\;keV X-ray emission (blue), scaled to emphasize the extent of the X-ray emission with respect to the low-ionization regions.  The NW arc is present in both components.  X-ray emission is enhanced in the outer cone NE and outer cone SW.
}
\label{fig:two-comp}
\end{figure*}

Using the PSF-deconvolved images, we examine the morphology and optical feature correspondence on the smallest scales in the {\it Chandra} images, $\sim0.2\arcsec$.  
In order to compare and contrast the morphology of regions with the strongest X-rays, [\ion{O}{3}]/H$\alpha$ and [\ion{S}{2}]/H$\alpha$, we show three-color images in  Figs. \ref{fig:shock-ion-xray} and \ref{fig:shock-ion-xray-panel}.  Different features become clearer in less complicated figures, or  more visible with different color stretches, so we also present two-color images comparing X-rays to [\ion{O}{3}]/H$\alpha$ and to [\ion{S}{2}]/H$\alpha$ in Fig. \ref{fig:two-comp}.

In general, the fainter extended X-ray regions correspond to the entire H$\alpha$-bright region in our optical line ratio maps (see especially Fig. \ref{fig:two-comp} center and right).  This larger-scale emission extending to $\sim6\arcsec$ is notably strong in the outer cone SW, where [\ion{O}{3}]/H$\alpha$  is relatively high, and in the outer cone NE where [\ion{O}{3}]/H$\alpha$ is strong and H$\alpha$ falls under our threshold ($1.2\times10^{-17}\,$\cgsbrid).  We also observe that the NW arc  described in \S\ref{xmorph} at $\sim2\arcsec$ NW of the nuclear ridge C corresponds to a  [\ion{S}{2}]/H$\alpha$ feature which is partially defined by surrounding regions of weak H$\alpha$.  

We see that the $\sim2\arcsec$ X-ray spur in the SE cross-cone noted in \S\ref{xmorph} is associated with a [\ion{S}{2}]/H$\alpha$ spur.  The base of this feature (Fig. \ref{fig:shock-ion-xray-panel}) falls approximately at the position angle of the pc-scale maser disk ($\rm{P.A.}\sim-34\degr$; Fig. \ref{fig:shock-ion-xray-panel}, left) studied by \cite{Kondratko08}.  

On smaller scales, we examine the correlation between X-rays and the line-emitting S-shaped arms in finer detail.  Although these X-ray arms are confined to the ionization cones like  [\ion{O}{3}]/H$\alpha$, in some cases they trace complementary regions.  For example, in Fig.  \ref{fig:shock-ion-xray-panel}, the brightest X-ray emission coincides (like [\ion{O}{3}]/H$\alpha$) with the N-S nuclear ridge `C' (Fig. \ref{fig:shock-ion-xray-panel}, right). 

\begin{figure} 
\centering
%\noindent
%\includegraphics[width=0.32\textwidth]{ion-xray_zoom_sat.jpg}\hspace{0.01\textwidth}%
\includegraphics[width=0.45\textwidth]{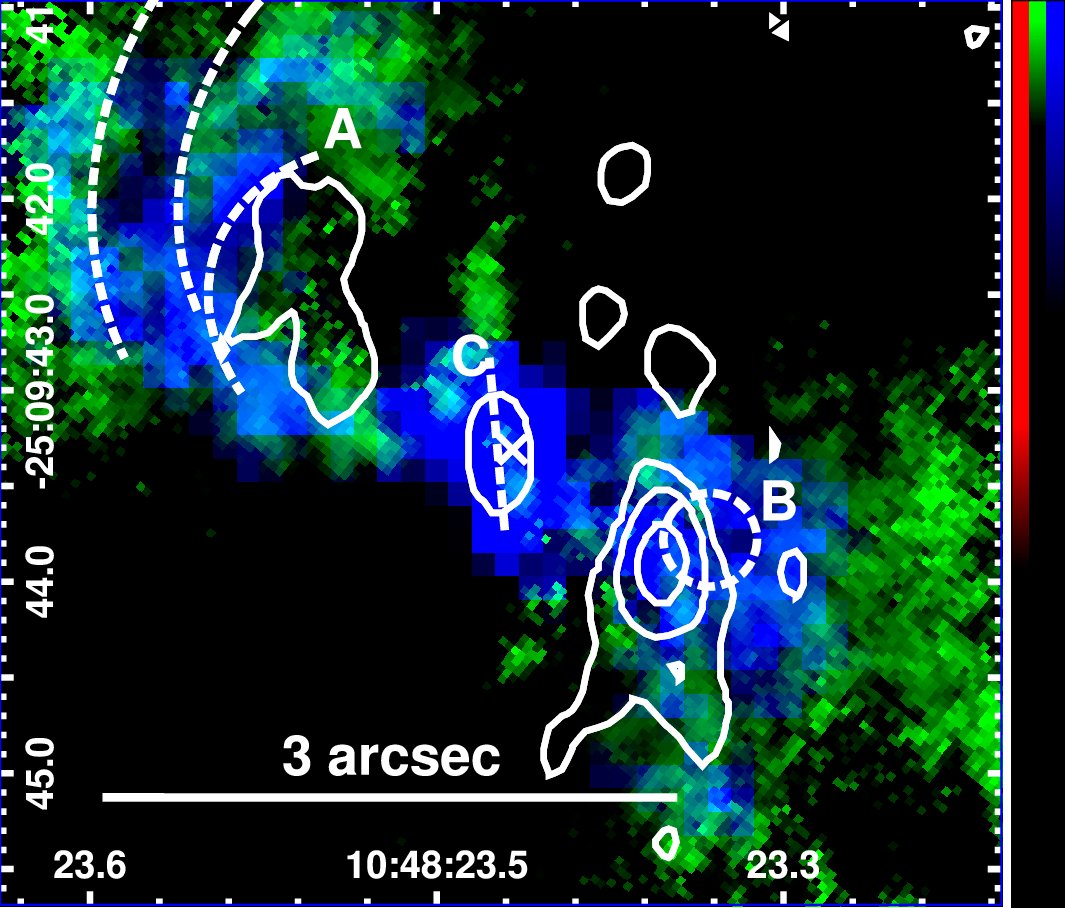}%\hspace{0.01\textwidth}%
\caption{Detailed [\ion{O}{3}]/H$\alpha$ (green) and 0.3-8\;keV X-ray emission (blue) maps of the inner ionization cones, blue-saturated to emphasize regions of strong X-rays.  Like  [\ion{O}{3}]/H$\alpha$ (green), the strongest X-rays are confined to ionization cones and form S-shaped arms, including the linear [\ion{O}{3}]/H$\alpha$ structure at the nucleus.  But on scales of $<0.5\arcsec$, X-rays typically trace different regions from [\ion{O}{3}]/H$\alpha$.  Contours are from radio imaging as in Figs. \ref{fig:ratio} and \ref{fig:shock-ion}.  X-rays trace the NE arm, particularly following ridge A, identified in [\ion{S}{2}]/H$\alpha$ (Fig. \ref{fig:ratio}), as well as the central ridge C traced by strong[\ion{O}{3}]/H$\alpha$ across the nucleus (X).  The $\sim0.3\arcsec$ ($\sim80\,$pc) gap described in Fig. \ref{fig:shock-ion} is indicated with longer dashed white curves.  X-rays show sub-arcsecond structure, including a knot of enhanced emission nearly coincident with the peak of the SW radio lobe.  Other X-ray emission surrounds [\ion{S}{2}]/H$\alpha$ knot B, and traces the edge of a cavity in the inner [\ion{O}{3}]/H$\alpha$ structure.
} 
\label{fig:ion-xray}
\end{figure}

\begin{figure} 
\noindent
\includegraphics[width=0.45\textwidth]{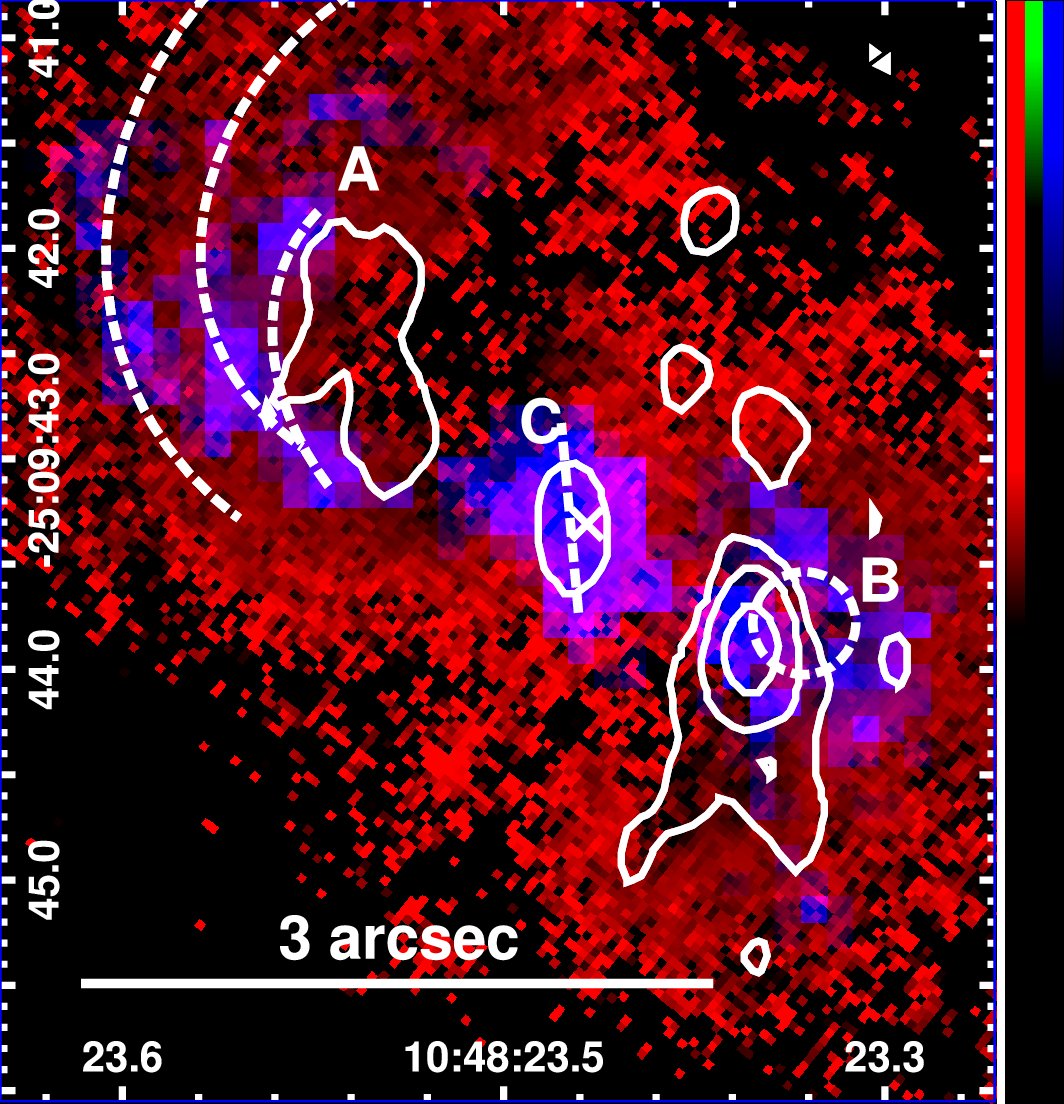}\par
\caption{Detailed [\ion{S}{2}]/H$\alpha$ (red) and 0.3-8\;keV X-ray emission (blue) maps of the nucleus and emission line arms, with X-rays in the arms saturated for emphasis.  X-rays follow the arms, but avoid some regions of higher  [\ion{S}{2}]/H$\alpha$.  Radio contours  (solid) and features are marked as noted in Figs. \ref{fig:ratio} and \ref{fig:shock-ion}.   X-rays trace the outside of curve A, which surrounds an arc of enhanced [\ion{S}{2}]/H$\alpha$ and the edge of the NE radio lobe.  X-rays also trace the edges of the [\ion{O}{3}]/H$\alpha$ gap which shows enhanced [\ion{S}{2}]/H$\alpha$.  X-rays strongly correspond to the nuclear ridge C, as well as the cavity of weak  [\ion{S}{2}]/H$\alpha$ in the SW cone.  As in Fig. \ref{fig:ion-xray}, regions of enhanced X-rays are consistent with the peak  radio emission of the SW lobe, and surround knot B, which shows enhanced  [\ion{S}{2}]/H$\alpha$.  Weaker radio emission fills the southern [\ion{S}{2}]/H$\alpha$ cavity.}
\label{fig:shock-xray}
\end{figure}

In the ionization cones the X-rays are strongest within $\sim3\arcsec$, and are strongest in regions of relatively low  [\ion{O}{3}]/H$\alpha$, whereas [\ion{O}{3}]/H$\alpha$ dominates the cones at larger radii.  Like [\ion{S}{2}]/H$\alpha$ and [\ion{O}{3}]/H$\alpha$, an X-ray curve follows curve A in the NE inner cone (Figs. \ref{fig:shock-ion-xray-panel} center, \ref{fig:ion-xray}, \ref{fig:shock-xray}), but at a different radius from the NE radio lobe, lying between the [\ion{S}{2}]/H$\alpha$ and [\ion{O}{3}]/H$\alpha$ features.  At larger radii ($\sim3\arcsec-4\arcsec$ NE of the nucleus), X-rays may also trace the interface between [\ion{S}{2}]/H$\alpha$ and [\ion{O}{3}]/H$\alpha$ features along the previously-mentioned [\ion{O}{3}]/H$\alpha$ gap.

In the SW cone, the brightest X-rays describe a structure that is less well-described as `S-shaped'.  But the SW cone does contain a clumpy semi-circular structure which surrounds [\ion{S}{2}]/H$\alpha$ knot B and is present within a cavity of relatively weak  [\ion{O}{3}]/H$\alpha$  (Figs. \ref{fig:shock-ion-xray-panel} right, \ref{fig:ion-xray}, \ref{fig:shock-xray}).  One of these X-ray knots is spatially consistent with the peak of radio emission from the SW radio lobe, and may continue south along a north-south [\ion{O}{3}]/H$\alpha$ structure associated with a radio filament.  The remaining knots occupy a space in the western side of the cavity, between [\ion{S}{2}]/H$\alpha$ knot B and surrounding [\ion{O}{3}]/H$\alpha$ structures.

In the outer clouds SW, the [\ion{O}{3}] and X-ray emission are aligned but not co-spatial.  Rather, the X-ray emission is found $\sim1\arcsec$ NW of the [\ion{O}{3}] region (i.e. interior to it).  The nearly-cospatial X-ray structure appears to be counter-clockwise along the outer edge of the central F438W stellar structure, described in \S\ref{omorph}.

\label{s2ha-text}

\subsection{Radio Morphology and Comparison to Other Wavelengths}

The morphology of the X-ray emission and optical emission are strongly defined in relation to the radio emission.  The main radio features consist of the nuclear emission, a NE radio lobe ($r\sim1.3\arcsec$ from the nucleus), and a brighter SW radio lobe ($r\sim0.95\arcsec$) with two $\sim1\arcsec$ filaments extending south of the SW lobe hotspot.  In continuum-subtracted narrow line emission, the S-shaped arms curve around these two radio lobes.  

A ridge of enhanced  [\ion{S}{2}]/H$\alpha$  that traces the edge of the NE radio lobe (marked A in Fig. \ref{fig:ratio}), and a knot immediately east of the SW radio lobe (marked B in Fig. \ref{fig:ratio}).  In the SW cone, the cavity west of knot B is dominated by extended filaments of radio emission connected to the SW jet lobe.  The nuclear radio emission is consistent with the peak of nuclear X-ray emission, and occupies a region of weak [\ion{S}{2}]/H$\alpha$ corresponding with the nucleus.  The nuclear radio emission is cospatial with the only strong [\ion{O}{3}]/H$\alpha$ region in the inner $\sim0.3\arcsec$.

The positions of these radio structures are marked using contours in the three-color X-ray and line ratio images (Figs. \ref{fig:ratio} and \ref{fig:shock-ion-xray-panel}), as well as in two-color images which compare [\ion{O}{3}]/H$\alpha$ vs. [\ion{S}{2}]/H$\alpha$  (Fig. \ref{fig:shock-ion}), [\ion{O}{3}]/H$\alpha$ vs. X-rays (Fig. \ref{fig:ion-xray}), and [\ion{S}{2}]/H$\alpha$ vs. X-rays (Fig. \ref{fig:shock-xray}).

\begin{figure*} 
\noindent
\includegraphics[width=0.32\textwidth]{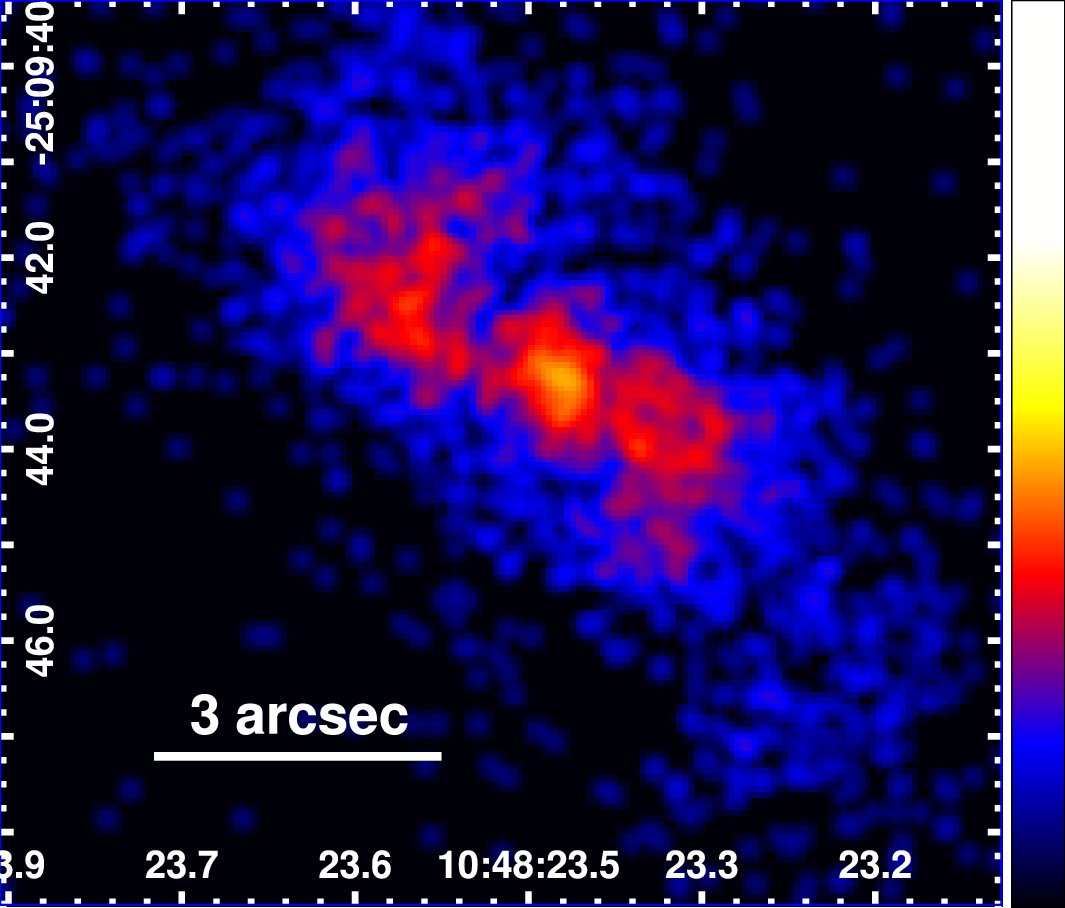}\hspace{0.01\textwidth}%
\includegraphics[width=0.32\textwidth]{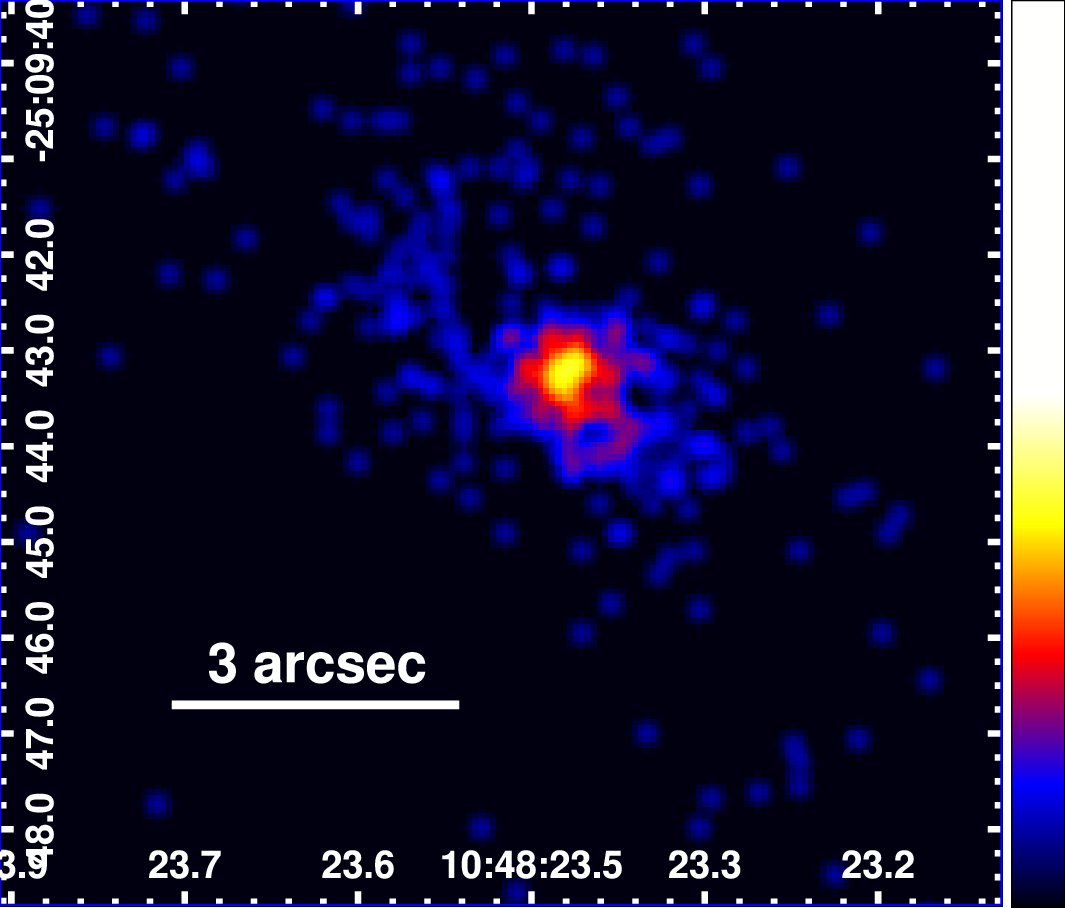}\hspace{0.01\textwidth}%
\includegraphics[width=0.32\textwidth]{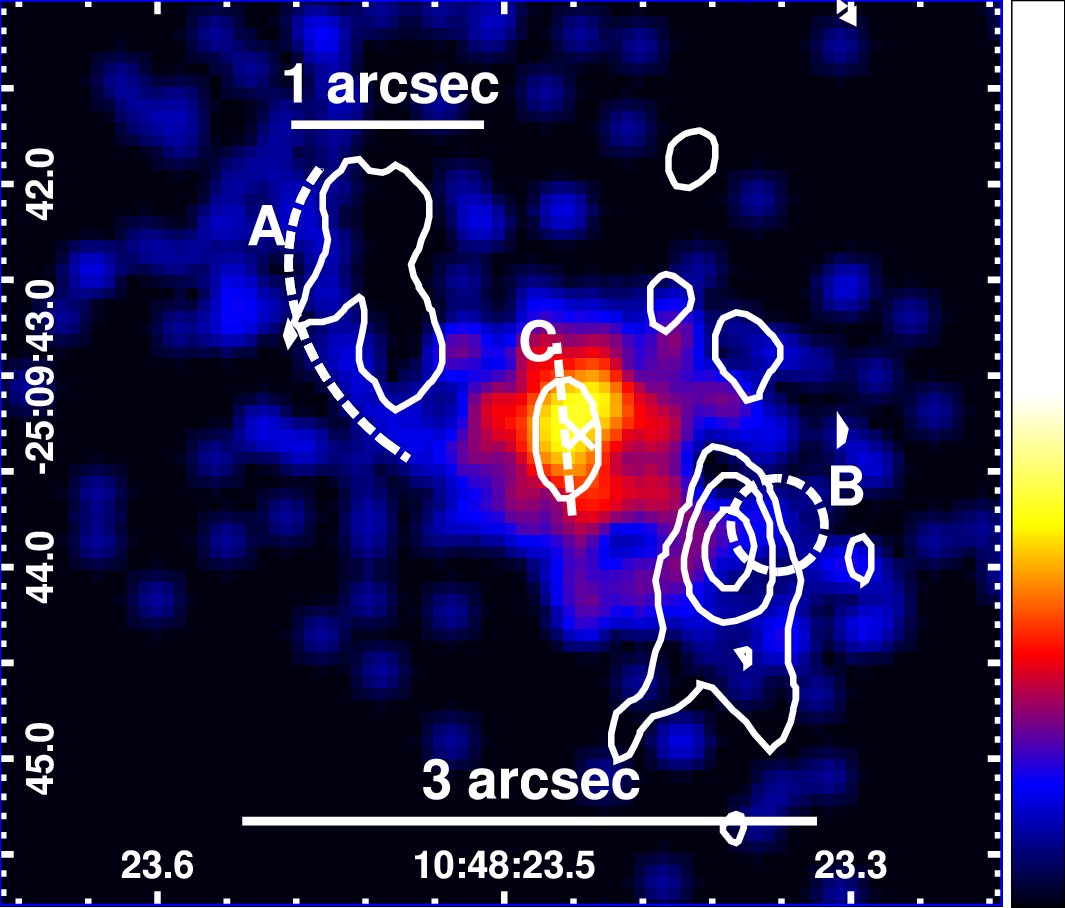}\par
\caption{{\bf Left:} Nucleus and ionization cones at soft ($0.2-2\;$keV) photon energies, binned to 1/8 native pixel size [$0.062\arcsec$] and smoothed with a Gaussian of FWHM 3 image pixels [$0.186\arcsec$].  {\bf Center:}  As before, but at hard ($2-8\;$keV) energies.  Even at hard energies, the emission is significantly extended in the direction of the cones, with hints of a hard ridge on the SW edge of the SW outer cone.  {\bf Right:} magnification of hard ($2-8\;$keV) X-rays, as in the center panel, with radio contours and labels overlaid, as before.  A filament of hard X-ray emission traces the NE radio lobe and curve A, but hard X-rays are conspicuously absent from the lobe itself.} 
\label{fig:xray-bands}
\end{figure*}

We take a first approach at energy-resolved X-ray morphology by examining smoothed images.  We will pursue more detailed energy-resolved X-ray analysis in a subsequent paper.  We compare X-ray emission at $0.2-2\;$keV (soft) and $2-8\;$keV (hard) using images binned at 1/8 ACIS native pixel resolution and smoothed with a Gaussian kernel with a radius of 3 image pixels, as shown in Fig. \ref{fig:xray-bands}.  X-ray emission is extended in the soft band, but we also see extended emission in the hard band in addition to the hard point source previously studied by \cite{Fabbiano11} and \cite{Koss15}.  Accounting for contamination from the PSF of the central source, we detect hard emission at $\simgreat17\sigma$ in each of the ionization cones, assuming two annular wedges covering $1\arcsec<r<5\arcsec$.  In the NE inner cone, we also observe a tenuous hard X-ray ridge associated with ridge A, along with a conspicuous absence of X-ray emission from within the NE radio lobe.  Measured at $2-4\,$keV to reduce contamination from the AGN PSF, we detect ridge A in a $0.4\arcsec\times2\arcsec$ box at $9.5\sigma$ above background and find that $2-4\,$keV X-ray emission from ridge A must be a factor of $>3.8$ brighter than the undetected radio-emitting region at $2\sigma$.
In the SW inner cone, the extended hard emission appears associated with the SW radio lobe, or between the lobe and the nucleus.

\subsection{Regional Diagnostics}

\subsubsection{BPT Diagrams} 

\begin{figure*} 
\centering
\noindent
\begin{minipage}{0.4\textwidth}
\includegraphics[width=\textwidth]{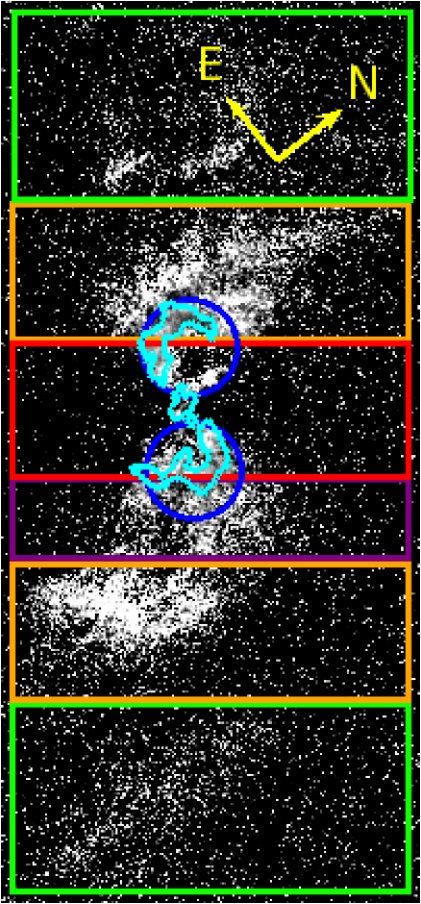}
\end{minipage}\hspace{0.05\textwidth}%
\begin{minipage}{0.55\textwidth}
\includegraphics[width=\textwidth,angle=0,origin=c]{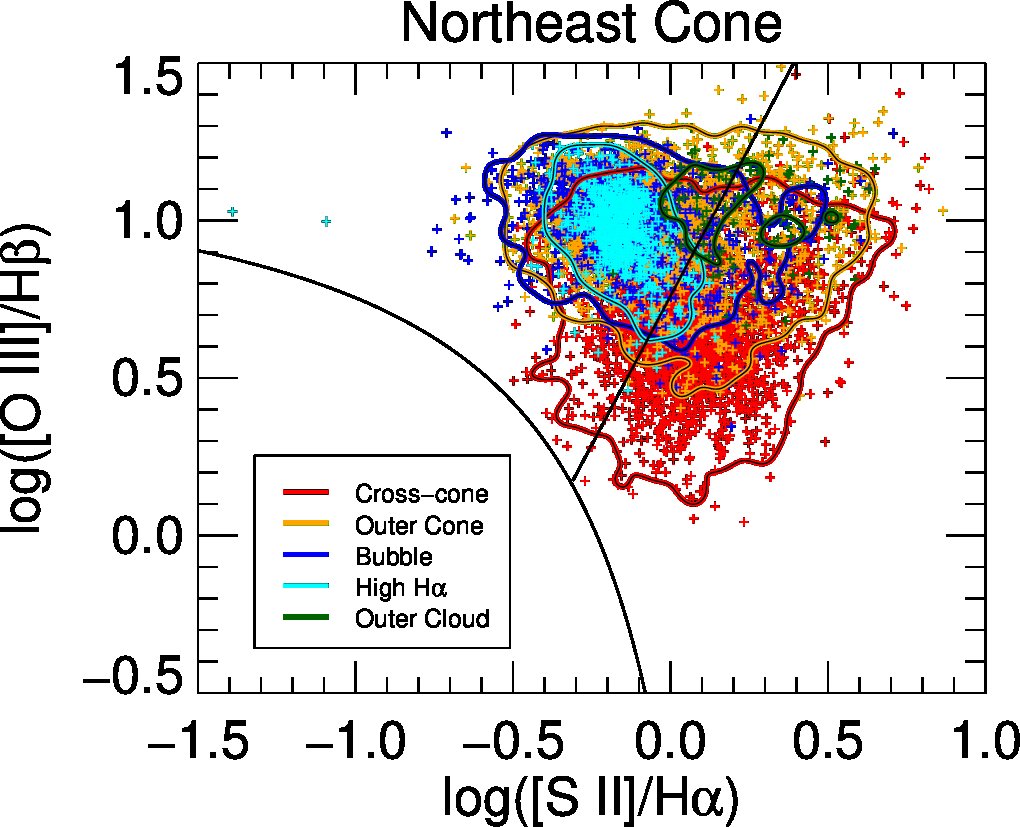}\par
\includegraphics[width=\textwidth,angle=0,origin=c]{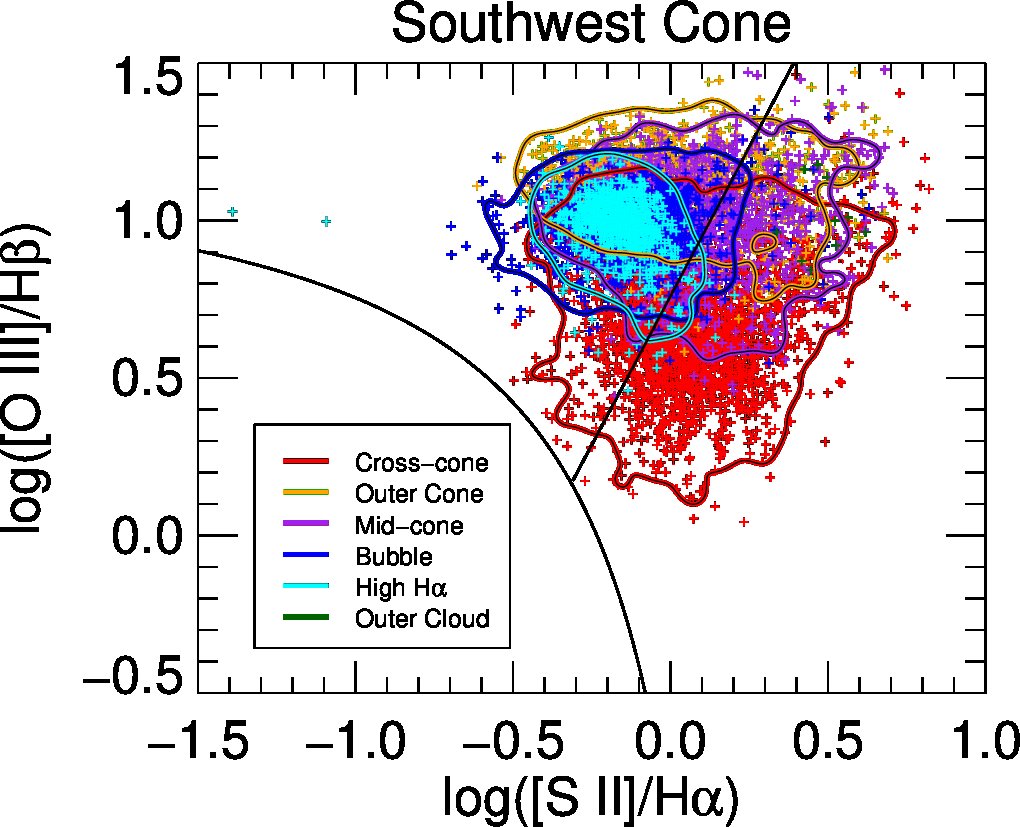}
\end{minipage}
\caption[Bicone line map and BPT diagrams]{{\bf Left:} Line ratio map of [\ion{O}{3}]/[\ion{S}{2}] from the central $19\arcsec\times8\arcsec$ of NGC 3393.  The map is divided into colored boxes.  Each color indicates different sub-regions from which BPT diagrams (right) were extracted.  Here, green represents the outer clouds,  orange represents the outer cones, and red is the cross-cone region (and depicted in both NE cone and SW cone BPT diagrams).  Cyan contours indicate the regions of brightest H$\alpha$ emission.  Blue circles indicate bubbles of high-ionization entirely enclosed within the large cocoon of elevated [\ion{S}{2}] /H$\alpha$ (seen e.g. in Figs. \ref{fig:ratio} and \ref{fig:two-comp}).  Only the SW cone has a mid-cone region (purple) defined by prominent regions where strong  [\ion{S}{2}] /H$\alpha$ and  [\ion{O}{3}] /H$\alpha$ co-exist, which is also easily separated from both the inner cone and the high-[\ion{O}{3}] /H$\alpha$ regions of the outer cones.
\\

{\bf Top-right:}  BPT diagram of individual pixels along the NE cone.  Black lines mark different BPT regions, with star formation at the lower left of each plot, Seyfert-like activity at the upper-left, and LINER-like activity on the right.  Colors indicate sub-regions from the  [\ion{O}{3}]/[\ion{S}{2}]  in this figure (left).  Contours emphasize parameter space with a large concentration of pixels, to clarify areas with significant overlap between multiple extraction regions.  {\bf Bottom-right:}  BPT diagram, as before, but for the SW cone.  The cross-cone region is common to both diagrams.} 
\label{fig:BPT}
\end{figure*}

Direct comparison of the relative strengths of multiple narrow emission lines is a well-established method of distinguishing between star formation and different forms of nuclear black hole activity as competing processes for the origin of those emission lines.  Such diagnostics have an empirical basis \citep{BPT81} and have been supported by various subsequent theoretical models \citep[e.g.][]{ED85,Kewley01}.  Although this is typically used to distinguish the dominant forms of emission in different galaxies, it can also be used \citep[as in][for NGC 3393 with ground-based spectra]{Cooke00} to examine the role of AGN photoionization in different portions of a single galaxy.  In Fig. \ref{fig:BPT}, we plot a BPT \citep{BPT81} diagram, using [\ion{O}{3}]/H$\beta$ on the vertical axis vs. [\ion{S}{2}]/H$\alpha$ for each $0.04\arcsec\times0.04\arcsec$ WFC3 pixel in the central $19\arcsec\times8\arcsec$.  Since we do not have direct measurements of H$\beta$ in our line emission maps, we assume H$\alpha$/H$\beta$$\sim3.0$, as is typical in AGN.  In order to distinguish between emission characteristic of star formation, Seyfert galaxies, and Low Ionization Nuclear Emission Regions (LINERs), we use the \cite{Kewley06} formalism for  BPT diagrams based on  [\ion{O}{3}] and [\ion{S}{2}], using the ratio map images we created from the WFC3 data.

We use colors to differentiate the sub-region of origin for each ratio map pixel.  The locations of these sub-regions are also marked by lines of the same color on the [\ion{O}{3}]/[\ion{S}{2}] ratio map of the ionization cones in Figure \ref{fig:BPT}, and are described in the captions as the cross-cone, mid-cone, outer cones, outer clouds, high-H$\alpha$ regions, and high-ionization bubbles.  Contours indicate surfaces of constant pixels\;dex$^{-2}$, to distinguish the most densely occupied portions of parameter space where there may be significant overlap between regions.

\begin{figure*} 
\centering
\noindent
\begin{tabular}{ccc}
%\includegraphics[width=0.45\textwidth]{bpt_contour_ne_ocloud.jpg} &
%\includegraphics[width=0.45\textwidth]{bpt_contour_ne_bubble.jpg} \\
%\includegraphics[width=0.45\textwidth]{bpt_contour_ne_ocone.jpg} &
%\includegraphics[width=0.45\textwidth]{bpt_contour_ne_hacontours.jpg} \\
%&
%\includegraphics[width=0.45\textwidth]{bpt_contour_cross.jpg}\\
%\vspace{-0.1in}
\includegraphics[width=0.31\textwidth]{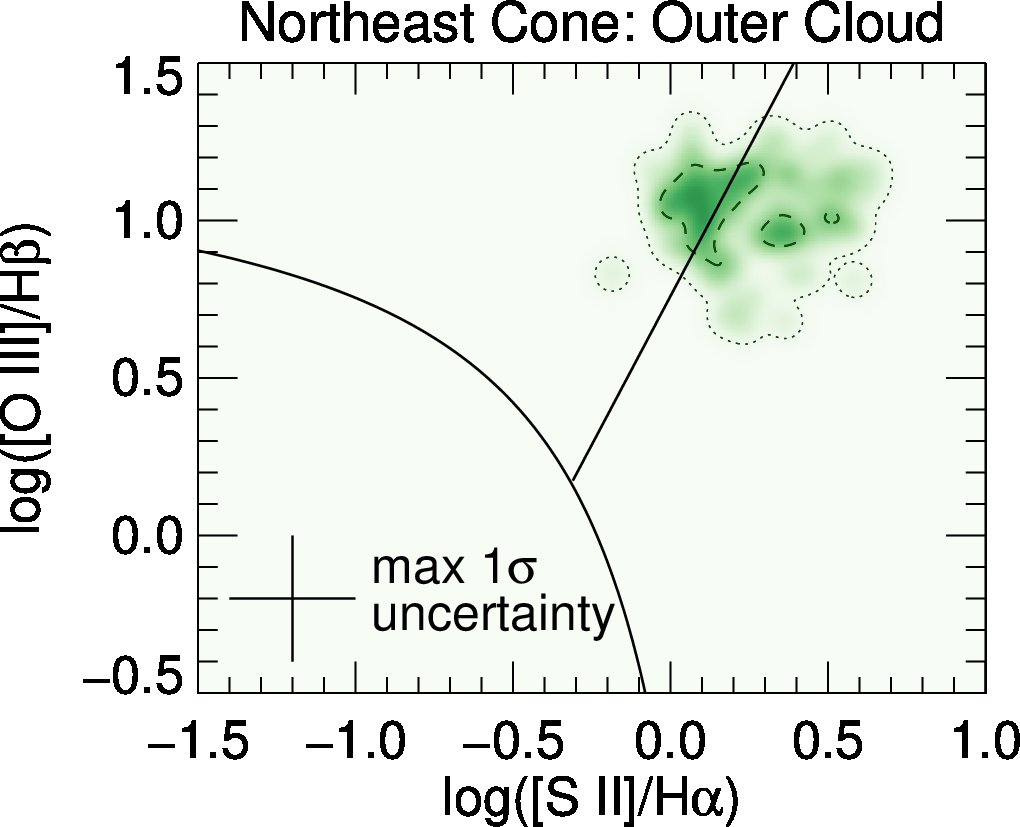} &
\includegraphics[width=0.31\textwidth]{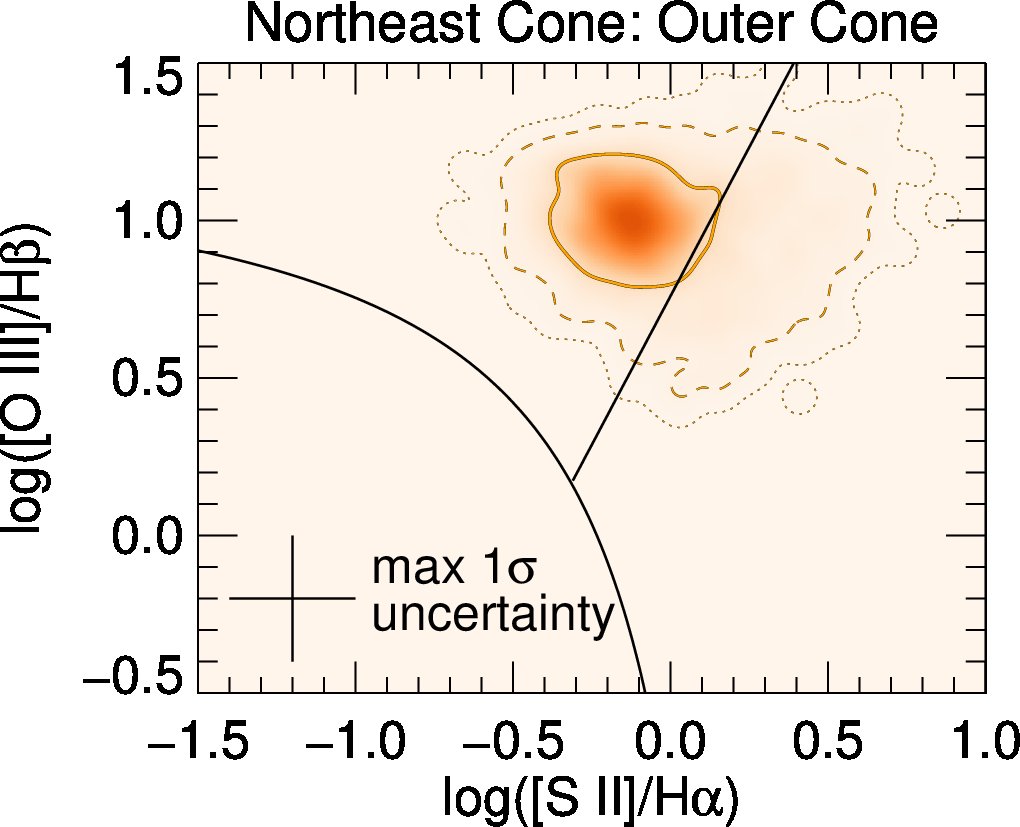} &
\raisebox{0.6\height}{\includegraphics[width=0.32\textwidth]{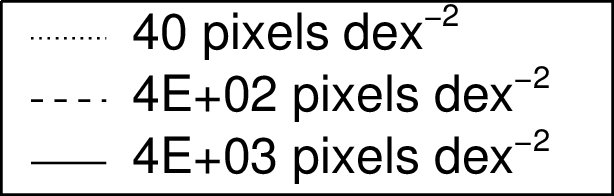} }
\\
%\vspace{-0.1in}
\includegraphics[width=0.31\textwidth]{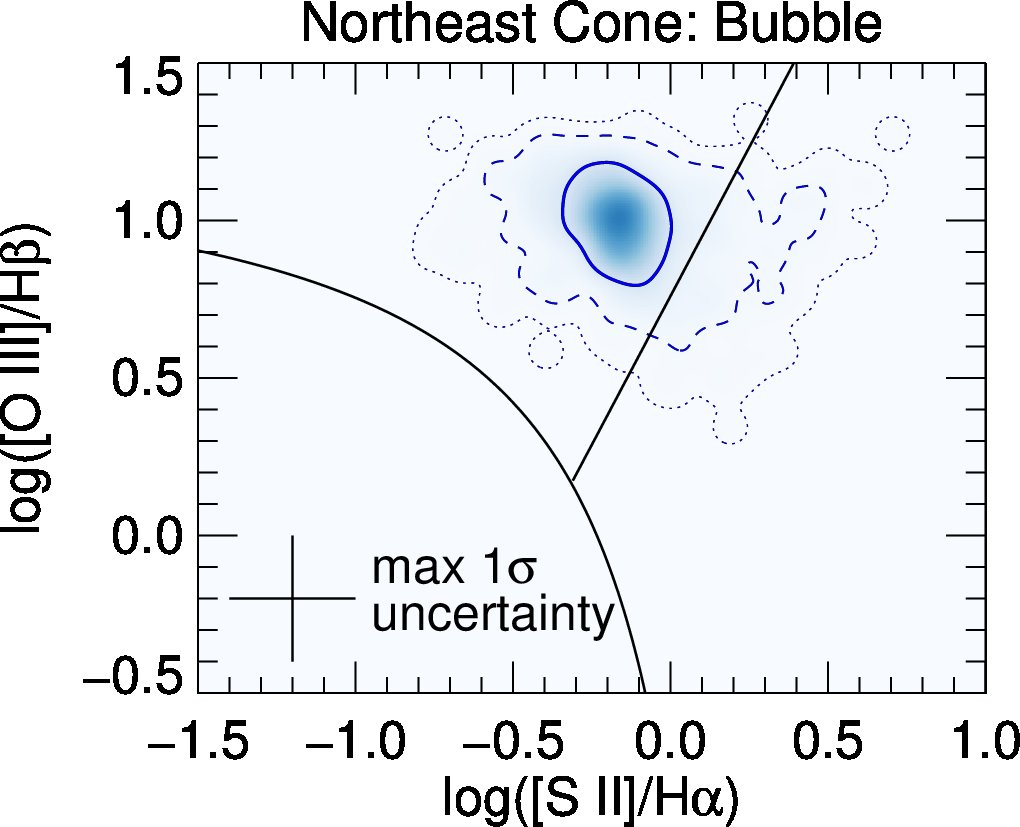} &
\includegraphics[width=0.31\textwidth]{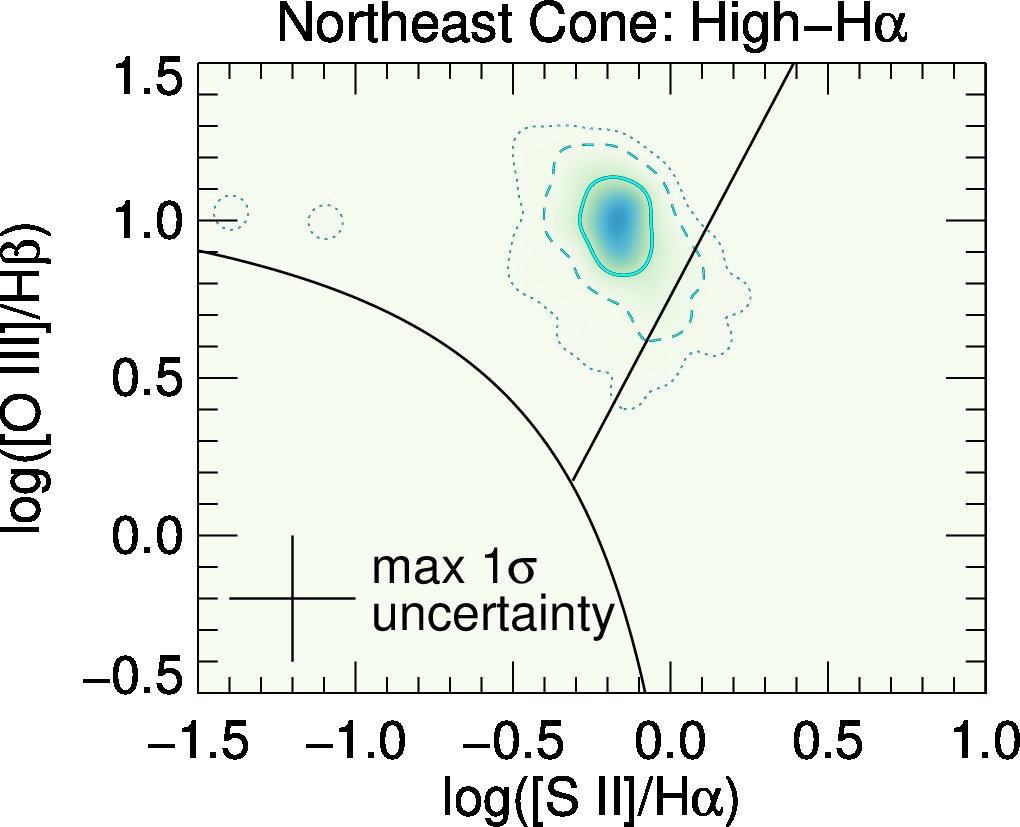}&
\\
%\end{tabular}
%\end{centering}
%\label{fig:BPTconNE}
%\end{figure*}
%\begin{figure*} 
%\begin{centering}
%\noindent
%\begin{tabular}{ccc}
%\includegraphics[width=0.45\textwidth]{bpt_contour_sw_ocloud.jpg} &
%\includegraphics[width=0.45\textwidth]{bpt_contour_sw_bubble.jpg} \\
%\includegraphics[width=0.45\textwidth]{bpt_contour_sw_ocone.jpg} &
%\includegraphics[width=0.45\textwidth]{bpt_contour_sw_hacontours.jpg} \\
%\includegraphics[width=0.45\textwidth]{bpt_contour_sw_mcone.jpg} &
%\includegraphics[width=0.45\textwidth]{bpt_contour_cross.jpg}\\
%\vspace{-0.1in}
\includegraphics[width=0.31\textwidth]{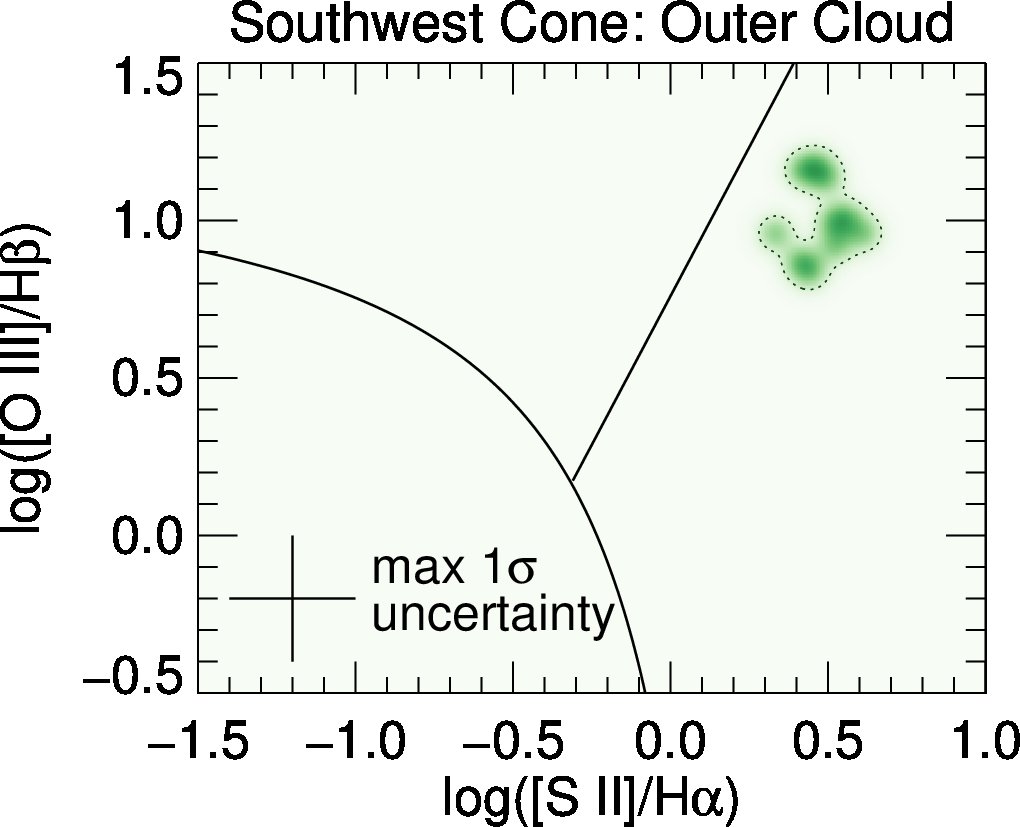} &
\includegraphics[width=0.31\textwidth]{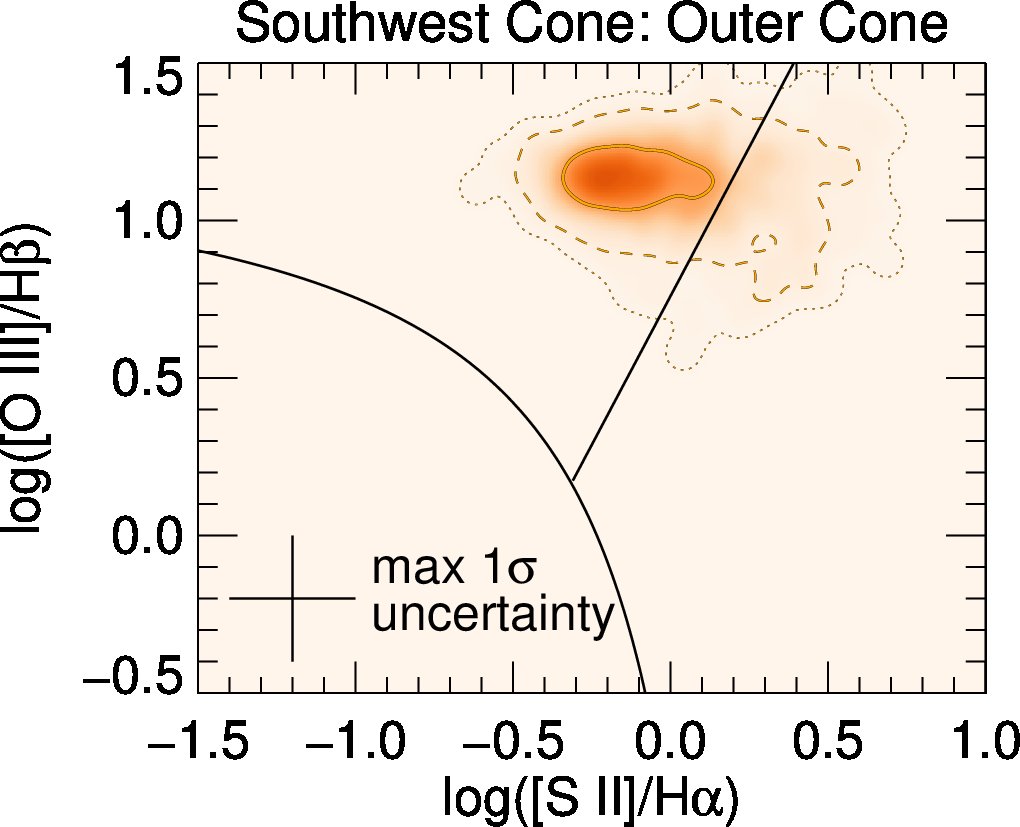} &
\includegraphics[width=0.31\textwidth]{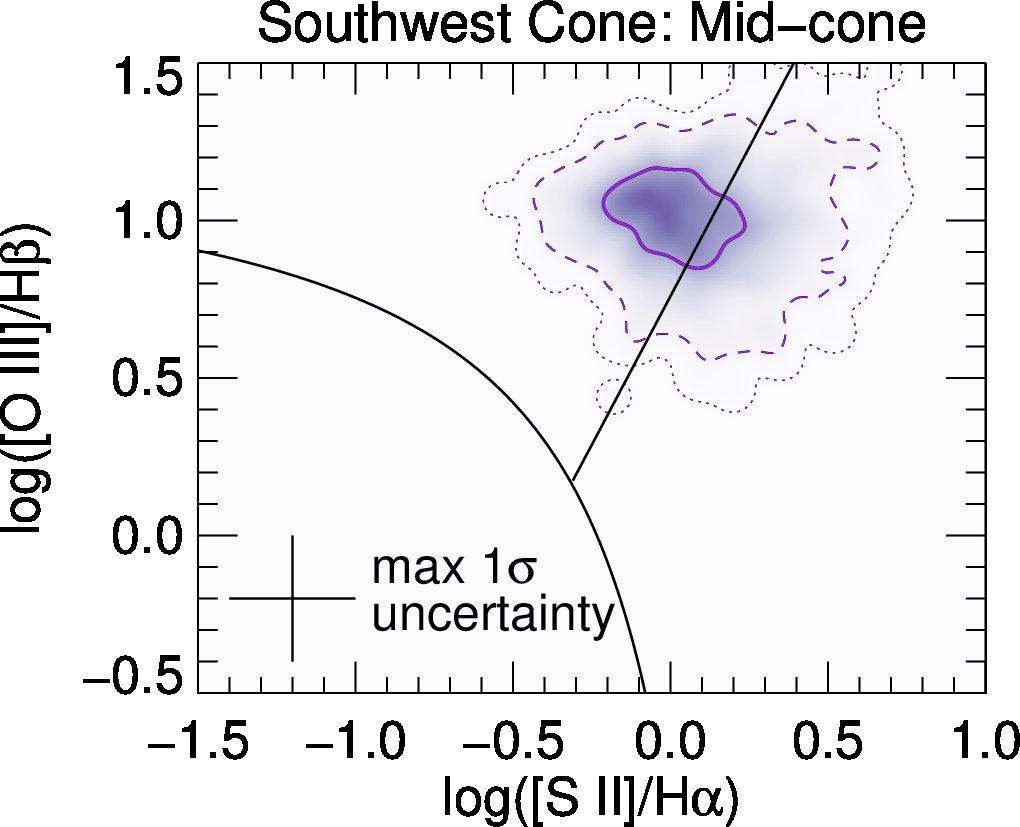} \\
%\vspace{-0.1in}
\includegraphics[width=0.31\textwidth]{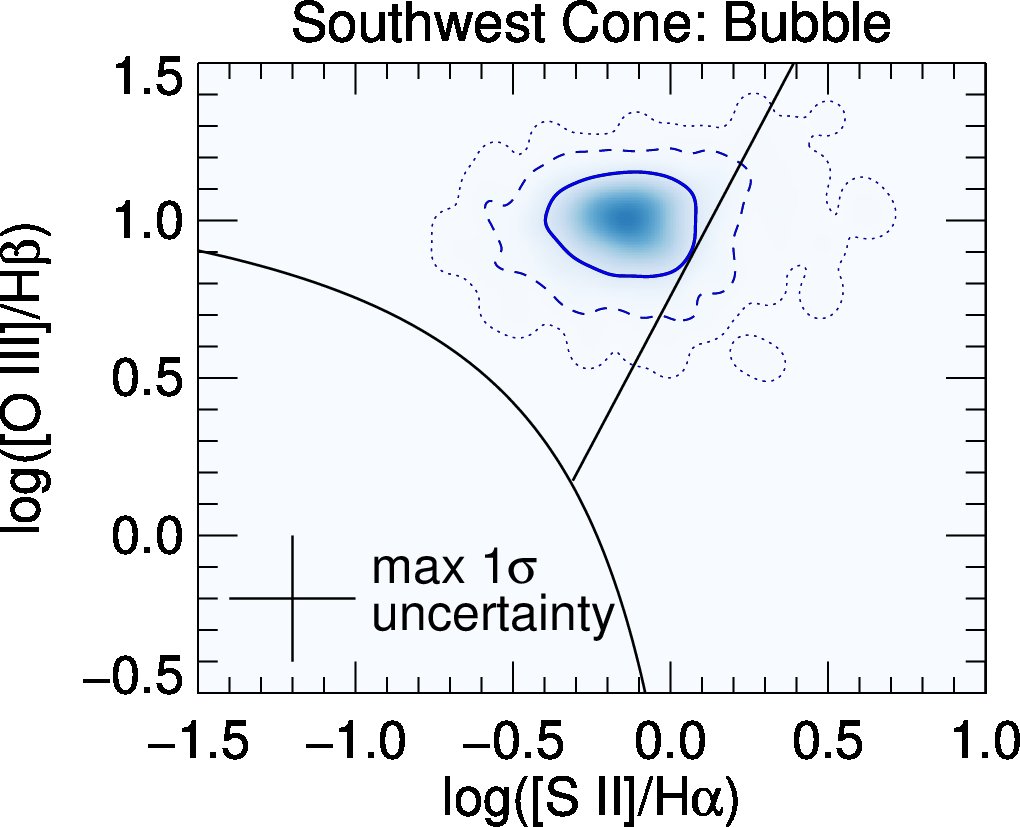} &
\includegraphics[width=0.31\textwidth]{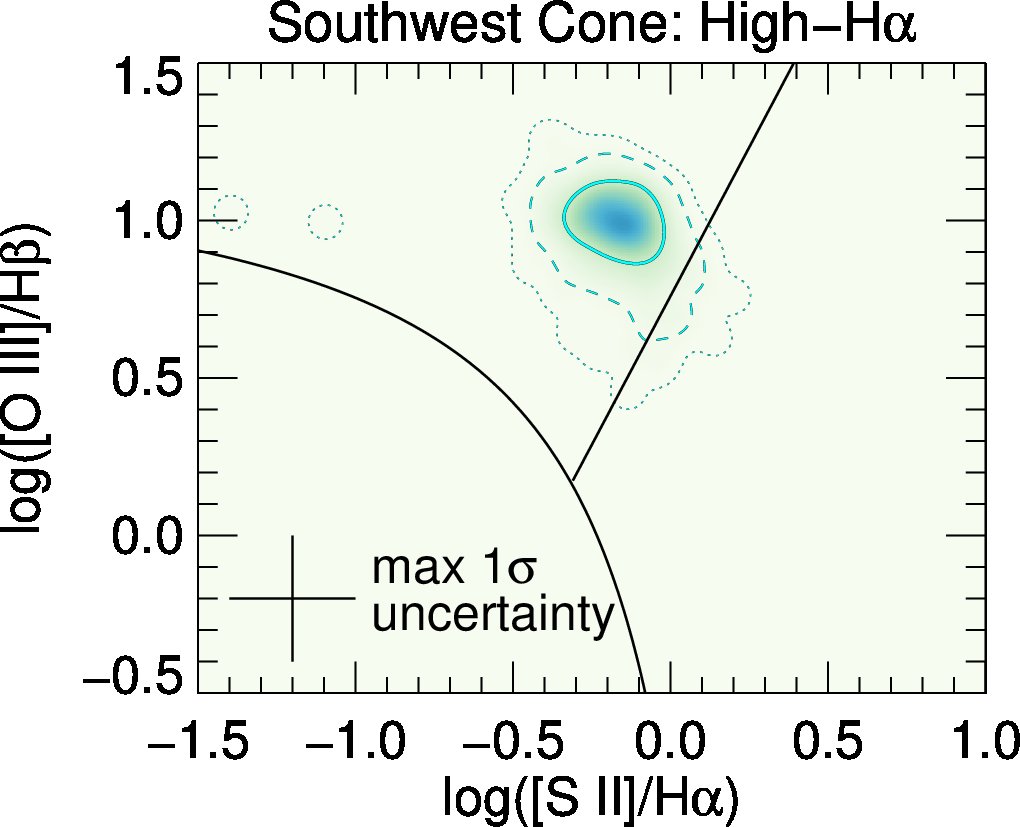} &
\includegraphics[width=0.31\textwidth]{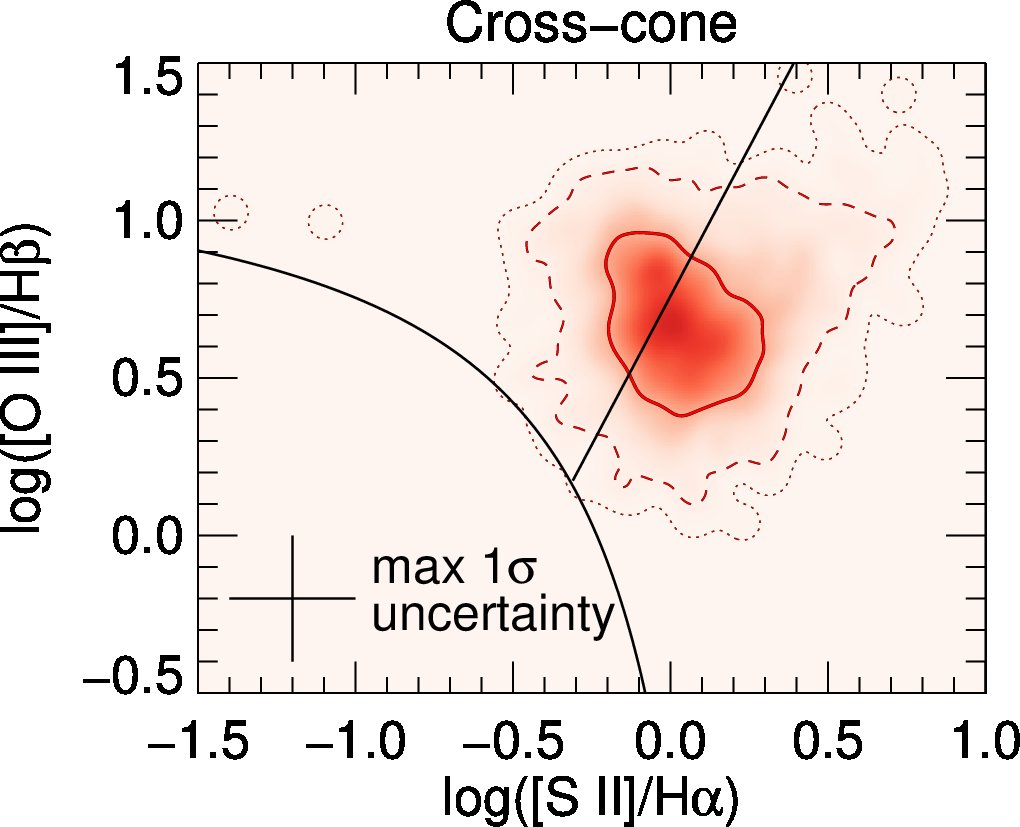}\\
\end{tabular}
\caption{BPT diagrams of individual sub-regions in the cones, with the same color representation as in Figure \ref{fig:BPT}.  Different contour levels (indicated in the top right panel) represent the number density of WFC3 pixels in the parameter space of the diagram, with large values corresponding to more WFC3 pixels and therefore a larger fraction of the region.  Color shading has the same representation as contours, with darker regions indicating large parameter space density, and is smoothed for clarity.  {\bf Row 1, left-to-right:} the outer cloud and outer cone of the NE cone.  {\bf Row 2, left-to-right:} The NE bubble and high-H$\alpha$ region.  The bubble includes high-H$\alpha$ pixels, but does not overlap spatially with other extraction regions. {\bf Row 3, left-to-right:} the outer cloud, outer cone and mid-cone of the SW cone.  {\bf Row 4, left-to-right:} The SW bubble, high-H$\alpha$ region, and cross-cone.} 
\label{fig:BPTcone}
\end{figure*}

For clarity, we plot each of these sub-regions separately for the NE and SW cones (Figure \ref{fig:BPTcone}).  In each plot, the color corresponds to the same color used to designate regions in Figure \ref{fig:BPT}.

Examining Figure \ref{fig:BPT}, we notice several noteworthy trends.  First, the high-H$\alpha$ and bubble regions are almost entirely within the Seyfert region of the BPT diagram.  Other regions have significant Seyfert components, but straddle the Seyfert/LINER dividing line.  The cross-cone region in particular is systematically more concentrated in the LINER region, with lower values of [\ion{O}{3}]/H$\alpha$ for its major parameter space locus, and higher values of [\ion{S}{2}]/H$\alpha$.  Very few pixels from any of these sub-regions can be attributed to star formation.  The outer clouds have a large fraction of their points in the LINER region, but also produce many of the highest-excitation ([\ion{O}{3}]/H$\beta$) points.  In the SW cone, we see a systematic trend in excitation moving between regions and away from the nucleus, such that the SW bubble has lower excitation than the SW mid-cone, which is lower-excitation than the outer cone.  Likewise, each region's locus extends more significantly towards low [\ion{S}{2}]/H$\alpha$ as the regions move away from the nucleus within the SW cone. 

The ability of this diagram to probe regions which are both outside the ionization cone and at large radii is limited by the fact that such regions are commonly deficient in at least one emission line, introducing uncertainty.

\subsubsection{Ionization vs. H$\alpha$} 

\begin{figure*} 
%\begin{centering}
%\noindent
\begin{minipage}{\textwidth}
\includegraphics[width=0.45\textwidth,angle=0,origin=c]{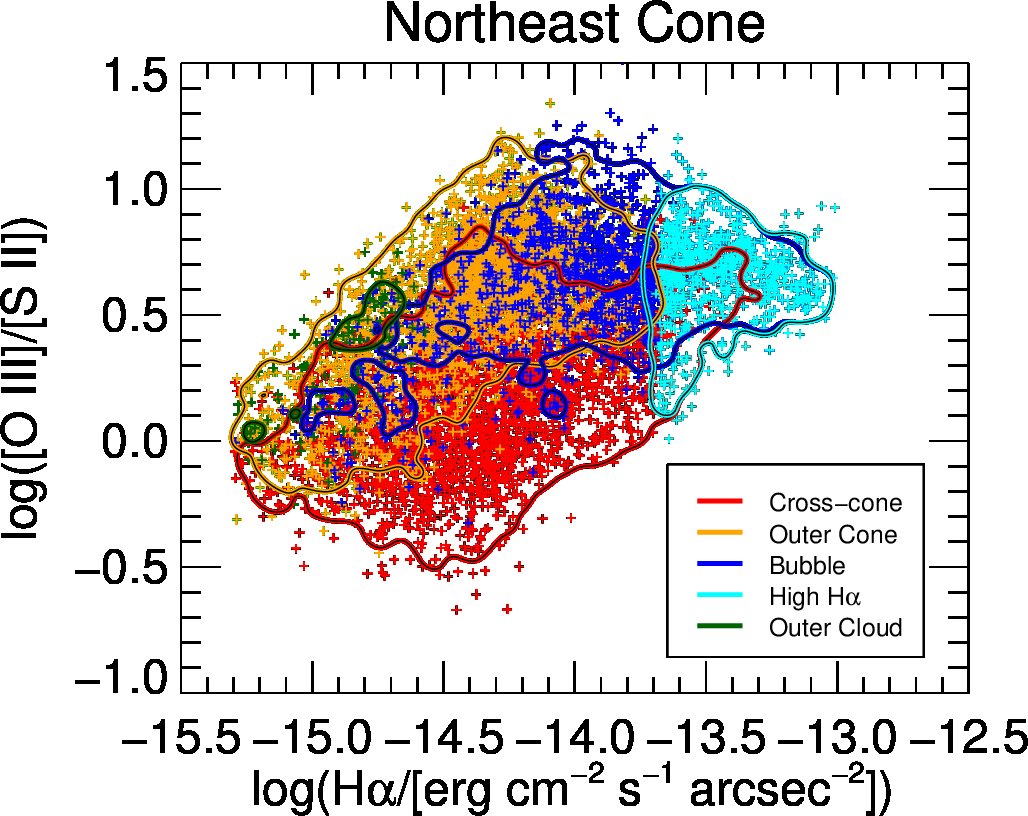}~
\includegraphics[width=0.45\textwidth,angle=0,origin=c]{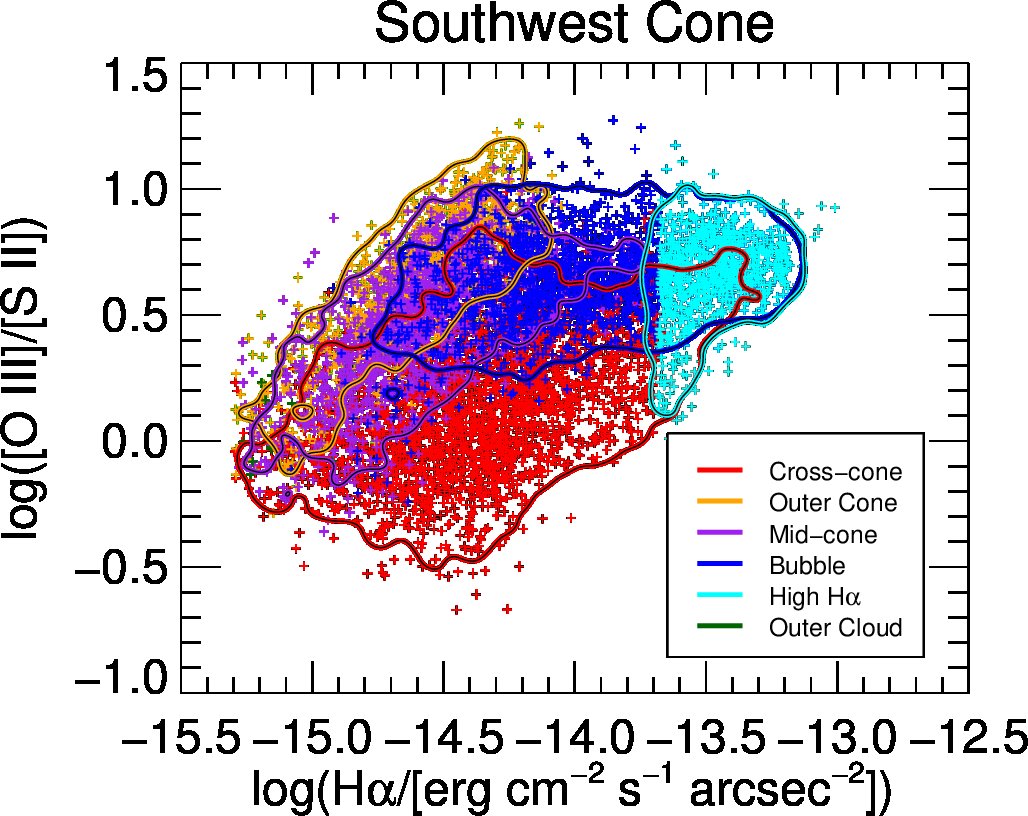}
\end{minipage}
%\includegraphics[angle=270,origin=c]{bpt_contour_ne.jpg}\par
%\includegraphics[angle=270,origin=c]{bpt_contour_sw.jpg}
%\end{centering}
\caption{Line ratio diagrams of the central $19\arcsec\times8\arcsec$ of NGC 3393, showing log([\ion{O}{3}]/[\ion{S}{2}]) (vertical) vs. log(H$\alpha$) (horizontal).   Colors indicate different regions, with the same locations as indicated in the [\ion{O}{3}]/[\ion{S}{2}] map in Fig. \ref{fig:BPT}.  The same cross-cone region is common to both diagrams.  As in Fig. \ref{fig:BPT}, contours are used to emphasize parameter space with a large concentration of pixels, to clarify areas with significant overlap between multiple extraction regions.  {\bf Left:} From the NE ionization cone. {\bf Right:}  From the SW ionization cone.} 
\label{fig:o3s2ha}
\end{figure*}

\begin{figure*} 
\centering
\noindent
\begin{tabular}{ccc}
%\includegraphics[width=0.45\textwidth]{o3s2_ha_ne_ocloud.jpg} &
%\includegraphics[width=0.45\textwidth]{o3s2_ha_ne_bubble.jpg} \\
%\includegraphics[width=0.45\textwidth]{o3s2_ha_ne_ocone.jpg} &
%\includegraphics[width=0.45\textwidth]{o3s2_ha_ne_hacontours.jpg} \\
%&
%\includegraphics[width=0.45\textwidth]{o3s2_ha_cross.jpg}\\
%\vspace{-0.1in}
\includegraphics[width=0.31\textwidth]{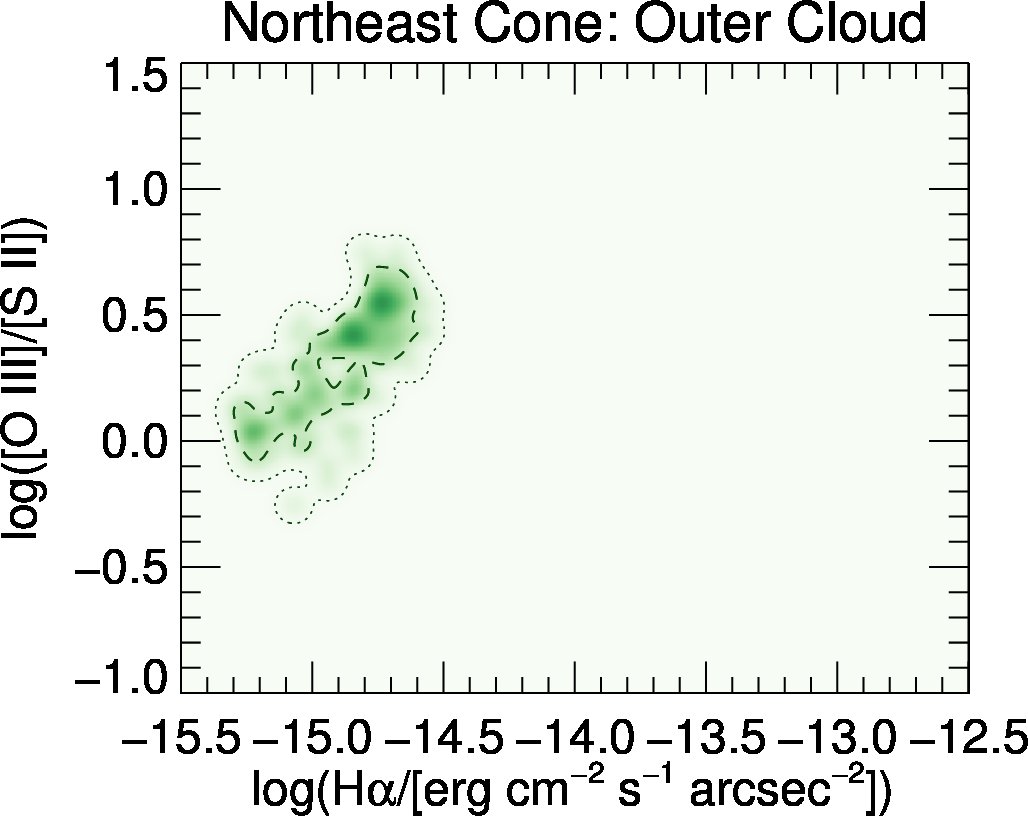} &
\includegraphics[width=0.31\textwidth]{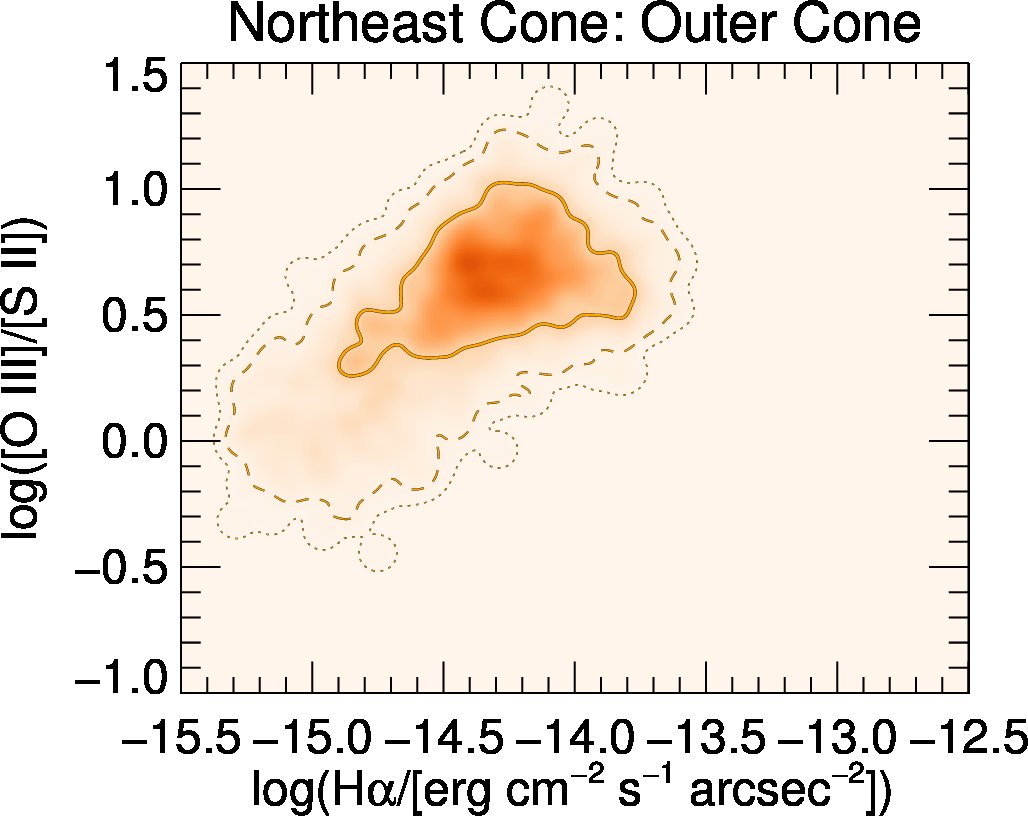} &
\raisebox{0.6\height}{\includegraphics[width=0.32\textwidth]{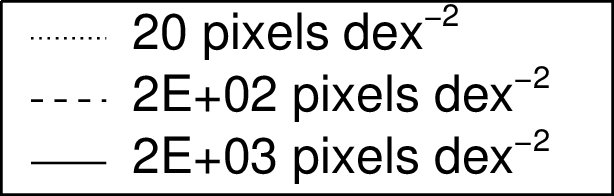} }
\\
%\vspace{-0.1in}
\includegraphics[width=0.31\textwidth]{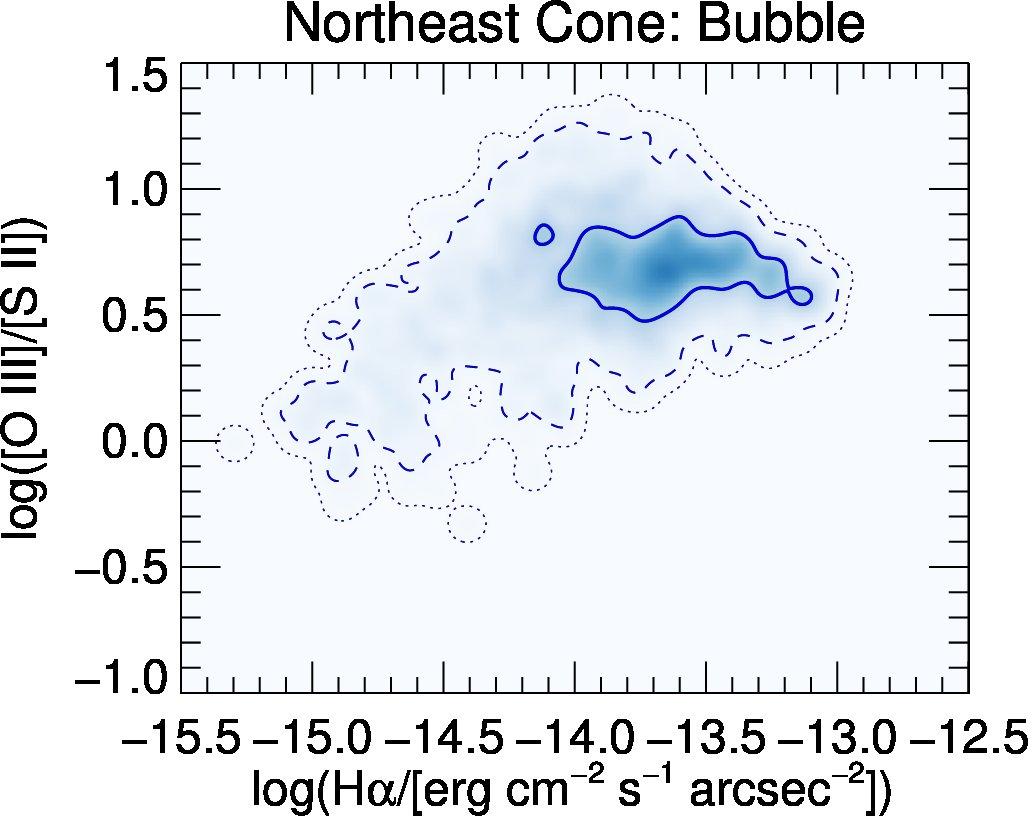} &
\includegraphics[width=0.31\textwidth]{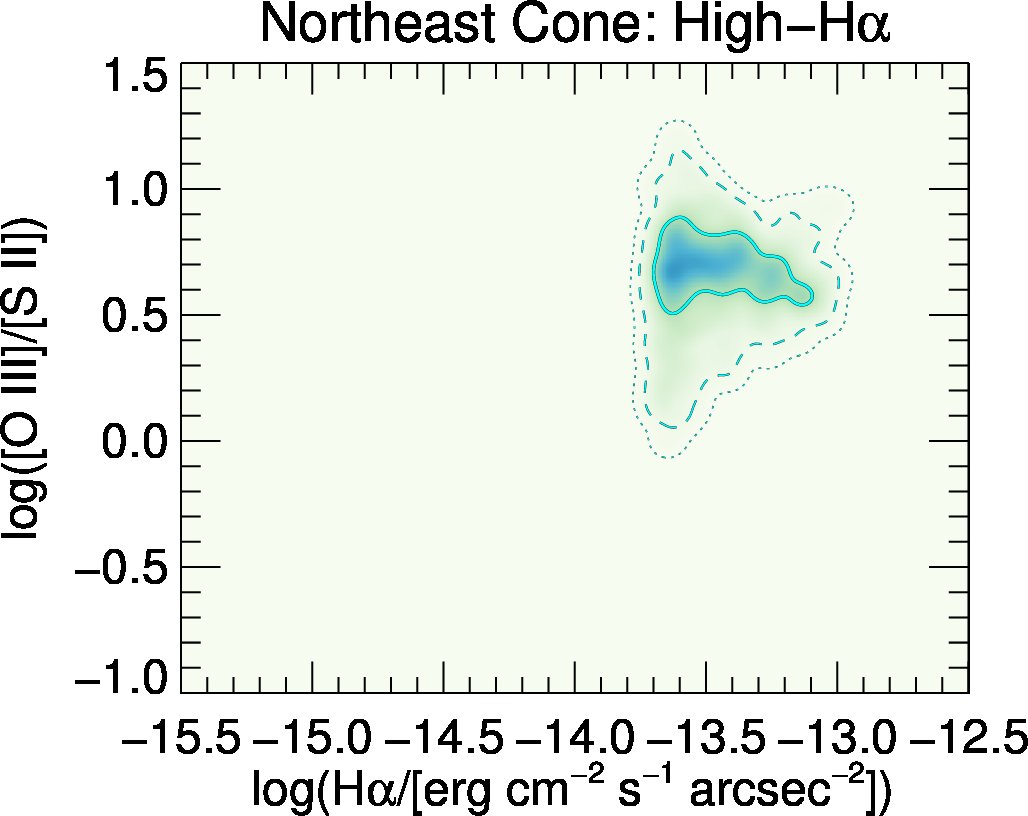} &
\\
%\end{tabular}
%\end{centering}
%\label{fig:o3s2ha-conNE}
%\begin{centering}
%\noindent
%\begin{tabular}{ccc}
%\includegraphics[width=0.45\textwidth]{o3s2_ha_sw_ocloud.jpg} &
%\includegraphics[width=0.45\textwidth]{o3s2_ha_sw_bubble.jpg} \\
%\includegraphics[width=0.45\textwidth]{o3s2_ha_sw_ocone.jpg} &
%\includegraphics[width=0.45\textwidth]{o3s2_ha_sw_hacontours.jpg} \\
%\includegraphics[width=0.45\textwidth]{o3s2_ha_sw_mcone.jpg} &
%\includegraphics[width=0.45\textwidth]{o3s2_ha_cross.jpg}\\
%\vspace{-0.1in}
\includegraphics[width=0.31\textwidth]{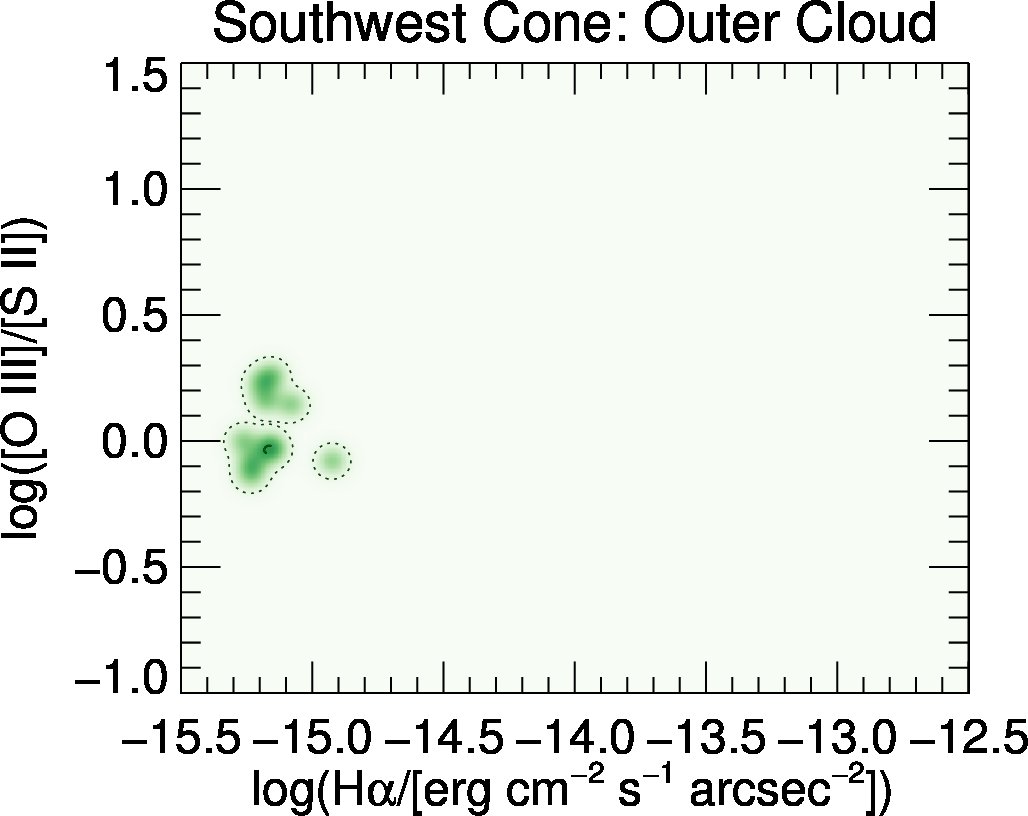} &
\includegraphics[width=0.31\textwidth]{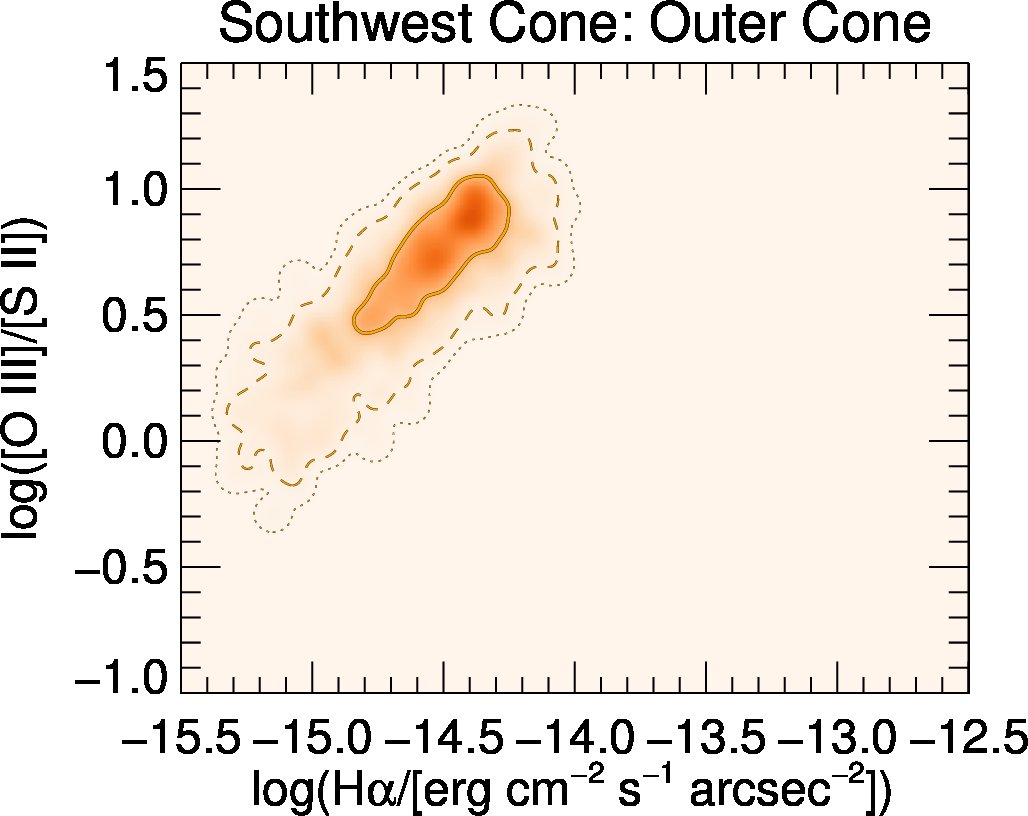} &
\includegraphics[width=0.31\textwidth]{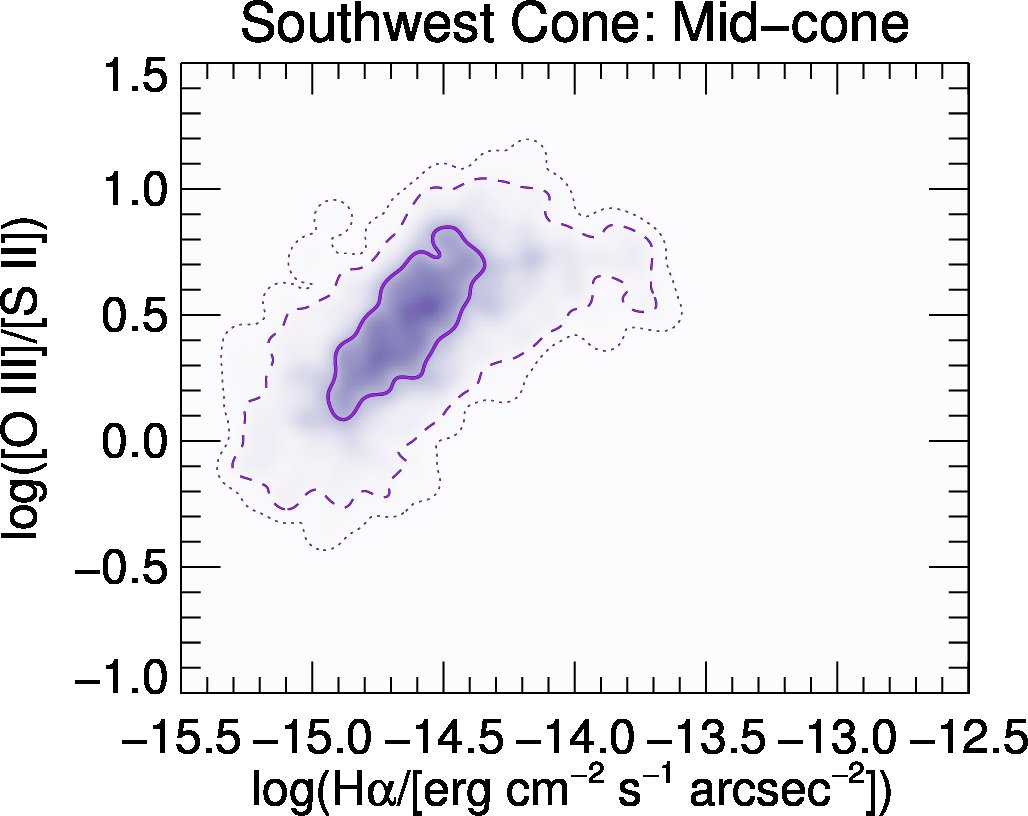}\\
%\vspace{-0.1in}
\includegraphics[width=0.31\textwidth]{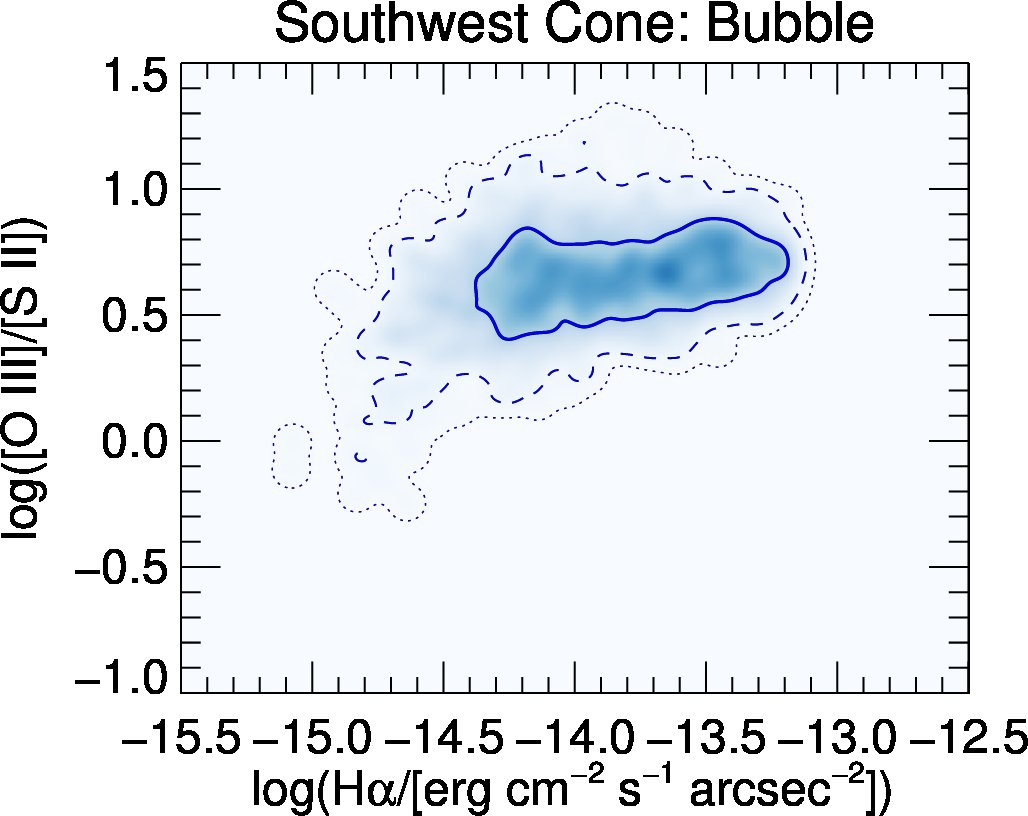} &
\includegraphics[width=0.31\textwidth]{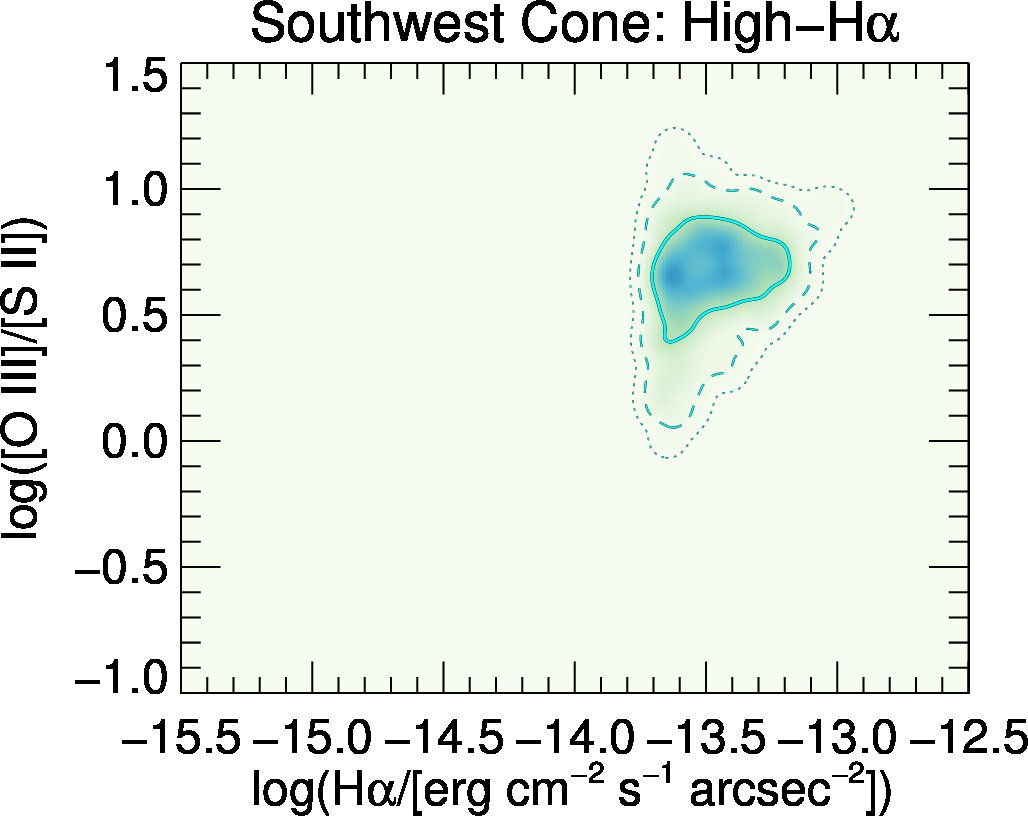} &
\includegraphics[width=0.31\textwidth]{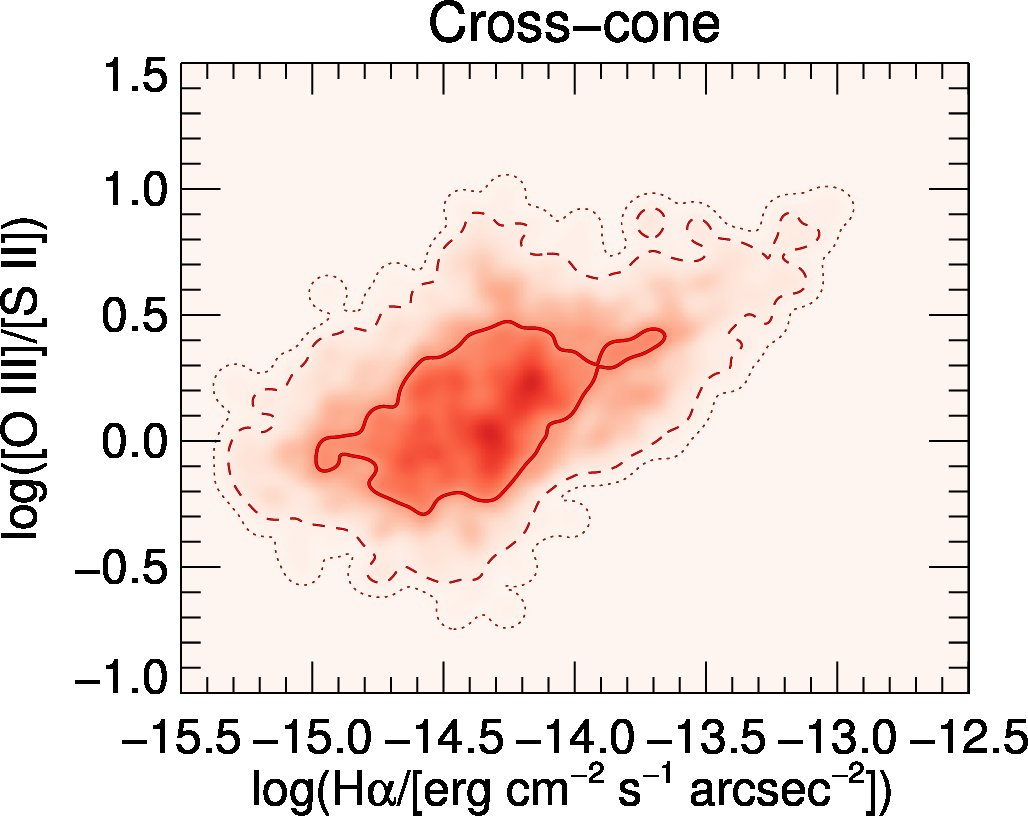}\\
\end{tabular}
\caption{Diagrams of log([\ion{O}{3}]/[\ion{S}{2}]) vs. log(H$\alpha$) for individual sub-regions of the cones, with the same color representation as in Figure \ref{fig:o3s2ha}.  Different contour levels (indicated in the top right panel) represent the number density of WFC3 pixels in the parameter space of the diagram, with large values corresponding to more WFC3 pixels and therefore a larger fraction of the region.  Color shading has the same representation as contours, with darker regions indicating large parameter space density, and is smoothed for clarity.  {\bf Row 1, left-to-right:} the outer cloud and outer cone of the NE cone.  {\bf Row 2, left-to-right:} The NE bubble and high-H$\alpha$ region.  The bubble includes high-H$\alpha$ pixels, but does not overlap spatially with other extraction regions. {\bf Row 3, left-to-right:} the outer cloud, outer cone and mid-cone of the SW cone.  {\bf Row 4, left-to-right:} The SW bubble, high-H$\alpha$ region, and cross-cone.} 
\label{fig:o3s2ha-cone}
\end{figure*}

In Figure \ref{fig:o3s2ha}, we also investigate relative ionization indicated by individual WFC3 pixels (in terms of [\ion{O}{3}]/[\ion{S}{2}]) as a function of H$\alpha$ strength, using the same spatial sub-regions defined for Fig. \ref{fig:BPT}.  Several trends are apparent from these diagrams.  First, within the cones themselves, most points in this region have $1\simgreat\;$log[\ion{O}{3}]/[\ion{S}{2}]$\;\simgreat 0$ regardless of distance from the nucleus or strength of H$\alpha$.  The cross-cone sub-region, however, has systematically lower [\ion{O}{3}]/[\ion{S}{2}] than all other sub-regions, and lower values of [\ion{O}{3}]/[\ion{S}{2}] become more common with decreasing H$\alpha$.  The notable exceptions in the cross-cone region include several high-[\ion{O}{3}]/[\ion{S}{2}], high H$\alpha$ points associated with the nucleus (labeled `C' in Fig. \ref{fig:ratio}).  For clarity, we plot each of the sub-regions separately for both the NE and SW cones (Fig. \ref{fig:o3s2ha-cone}).

In both cones, the brightest H$\alpha$ of a given sub-region's spatial frequency contour decreases with distance from the nucleus.  In the SW cone, we see a systematic evolution in the spread of  [\ion{O}{3}]/[\ion{S}{2}] as we move away from the nucleus.  For a given H$\alpha$, the lower contour bound of the outer cone typically has higher [\ion{O}{3}]/[\ion{S}{2}]  than the mid-cone, which is higher than the bubbles.  The same is true for the upper bound of each region's contour.  

For both cones, the outer clouds occupy a wide range of [\ion{O}{3}]/[\ion{S}{2}].  Although this sub-region is typically very weak in H$\alpha$, it contains many of the most extreme outliers for both low and high  [\ion{O}{3}]/[\ion{S}{2}].

\subsubsection{Radial Trends in Line Ratio Maps} 

\begin{figure*} 
\centering
\noindent
\begin{minipage}{0.4\textwidth}
\includegraphics[width=\textwidth]{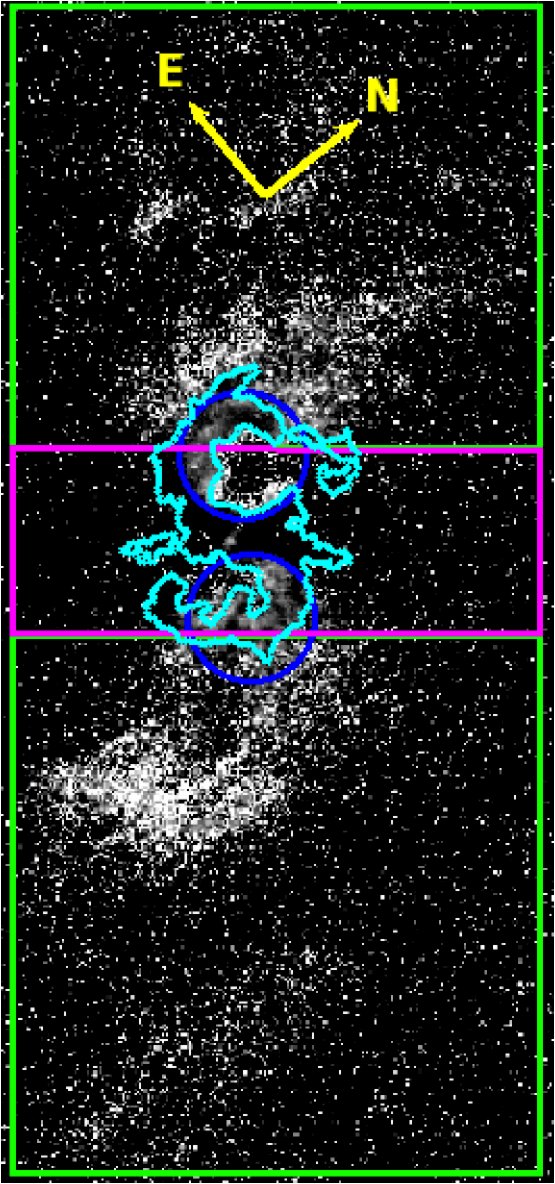}
\end{minipage}\hspace{0.05\textwidth}%
\begin{minipage}{0.55\textwidth}
\includegraphics[width=\textwidth,angle=0,origin=c]{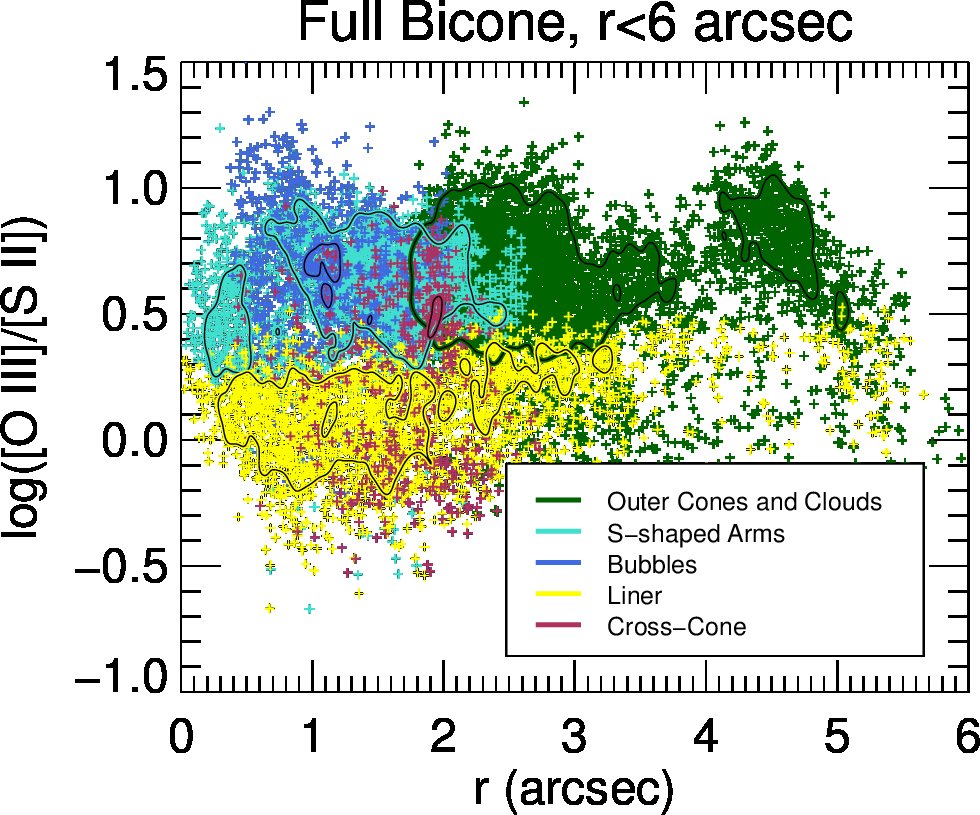}\par
\includegraphics[width=\textwidth,angle=0,origin=c]{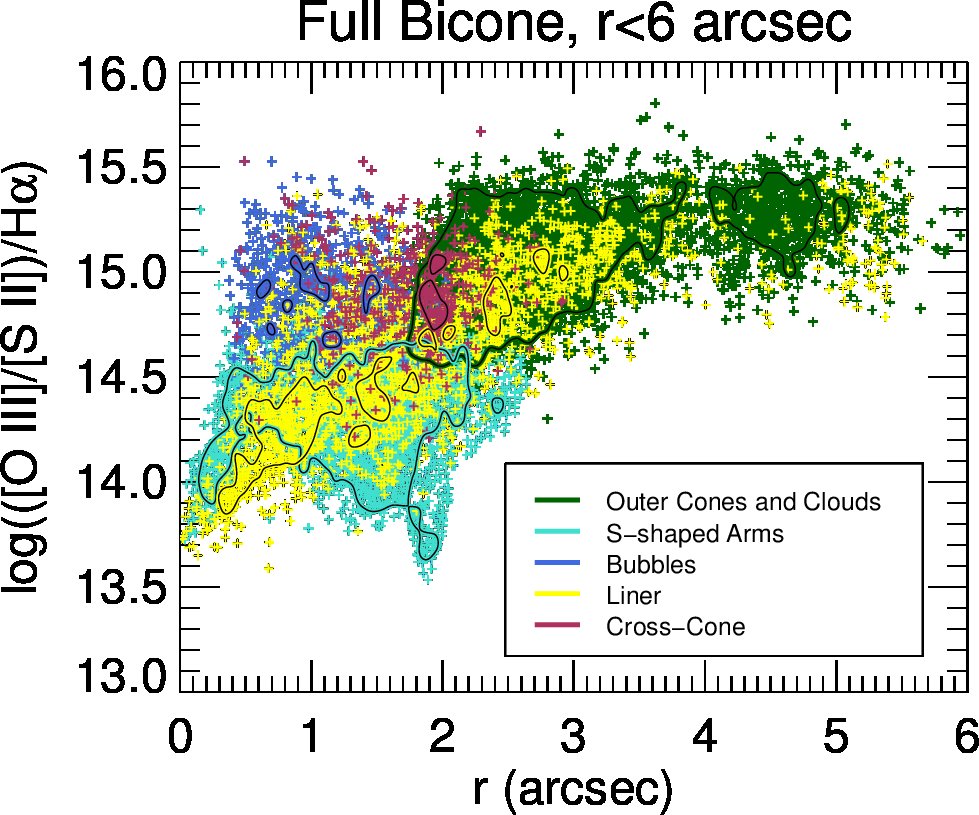}
\end{minipage}
\caption{{\bf Left:} Line ratio map of [\ion{O}{3}]/[\ion{S}{2}] from the center of NGC 3393, as in Fig. \ref{fig:BPT}.  Colored lines indicate borders of different sub-regions from which line ratio profiles are extracted.  The color scheme is similar to figures on the upper-right and lower-right.  LINER regions are not shown, but are selected with the same criteria as in \cite{Maksym16}.  Each pixel is graphed on the upper-left or upper-right according to an unique region identity, and is excluded from all others with lower priority.  In order of descending priority, these classifications are: LINER (yellow), S-shaped arms (cyan), bubbles (blue), outer cones and clouds (green), and cross-cone (magenta).  As in \cite{Maksym16}, all non-LINER regions above the minimum significance threshold are effectively Seyfert-like.
{\bf Top-right:} Radial profile plot of log([\ion{O}{3}]/[\ion{S}{2}]) vs. radius in arcseconds.  Each WFC3 pixel is plotted according to its distance from the nucleus.  Note that the log([\ion{O}{3}]/[\ion{S}{2}]) almost strictly describes the LINER/Seyfert dichotomy.  Exceptions are likely due to reddening.  {\bf Bottom-right:}  Radial profile plot of log(([\ion{O}{3}]/[\ion{S}{2}])/H$\alpha$) vs. radius in arcseconds. The selection criteria for the S-shaped arms is a region of preferentially low log([\ion{O}{3}]/[\ion{S}{2}]), as is evident here.  The inner $\sim1\arcsec$ shows a systematic offset between the arms and LINER pixels, consistent with dilution.  Comparison with the top-right plot shows that the vertical offset for the arm pixels is almost entirely due to H$\alpha$, and is consistent with compression of the ambient ISM by the outflows.}
\label{fig:o3s2harad}
\end{figure*}

Regions outside the S-shaped arms are characterized by lower densities \citep[approaching $n\sim10^{2}\,\rm{cm}^{-3}$ outside the S-shaped arms, vs. $n\sim10^{3}\,\rm{cm}^{-3}$ inside them;][]{Cooke00}.  In the outer regions, [\ion{O}{3}]/[\ion{S}{2}] appears to rise with increasing H$\alpha$, whereas [\ion{O}{3}]/[\ion{S}{2}] flattens or saturates in the brightest H$\alpha$ regions.  To examine radial trends of [\ion{O}{3}]/[\ion{S}{2}], we plot  log([\ion{O}{3}]/[\ion{S}{2}]) vs. radius in Fig. \ref{fig:o3s2harad} (top).  Here, we use a different set of regions: LINER \citep[as in ][]{Maksym16}, the S-shaped arms (selected from contoured regions such that log([\ion{O}{3}]/[\ion{S}{2}])/H$\alpha<14.7$), the bubbles (enclosed by the S-shaped arms), outer cones and clouds (boxes covering the ionization cones at $r\simgreat1.5\arcsec$), and the cross-cone, where brightnesses are in \cgsbrid.

Although the dichotomy between Seyfert-like and LINER pixels is typically well-described by Fig. \ref{fig:o3s2harad} (top) by position on the [\ion{O}{3}]/[\ion{S}{2}] axis, there are several Seyfert-like pixels with [\ion{O}{3}]/[\ion{S}{2}] characteristic of the LINER pixels.  This population of apparently low-[\ion{O}{3}]/[\ion{S}{2}] Seyfert-like pixels could be an artifact of the choice to approximate $\rm{H}\beta\sim\rm{H}\alpha/3$  for BPT diagnostics, under the assumption that extinction is negligible \citep[see][]{Maksym16,Cooke00}.  For comparison, the \cite{CalzettiRed} reddening law predicts $\delta$[\ion{O}{3}]/[\ion{S}{2}]\,$\sim-0.5$ for $E(B-V)\sim1.0$ \citep[$\sim7\times10^{21}\,\rm{cm}^{-2}$, ][]{GO09}, which is consistent with dust lanes that are evident from comparison of {\it HST} UV and optical images.  Scatter in Fig. \ref{fig:o3s2harad} is likely due to local variations in cloud density.

 In Fig. \ref{fig:o3s2ha}, the ``saturated" high-H$\alpha$ regions are associated with the arms, which may be influenced by shocks, whereas in other regions [\ion{O}{3}]/[\ion{S}{2}] appears to rise with H$\alpha$.  [\ion{O}{3}]/[\ion{S}{2}] in a given ENLR may vary with the locally incident flux from the AGN.  We therefore normalize [\ion{O}{3}]/[\ion{S}{2}] by H$\alpha$ to remove this dependency, using H$\alpha$ as a proxy for the rate of ionizing photons available to power the emission.  In Fig.  \ref{fig:o3s2harad} (bottom) we plot log(([\ion{O}{3}]/[\ion{S}{2}])/H$\alpha$) vs. radius with regions defined as in Fig. \ref{fig:o3s2harad} (top) .  While most of the outer regions have relatively flat ([\ion{O}{3}]/[\ion{S}{2}])/H$\alpha$, 
pixels in the arms (cyan $r\simless2.5\arcsec$) are typically $~0.5-1$\,dex below the bubbles, cross-cone and outer regions.  Since we do not see trend in  Fig. \ref{fig:o3s2harad}, [\ion{O}{3}]/[\ion{S}{2}] in the arms must be small relative to H$\alpha$ as compared to other regions.  At small radii, LINER pixels tend to trail Seyfert-like arm pixels with radius, which is consistent with \cite{Maksym16}.

\subsubsection{Selection Effects Due to Surface Brightness Thresholds} 

\begin{figure*} 
\centering
\noindent
\begin{tabular}{cc}
\includegraphics[width=0.45\textwidth]{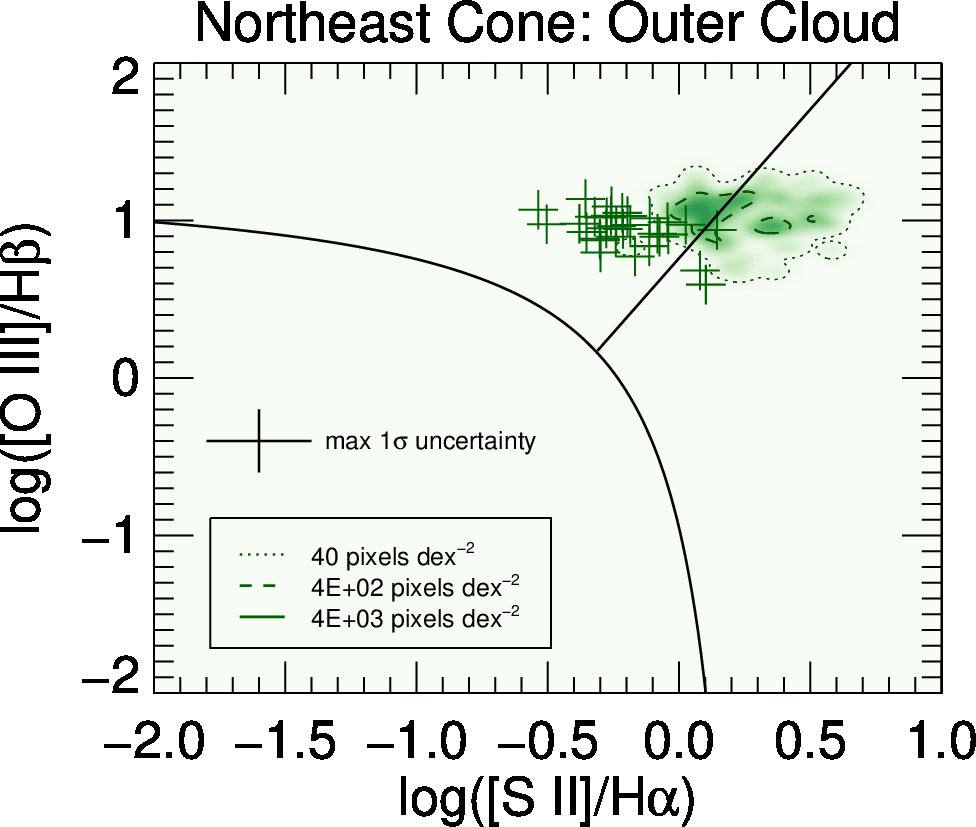} &
\includegraphics[width=0.45\textwidth]{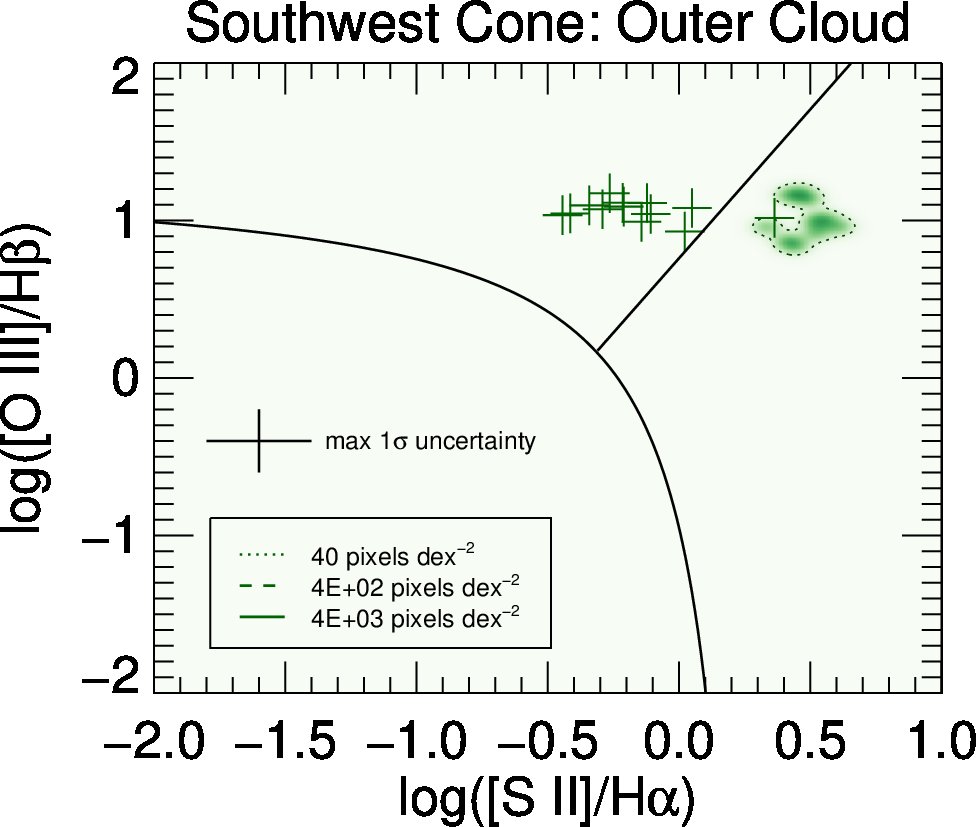} \\
\includegraphics[width=0.45\textwidth]{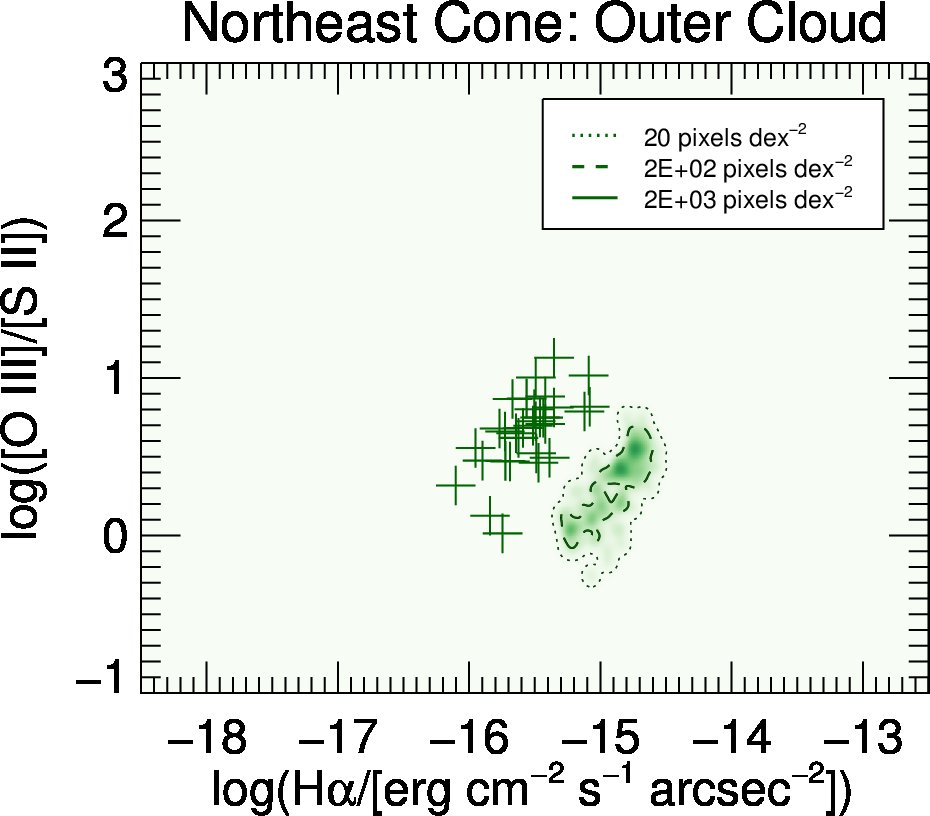} &
\includegraphics[width=0.45\textwidth]{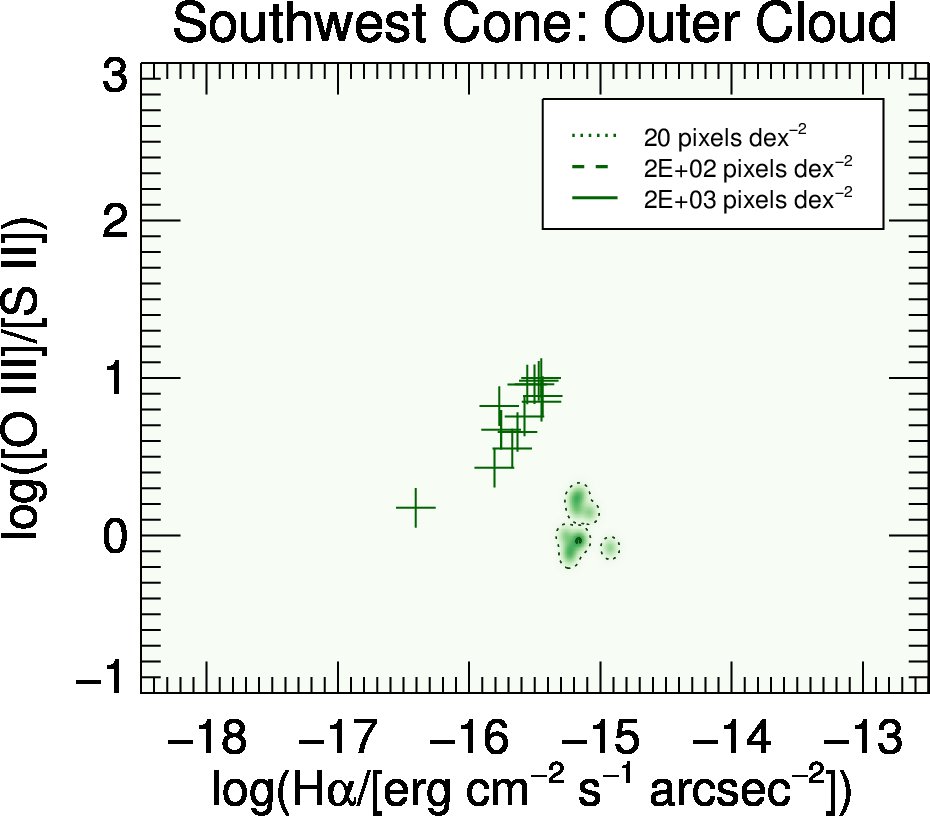} \\
\end{tabular}
\caption{As in Figs.  \ref{fig:BPTcone} (top) and \ref{fig:o3s2ha-cone}  (bottom), but for the ``outer clouds" regions of the SW and NE cones only.  Crosses have been added to represent data which have been spatially rebinned by $\times20$ in each chip axis, in order to permit analysis of low-surface-brightness features which are evident to the eye but rejected by our unbinned brightness threshold.} 
\label{fig:rebin}
\end{figure*}

In order to investigate selection effects in our line ratio map analysis, we reduce the the effective signal-to-noise threshold by rebinning the ``outer clouds" regions in Fig. \ref{fig:BPT} by a factor of 20 in each image dimension.  This region has faint extended structure which is visible by eye in H$\alpha$, [\ion{O}{3}] and [\ion{S}{2}] maps.  We then plot resolved BPT diagrams and [\ion{O}{3}]/[\ion{S}{2}] vs. H$\alpha$ in Fig. \ref{fig:rebin}.  In the BPT diagrams, rebinning produces more data points in the Seyfert region.  In [\ion{O}{3}]/[\ion{S}{2}] vs. H$\alpha$, re-binning fills in a low-H$\alpha$, high-[\ion{O}{3}]/[\ion{S}{2}] region which was previously unpopulated at high resolution and a high signal-to-noise threshold.  This discrepancy between bin sizes may be explained by a selection effect resulting from the weakness of [\ion{S}{2}].  All plausible models predict some correlation of both [\ion{O}{3}] and [\ion{S}{2}] with H$\alpha$.  But emission from [\ion{S}{2}] is expected to generally be weaker than [\ion{O}{3}], and the depth of our [\ion{S}{2}] images is less than both H$\alpha$ and [\ion{O}{3}].  We therefore expect the higher resolution binning to preferentially select the brightest [\ion{S}{2}] regions while excluding higher-ionization regions with bright [\ion{O}{3}] but weak [\ion{S}{2}].  Larger bin sizes will also have the effect of smearing the [\ion{S}{2}]-bright regions, such that they are not seen in the rebinned data.

\subsection{Decomposition of Near-Ultraviolet Line and Continuum Emission}\label{UVdecom}

\begin{figure*} 
\centering
\noindent
\begin{tabular}{ccc}
\includegraphics[width=0.31\textwidth]{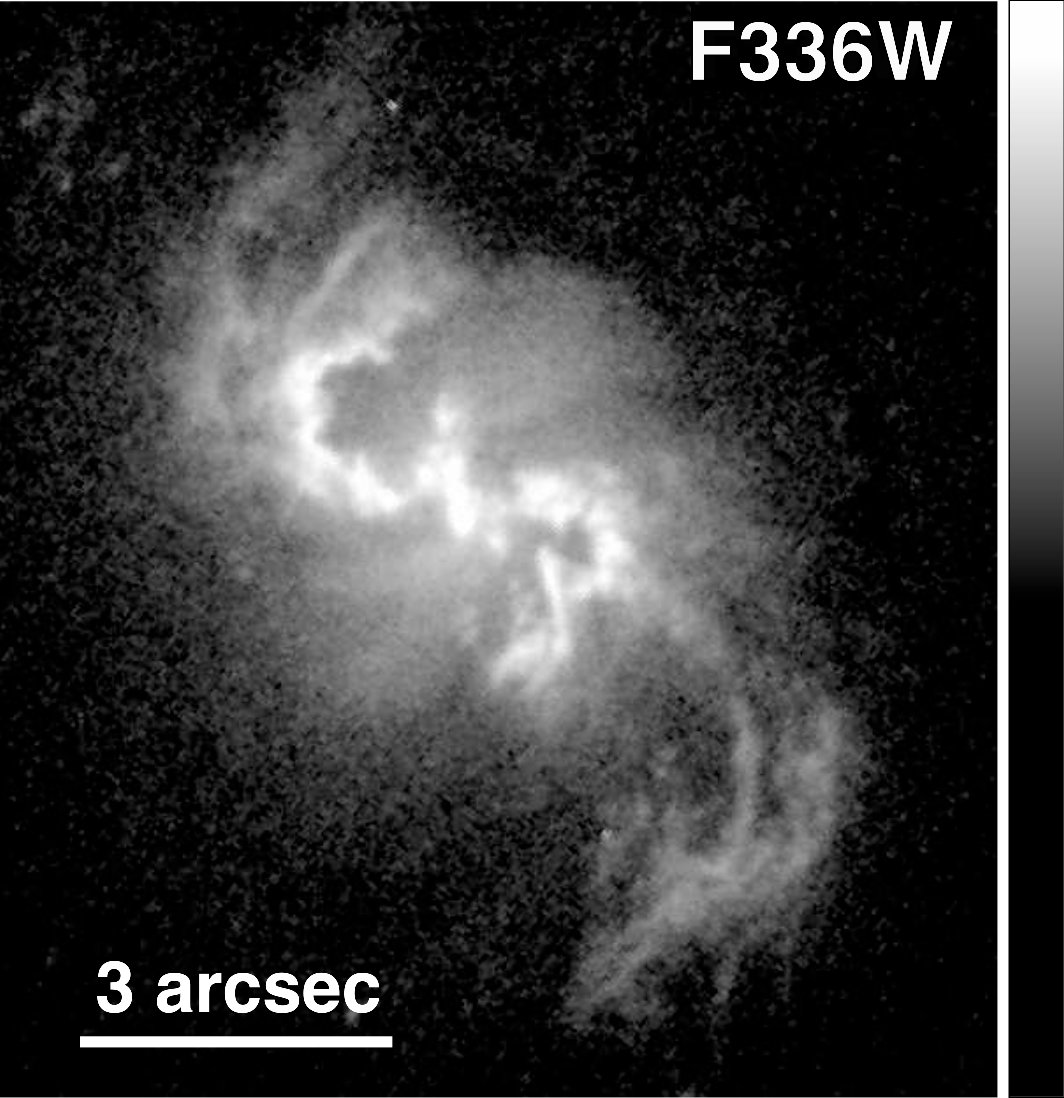} &
\includegraphics[width=0.31\textwidth]{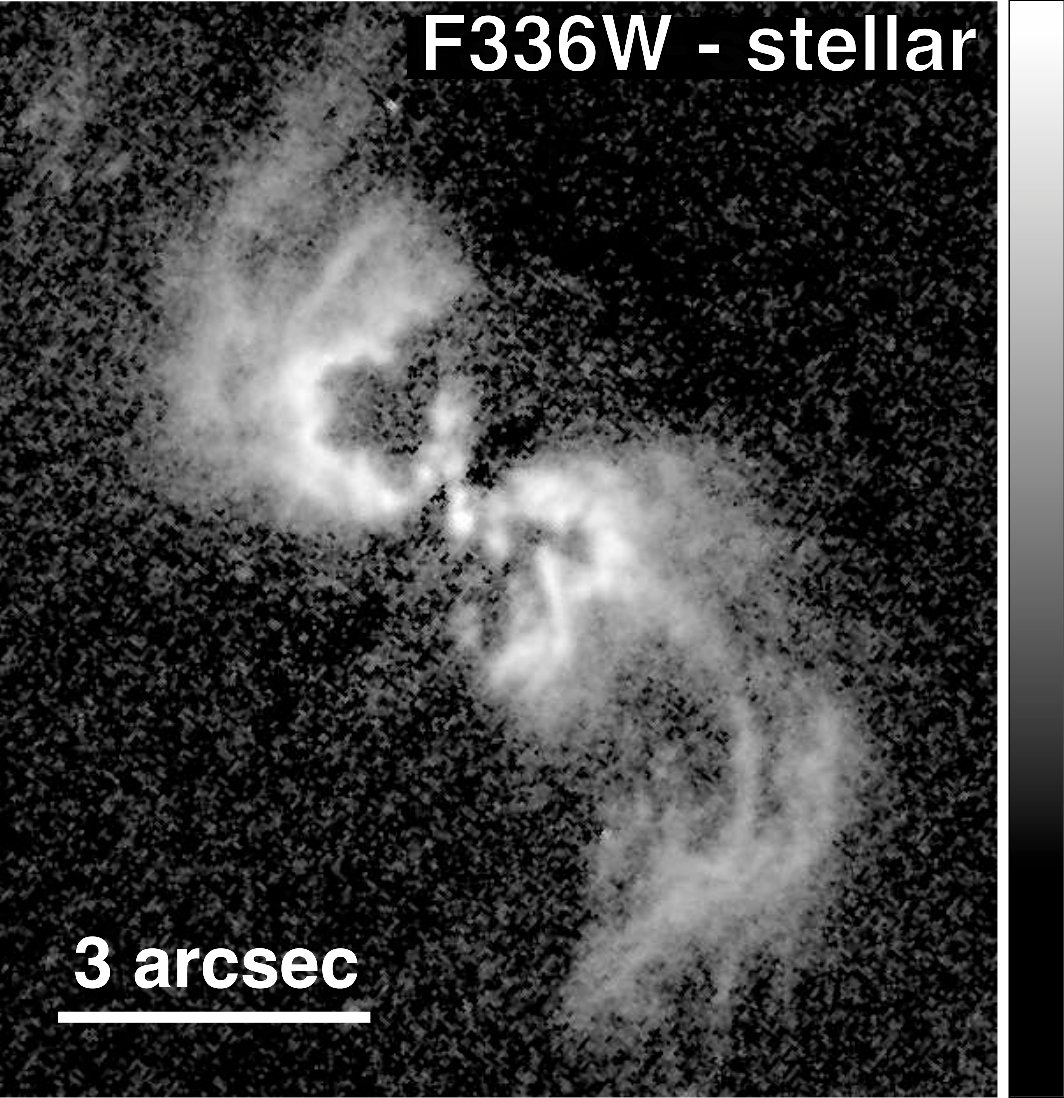} &
\includegraphics[width=0.31\textwidth]{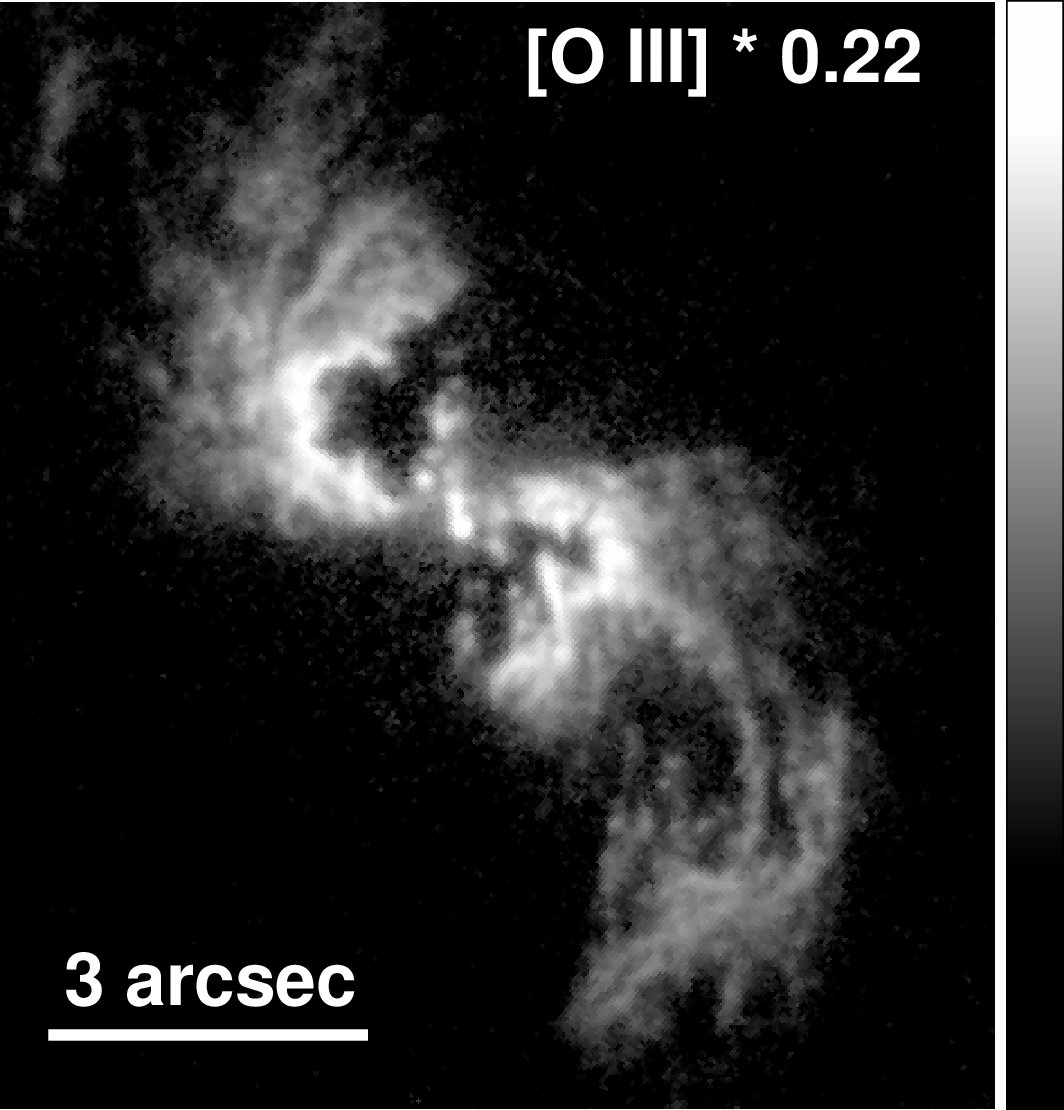} \\
\includegraphics[width=0.31\textwidth]{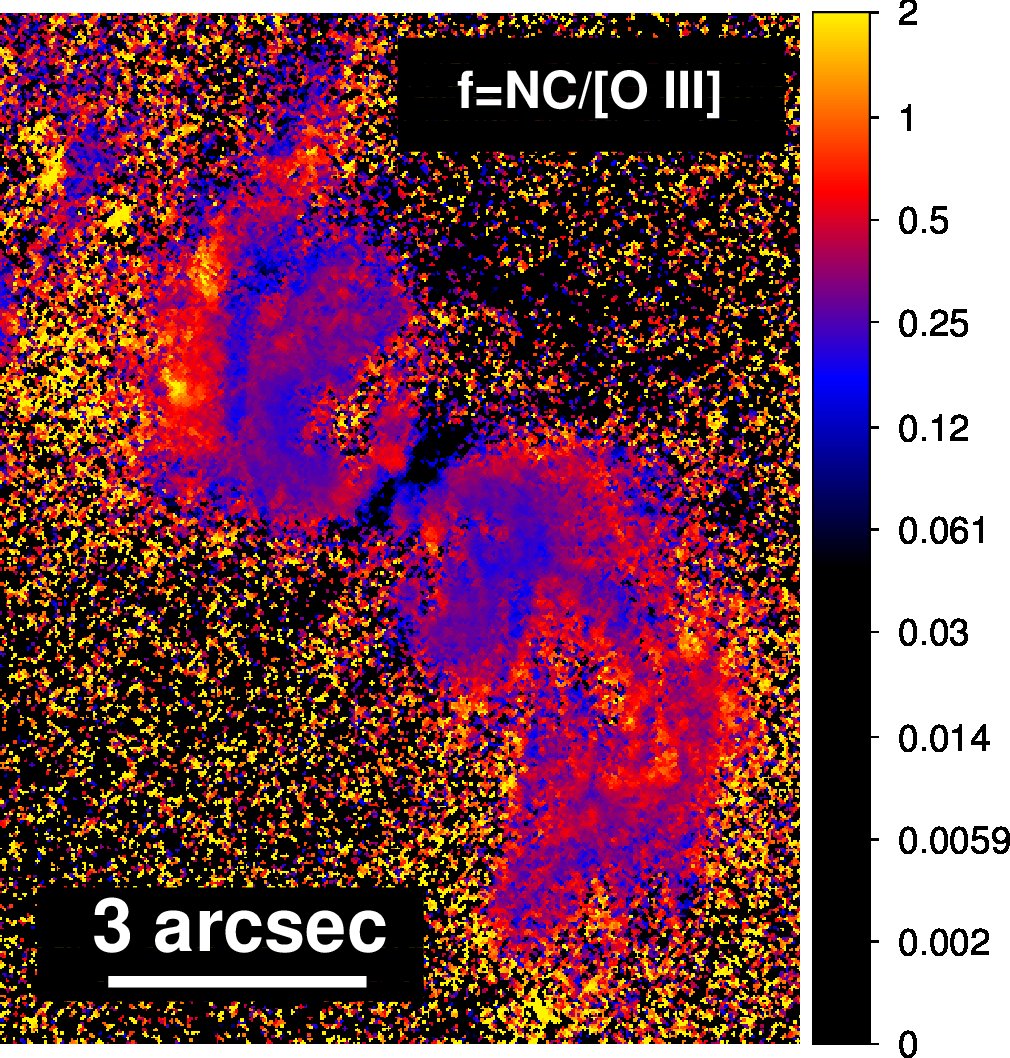} &
\includegraphics[width=0.31\textwidth]{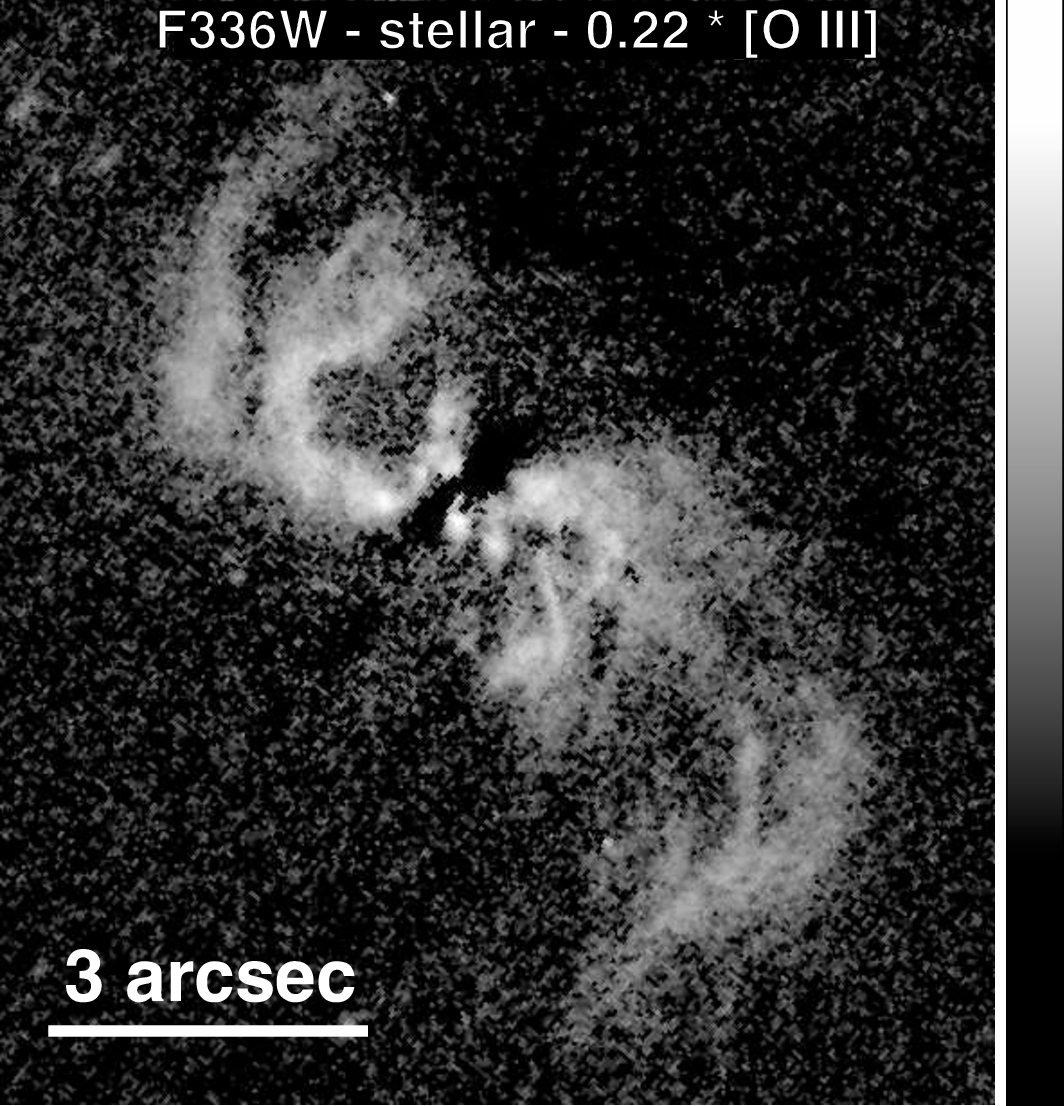} &
\includegraphics[width=0.31\textwidth]{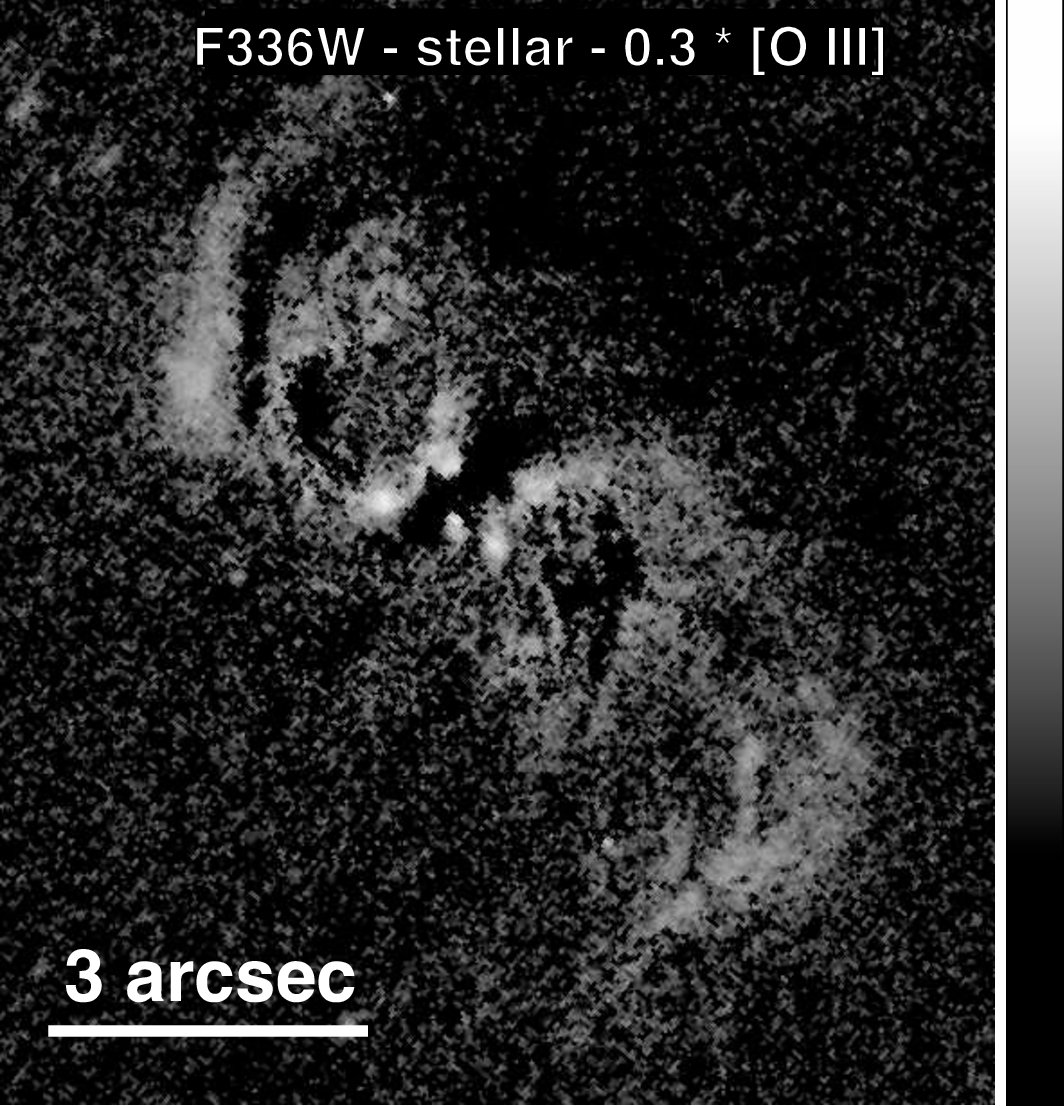} \\
\end{tabular}
\caption{Decomposition of the nuclear NUV based on \cite{MM09}.
\newline{\bf Top Panels:} Left: F336W image of the NGC 3393 nucleus.  Center: As in the left panel, but subtracting a stellar continuum (SC) component scaled from the [\ion{O}{3}] continuum to match the cross-cone stellar light at $r<2\arcsec$.  Right: [\ion{O}{3}]$\times0.22$, representing the nebular continuum in F336W by assuming $f=0.22$. 
\newline {\bf Bottom Panels:} Left: map of $f$ required such that F336W-SC-[\ion{Ne}{5}]-$f\times F_{[O III]}=0$.  Center:  F336W-SC-$f\times F_{[O III]}=0$ for $f=0.22$.  Small regions in both the NE and SW cones begin to show negative residuals (black). Right: F336W-SC-$f\times F_{[O III]}$ for $f=0.30$.  Positive residuals are much-reduced, but there are strong negative residuals at multiple locations within the bicone.
} 
\label{fig:NUVdecomp}
\end{figure*}

\cite{Koss15} noted that the extended biconical NUV emission observed in F336W was likely due to some combination of nebular continuum and [\ion{Ne}{5}]$\lambda\lambda3346,3426$, in accordance with \cite{MM09}. \cite{MM09} had previously used WFPC F330W and [\ion{O}{3}]$\lambda5007$ narrow filter images to investigate the relative contributions of stellar emission, the nebular continuum, line emission, and scattered light to the near ultraviolet emission of 15 galaxies, including NGC 3393.  The pre-COSTAR NGC 3393 data were noisy and required deconvolution which introduced artefacts.  We therefore follow a similar procedure to investigate the relative contributions of the nebular continuum and line emission.  We display the results of this analysis in Fig. \ref{fig:NUVdecomp}.   Differences in bandpass and central wavelength between WFPC F330W and WFC3 F336W are small enough to be negligible.  We performed our image scaling and subtraction by using AstroDrizzled flux density images which have been multiplied by the bandwidth keyword PHOTBW to obtain the total band-specific flux.

Like \cite{MM09}, we assume that [\ion{Ne}{5}] is the primary contributor to line emission in F336W and assume $F_{[O\,III]}/F_{[Ne\,V]}=23.5$ from the \cite{Cooke00} analysis of {\it HST} FOS data.   \cite{MM09} report $F_{[O\,III]}/F_{[Ne\,V]}=16$ from analysis of CTIO data by \cite{sb95}, although we calculate 27.5) but all values imply a modest contribution of [\ion{Ne}{5}] relative to the nebular continuum.  For comparison, for which \citealt{MM09} derive $F_{[O\,III]}/F_{[Ne\,V]}\sim16.9$ for their set of Seyfert 2 galaxies.  

Unlike \cite{MM09}, we directly infer the stellar continuum contribution to F336W by rescaling the [\ion{O}{3}]$\lambda5007$ continuum.  To rescale the continuum assume that the extended UV light in the cross-cone within 2\arcsec\ of the nucleus is dominated by light from a uniform stellar population.  As is evident from Fig. \ref{fig:NUVdecomp} (top, center), the stellar component (SC) is well-represented by a scale factor such that $F336W-F547M\sim1.5$ STMAG, which is compatible with an old ($\simless10$\,Gyr) nuclear stellar population according to Fig. 3 of \cite{MM09}.

Similarly to \cite{MM09} (but accounting for our inferred SC), we infer $f=F_{NC}/F_{[O\,III]}$ where $F_{NC}$ is the flux of the nebular continuum (NC) by varying $f$ to minimize image residuals for F336W-[\ion{Ne}{5}]-$F_{SC}$.

Like \cite{MM09}, we find a plausible range of $0.5\simgreat f\simgreat 0.2$, such that different values of $f$ minimize the residuals of different structures.   This range can be seen in Fig. \ref{fig:NUVdecomp} (bottom, left).  By visual inspection, \cite{MM09} determine a best-fit $f=0.35\pm0.05$, which is consistent with their prediction of $f=0.30$.  We find that although $f\simgreat0.30$ accounts for the majority of extended nuclear F336W emission, it simultaneously introduces strong negative residuals (black in Fig. \ref{fig:NUVdecomp}, bottom right) without eliminating the strongest positive residuals.  Unlike in \cite{MM09}, these cannot be attributed to deconvolution artefacts.  Using the same terminology in Fig. \ref{fig:BPT}, we see that $f\sim0.2$ produces negative residuals at the centers of the bubbles, as well as in lanes perpedicular to the ionization cone, both across the nucleus (in the cross-cone) and in the outer cone NE.  In the outer cones, $f\sim0.5$ fails to eliminate positive residuals of regions in the outer cones NE and SW, as well as for several bright knots in the S-shaped arms at $r\lesssim0.5\arcsec$ from the nucleus.  We will discuss the physical implications of these results in \S\ref{UVOIRDiscuss}.

\section{Discussion}\label{sec-disc}

This work builds upon previous work by \cite{Cooke00}, \cite{Bianchi06}, \cite{Zeng09} and \cite{Koss15}.  \cite{Cooke00} used spectroscopy and {\it HST} narrow-line images to investigate the ENLRs of NGC 3393, and found evidence of S-shaped cloud structures, which appeared to be shaped by outflows associated with radio jets.  \cite{Bianchi06} observed spatially resolved X-ray emission with {\it Chandra}, and \cite{Koss15} supported these associations with new and deeper {\it HST} and {\it Chandra} observations, which we also use here.  We go beyond this previous analysis to investigate the spatial correlation between narrow line emission, radio jets, and extended X-ray structures in greater detail.  Our aim is to better determine the origin of those features, and the roles that the jets and ionization cones of NGC 3393 play in the excitation of the ENLRs of NGC 3393.

\subsection{Discriminating Photoionization from Shock Excitation}

\cite{Cooke00} used pre-COSTAR Planetary Camera images to investigate the relative roles of shocks and photoionization in excitation of the NGC 3393 ENLR.  They determine that the kinetic energy of the outflows is sufficient to power the line emission, but that other tests such as correlations between local velocity dispersion, surface brightness, excitation, and relevant abundances are more consistent with photoionization by a central source.  The implication would be that the radio jets have shaped the emission line arms, but ionizing flux from the AGN is necessary to power the line emission.  This is notably different from other interpretations of similar objects based on X-ray data, e.g. by \citet[NGC 4151]{Wang11a} and \citet[Mkn 573]{Paggi12}, who show that X-ray emission is consistent with shock-driven excitation of the associated optical line arcs in the regions of radio jet impact.  In both NGC 4151 and Mkn 573, however, photoionization is dominant in most of the ionization cones \citep{Wang11c,Paggi12}.  We will examine these diagnostics for NGC 3393 in Paper III (Maksym et al., {\it in prep}).

\subsubsection{Constraints from Optical Line Mapping and UV Emission}\label{UVOIRDiscuss}

\cite{Cooke00} noted that their uncertain $1.2\sigma$ detection of mid-ultraviolet (MUV) emission associated with the arms, if real, could be consistent both with shocks and with scattering of AGN light, but that unambiguous nondetection would strongly disfavor the shock scenario.  Our unambiguous detection therefore permits both shock and scattered light interpretations.  

\cite{Koss15} suggest the F336W probably corresponds to [\ion{Ne}{5}] and the continuum.  Using methods similar to \cite{MM09}, we confirm in \S\ref{UVdecom} that the data are compatible with the  \cite{Koss15} interpretation.  Most of the extended emission is subtracted by this procedure, supporting supporting an important role for the line and continuum gas emission.  But the exact value of $f=F_{NC}/F_{[OIII]}$ needed to account for the nebular continuum varies widely with position in the complex biconical ENLR.  We suggest several possible explanations for this variation.

The strong negative residuals at $f=0.2$ may be due to a nuclear dust lane (in the plane of the maser disk), and to dust lanes surrounding the S-shaped arms in the outer cones.  The latter dust lanes may either guide the radio outflows, or be shaped by them.  The positive residuals in the outer cone which remain when $f=0.3-0.5$ may be due to scattering by dust, or possibly recent star formation outside the dust lanes.  Positive residuals in the outer cones may be due to scattered light (such as by dust), or star formation on the outsides of dust lanes.  Polarization measurements would be necessary to test a dust scattering hypothesis. 

Positive F336W residuals near the nucleus may be due to spatial variation of the ionization parameter, possibly resulting in a locally-enhanced [\ion{Ne}{5}] contribution relative to the FOS measurement.  The \cite{MM09} method assumes uniform $T=10^4\,$K, but this assumption may not be valid: inspection of [\ion{O}{3}]$\lambda\lambda4363,4959,5007$ in the 2D STIS spectra used by \cite{sb95}  implies $T\sim2.5\times10^4\,$K at near the nucleus, and possibly in the S-shaped arms as well \citep{OF06}.  The spatial complexity of the NGC 3393 NLR suggests a need for a more thorough spatially resolved investigation of the temperature, density and kinematics of these STIS spectra.  

In F218W, the emission is likely a composite of the continuum and the strongest lines identified by \cite{Cooke00} between $\sim1900\,$\AA\ and $\sim2500\,$\AA\ with {\it IUE} and the {\it HST} FOS, including \ion{C}{3}]$\,\lambda1909$\AA, \ion{C}{2}]$\,\lambda\lambda 2328,2329$\AA\, and [\ion{Ne}{4}]$\,\lambda2425$\AA.  The extended nature of the F218W emission implies that these lines are entirely associated with the S-shaped arms, and most likely the FUV emission as well (given the absence of a UV point source is likely due to extinction, which decreases with wavelength in the UV).    

By setting a [\ion{O}{3}]/(H$\alpha$+[\ion{N}{2}])$<0.73$ threshold, \cite{Cooke00} infer a lower-ionization medium associated with the cross-cone region within $\sim2\arcsec$ of the nucleus.  Here, the inferred  [\ion{O}{3}]/H$\alpha$ and  [\ion{S}{2}]/H$\alpha$ maps roughly indicate the low-ionization cocoon surrounding the ionization cones which correspond to the LINER and Seyfert-like regions identified in \citep{Maksym16}.  Here we also identify numerous smaller-scale structures which appear physically distinct in each ratio map.

To first order,  [\ion{O}{3}]/H$\alpha$ may be used to represent photoionization, whereas [\ion{S}{2}]/H$\alpha$ may be attributed to shocks.  For example, narrow-line imaging of [\ion{S}{2}]/H$\alpha$ is commonly used as an indicator of shocks in supernova remnant studies \citep[e.g.][]{LeeLee14}, and a threshold of [\ion{S}{2}]/H$\alpha>0.4$ is a common selection criterion in supernova remnant searches.  Morphological identification of jet-driven fast shocks has been used in conjunction with resolved observations of AGN EELRs with [\ion{O}{3}]/H$\alpha$  and [\ion{S}{2}]/H$\alpha$ \citep[e.g. SDSS J224024.1Ð092748, ][]{Davies15}.

The strong [\ion{S}{2}]/H$\alpha$ along the angular boundaries of the ionization cones (Fig. \ref{fig:shock-ion-xray-panel}) might be interpreted as due to shocks, such as from bubble of ionized gas expanding into the ISM, but such a possible origin is complicated by the geometry.  In particular, dense material in the torus or at the base of the cone could partially shield the gas from lower-energy ionizing photons, resulting in low-ionization material at extreme angles.  Such a mechanism has been shown to explain cross-cone emission in NGC 4151 \citep{Kraemer08} and LINER-like emission in NGC 5252 \citep{Goncalves98}.  Alternately, the  [\ion{S}{2}]/H$\alpha$ cocoon might result partially from ionizing radiation reprocessed in the inner cones.  Additional narrow-line imaging observations are necessary to break this degeneracy in interpretation, such as [\ion{O}{2}] $\lambda$3727 to probe collisional excitation and [\ion{O}{3}] $\lambda$4363 for a temperature diagnostic.  We therefore proceed with  these caveats in mind.

In our examination of the [\ion{S}{2}]/H$\alpha$ maps of the NGC 3393 ENLR, we find that almost {\it all} ($\simgreat95\%$) of the area meeting our minimum H$\alpha$ criterion show [\ion{S}{2}]/H$\alpha>0.4$.  The validity of this criterion is dependent upon our modeling of [\ion{N}{2}], which \cite{Cooke00} show to be large (comparable to H$\alpha$), and may have significant spatial variations given the range of densities likely between the jet outflows and S-shaped arms.  But given the large values in our  [\ion{S}{2}]/H$\alpha$ maps, we would expect large regions of [\ion{S}{2}]/H$\alpha>0.4$ to remain even if  [\ion{N}{2}] varied spatially by as much as a factor of a few.

\subsubsection{Resolved X-ray Emission as a Signature of Shocks}

At least two aspects of the X-ray morphology support an origin for the line emission that is at least partially shock-driven.  First, in the NW cone, curve A is associated with a filament of X-ray emission which runs between, and parallel to, the [\ion{S}{2}]/H$\alpha$ curve at the edge of the NE radio lobe, and the [\ion{O}{3}]/H$\alpha$ filament at larger radius from the lobe.  This might be expected if the X-rays correspond to shocked gas leading the jet expansion, and line emission is stimulated by photons generated within the shocks \citep{Cooke00,DS95}.  Since the critical density of  [\ion{S}{2}] is low ($\rm{log}[N_{cr}]=3.2$), [\ion{S}{2}]/H$\alpha$ would correspond to the low-density post-shock, X-rays to the shock itself, and [\ion{O}{3}]/H$\alpha$ to the ionized precursor medium.

Second, peak radio emission from the SW radio lobe corresponds to a knot of strong X-ray emission.  \cite{Koss15} examined VLA and VLBI observations of NGC 3393 and proposed that the strong emission in the southwest jet is due to doppler beaming in the hotspot of an approaching jet.  The X-ray emission, combined with the north-south feature in the SW cone mid-cavity, suggests shock interaction with the surrounding medium.  Knot B also indicates strong [\ion{S}{2}]/H$\alpha$ and hence a possible role for shocks in this region.  Bulk motion from a jet in the direction of the observer could explain some of the irregular kinematics found by \cite{Fischer13}, who invoke an off-nuclear kinematic center to explain blueshifted material in this region.  The large velocities in this region ($\Delta v\sim400-600 \kms$; $\rm FWHM\simless1200 \kms$) are consistent with shock models \citep[see, e.g.,][]{Contini12}, although the large FWHM may be due to line splitting.

As can be seen in Fig. \ref{fig:xray-bands}, the extended X-ray emission associated with arc A, knot B, and the jet lobes is relatively hard, with significant emission above 2 keV.  As in recent studies of ESO428-G014 by Fabbiano et al (2017; {\it submitted}), the extended hard emission on $\sim$kpc scales suggests that the origin of $2-10\,$keV X-rays is more complex in compton-thick AGN than is generally assumed.  In the context of our other NGC 3393 data, the extended hardness suggests a possible role for shock origins of these features.   Fabbiano et al (2017; {\it submitted}) also note that an extended hard continuum can be enhanced by charged particles accelerated in a radio jet.  In Paper III of this series (Maksym et al, {\it in prep}), we will investigate these features via detailed study of spatially resolved X-ray spectroscopy and X-ray line emission morphology.

\subsubsection{Shock Compression and Filling Factors}

If line emission in the S-shaped arms is due to shocks, we expect the filling factors of ENLR gas to be consistent with this picture.  We assume $L_{\rm{H}\alpha}\sim n^{2}\alpha h \nu \epsilon V$ where $L_{\rm{H}\alpha}$ is the H$\alpha$ luminosity, $n$ is density, $h$ is Planck's constant, $\alpha$ is the recombination coefficient for H$\alpha$, $\nu$ is the frequency of H$\alpha$, $\epsilon$ is the filling factor, and $V$ is the volume.  We assume densities determined from the [\ion{S}{2}] doublet by \cite{Cooke00} ($\sim10^3\,\rm{cm}^{-3}$ in the arms and $\sim10^2\,\rm{cm}^{-3}$ in the outer regions), although we note that these values are limited by the angular resolution of their ground-based spectroscopy.  These densities also approach the useful limits of the [\ion{S}{2}] doublet at the low and high ends, suggesting more extreme values may be possible.

Under these density assumptions, the filling factors are consistent with shocks for a broad range of assumptions regarding gas cloud thickness and the precise location of the shock edge.  If the gas clouds extend $\tau_{\rm 100}\sim100\,pc$ in the observer's line-of-sight, we find $\epsilon={7.3\times10^{-5} \Sigma_{-16} n_{100}^{-2} {T_4} {\tau_{100}}^{-1} }$, where $\Sigma_{-16}$ is the H$\alpha$ surface brightness in units of $10^{-16}\,$\cgsbri, $n_{100}$ is density in $10^2\,\rm{cm}^{-3}$, and ${T_4}$ is temperature in $10^4\,$K.  For representative values inside and outside the S-shaped arms ($\Sigma_{-16}=[5, 1]$), such a density gradient ($n_{100}=[10, 1]$) produces a filling factor gradient ($\epsilon=[1.5,30]\times10^{-4}$) and a compression factor ($\epsilon_2/\epsilon_1=20$) consistent with shocks.  The larger filling factor is comparable to $\epsilon\sim3\times10^{-3}$ expected from \citep{Contini12} using [\ion{Ne}{3}]\,$\lambda15.5\mu$m, if we assume the mid-infrared [\ion{Ne}{3}] is also co-spatial with the optical, UV and X-ray emission at $r\sim400$\,pc ($\sim1.6\arcsec$).

In this scenario, the leading edge of the shock compresses pre-shock gas into a very small volume, producing a filling factor that is significantly lower than that of the circumnuclear ISM.  The filling factor in such pre-shock gas is already low in the NLR of typical AGN.  High spatial resolution spectroscopy of the [\ion{S}{2}] doublet could confirm the specific assumptions we make about the density distribution.  Also, given the limited density range of [\ion{S}{2}] the true compression ratio may be even larger than we infer.

Fig. \ref{fig:o3s2harad} provides a check on this scenario.  In the S-shaped arms, ([\ion{O}{3}]/[\ion{S}{2}])/H$\alpha$ is a factor of $\sim50$ below the outer regions ($r\simgreat3\arcsec$), and a factor of $\sim20$ below other regions at small radii ($r\simless3\arcsec$) comparable to the arms.  In Fig. \ref{fig:o3s2ha}, however, the S-shaped arms have comparable [\ion{O}{3}]/[\ion{S}{2}] Seyfert-like regions at larger radii.   The low ([\ion{O}{3}]/[\ion{S}{2}])/H$\alpha$ of the S-shaped arms must therefore result from the H$\alpha$ surface brightness.      The spatial variation in $H\alpha$ surface brightness here acts as a check on our inferences regarding the gas compression; if with a compression factor $x$ the gas volume decreases by a factor $x$, then emissivity also increases by $x^2$.  Thus, if we expect a compression factor of $\sim20$, we should see a similar variation in $H\alpha$.  Here, we see that the $H\alpha$ variation ($\sim20-50$) is small relative to the inferred compression factor.  A compression factor compatible with shocks is therefore reasonable.

\subsubsection{NGC 3393 in the Context of Previous Theoretical Models}

\cite{Contini12} modeled UV, optical and IR spectroscopy of NGC 3393 to investigate the roles of collision and merging in NGC 3393 relative to claims of a supermassive black hole binary by \cite{Fabbiano11}.  \cite{Contini12} found that though the AGN was likely the dominant source of [\ion{O}{3}] excitation, shocks of $100-600\;\rm{km\;s}^{-1}$ are necessary to explain their full dataset, and required pre-shock densities a factor of $\sim10$ higher than is typical of other Seyfert NLRs.   \cite{Contini12} suggests that the \cite{Levenson06} models of X-ray emission extended throughout inner $\sim$kpc of the ionization cone region are consistent with these shock models, with the highest-velocity shocks in the galaxy's central regions.  \cite{Contini12} also suggests some of the X-rays could originate from X-rays emitted downstream from the shocks.  Using our deeper {\it Chandra} data, we have produced deconvolved X-ray images which support the \cite{Contini12} models, but with the specific implication that the strongest X-ray emission comes from narrow ($r\simless0.2\arcsec$, 50\;pc) filaments and knots associated with nearby  [\ion{O}{3}]/H$\alpha$  and [\ion{S}{2}]/H$\alpha$  features suggestive of shocks and shock-illuminated material.  
Outside of these filaments, extended X-ray emission is still present, but with a surface brightness at least an order of magnitude less than in the filaments.  In our subsequent paper, Maksym et al. ({\it in prep}), we will investigate plausibility of these claims using resolved X-ray spectroscopy.

 \cite{Contini12} claim that because their models are consistent with the illumination of the leading surface of a shock rather than the receding side, matter must be infalling.  We note that such a scenario may be possible for material driven by an expanding jet of this nature, particularly when much of the shock front is parallel to the jet axis and some material may be driven sideways or back towards the center.  

\subsubsection{Cosmic Ray Heating}

Line-emitting gas in contact with radio plasma (such as is present in NGC 3393) may experience cosmic ray heating, such that a relativistic population of particles in the plasma produces anomalous line enhancements \cite{FM84}.  For example, models by \cite{Gagne14} required cosmic ray heating to explain enhanced [\ion{O}{2}]$\lambda3727$/[\ion{O}{3}]$\lambda5007$ ratios in the ENLR of the ``Teacup" galaxy, which surrounds bubbles of radio-emitting that span $\sim15\,$kpc.  Although the FOS spectra of the NE arm analyzed by \cite{Cooke00} point to much lower [\ion{O}{2}]$\lambda3727$/[\ion{O}{3}]$\lambda5007$ values ($\sim0.2$ in NGC 3393, vs. $\simgreat1$ in parts of the Teacup), the magnetic field strength, pressure and energy derived by \cite{Cooke00} are comparable to those found in the Teacup by \cite{Gagne14}, suggesting that cosmic ray heating could be energetically significant.  Spectral modeling beyond the scope of this paper would be useful to constrain the contribution of cosmic ray heating relative to other effects.

\subsubsection{Summary}

Although we find ambiguity in the interpretation of the low-[\ion{S}{2}]/H$\alpha$ cocoon with respect to alternate interpretations of shocks and reprocessed photons from the AGN, we now see a unified picture arising for the SW cone:

The bright knot from our PSF-deconvolved {\it Chandra} image corresponds to the radio hotspot described by \cite{Koss15} and the kinematic `center' suggested by \cite{Fischer13}.  The jet, coming from the true nucleus at the hard X-ray centroid and central radio source, collides with a locally overdense region of the ISM at $\simgreat1000\;\kms$.  The actual jet velocity may be much larger, since the FWHM is sampled for a region displaced from the center of the knot.  Since this motion is largely in the plane of the sky (which is nearly parallel to the plane of the galaxy), we do not see the strong signatures of this motion in the kinematics described by \cite{Fischer13} {\it until} the jet collides with the overdense material $\sim0.9\arcsec$ SW of the nucleus and begins to ablate it.  The collision shocks the gas at the jet-ISM interface, and the material from the jet itself splays outwards from the collision site and inflates a bubble described by a P-Cygni profile seen in the STIS spectroscopy of \cite{Fischer13} and the line ratio map cavity described \S\ref{sec-results}.  The gas at the collision site is shocked, producing, an X-ray knot, enhanced radio emission, and possibly the nearest regions of locally enhanced [\ion{O}{3}]/H$\alpha$  and [\ion{S}{2}]/H$\alpha$ as well.  Such shocks may explain the need for a shock component in the spectral models of \cite{Contini12}.

%XXX Mun?oz Mar?õn stis, UV 2009
%XXX COOKE -UV as sign of shocks?
%XXX why do X-rays follow leading edge? shocks? density? can we extrapolate X-rays to optical, get UV?
%XXX Yan \& Blanton 2011? shocks and LINER geometry? what about mid-IR? http://arxiv.org/pdf/1405.4159v1.pdf http://arxiv.org/pdf/1504.01127v1.pdf
% XXX O I? Gemini!  Kinematics, density too.

\subsection{Spiral and Cocoon Structures}

\subsubsection{The Origin of the S-shaped Arms}

\cite{Cooke00} question the origin of the S-shape for the arms, when radial jet motion might produce a figure-8 shape and the galaxy's spiral arms should sweep gaseous material onto the opposite side of the jet if they are trailing.  This asymmetry is obvious from a ratio of $\lesssim60$ between the bright and faint sides in H$\alpha$ surface brightness, and ratios of $\sim10-100$ in X-ray flux \citep[although in][the delineation between Seyfert-like regions in the bubbles and the surrounding LINER-like cocoon is basically symmetrical]{Maksym16}.  The default motion of jets and disk-wind bubbles through the nuclear ISM of a galaxy should be linear and produce line-emitting overdensities or shocks at the leading edge of the bubble-ISM interaction.  

If the cavities are predominantly formed by the motion of jet expansion, the apparent extent and position of the radio lobes are a mystery, as the cavities are at least a factor of $\sim2-3$ larger than the jet lobes.  If the structure of the jet follows the cavities, the small-scale complexity of the cavities at $r\simless1\arcsec$ also implies similar small-scale radio counterparts, which we do not observe.  The nature of the X-ray arc in the NW cross-cone is uncertain.  It may be part of a larger-scale ellipsoidal structure which surrounds the inner $\sim2\arcsec$ of the ENLR, or it may be a trailing portion of the NE S-shaped arm.

The S-shaped structure of the X-ray and optical line emission implies non-radial directionality either in the leading material, or a rotational component to the bubbles' motion.  The S-shape may result from entrainment by patterns in the ISM (such as spiral structure near the nucleus; Fischer, private communication).  If the galactic medium is denser on the non-emitting side, the rising jet lobes might preferentially expand into lower-density material and therefore transfer more kinetic energy in that direction.  In such a case, the elongation of the blobs could be due to lateral expansion relative to the ISM.

\subsubsection{Accretion Disk Precession}

Rather than local gas overdensities, the S-shaped structure could result from the rotation of the jets or winds.  Such rotation may occur if the outflow-launching disk precesses due to gravitational interaction, such as in a supermassive black hole binary \citep[SMBHB][]{Begelman80}.  Simulations of precessing accretion disks show that S-shaped structures arise naturally on smaller scales \citep{KP08}.
A prerequisite SMBH-SMBH or SMBH-disk angular momentum misalignment is plausible, given the plane of the megamaser accretion disk is perpendicular to the plane of the sky \citep{Kondratko08}, whereas the galaxy itself is nearly face-on.  We expect the outflow-emitting black hole to be precessing counterclockwise, based on the S-shaped orientation,

The radio plasma within the S-shaped structures is likely to be stable against rotational motion, since it is strongly magnetized \citep[$H\sim10^3\,\mu G$;][]{Cooke00} and even weak magnetic fields are sufficient to stabilize outflow bubbles in galaxy cluster simulations \citep[][]{JY05}.  The conditions of the ambient medium differ from galaxy clusters, with higher density ($n\sim100\,\rm{cm}^{-3}$) and only partial ionization ($kT\sim10^4$\,K). 

Detection of a bent radio jet would support a precession hypothesis \citep{Begelman80}, but there is no clear evidence for such a structure.  \cite{Koss15} examined VLBI observations of NGC 3393 at 2.3 GHz and 8.4 GHz, but only detected the SW hotspot, suggesting that most of the radio emission detected by VLA is resolved out by VLBA data.  Structure brighter than $\sim0.4$ mJy/beam is not found on scales of $\lesssim13\,$pc, suggesting that the radio bubbles should be treated as an expanding plasma rather than collimated structures.  

SMBHB-induced precession of the disk might produce the S-shape through deformation of the ISM by the rising bubbles.  Such an explanation might also be consistent with the SMBHB claim of \cite{Fabbiano11}.  The $0.5\arcsec$ ($\sim130\,$pc) separation from \cite{Fabbiano11} provides a relevant starting point for the possible impact of a SMBHB in this system, given a wide variety of possible SMBHB parameters.  At this separation, we expect disk precession to be dominated by the orbit, whose period (neglecting the effects of disk mass) is $P_{orb}= 3.93\times10^6 M_8 (a/(10^7 r_G))^{3/2} (5/(1+q+M_{star}/\Mbh))^{1/2}\,$yr  \citep[eq. 6, ][]{LC07}, where $M_8$ is the primary black hole mass \Mbh\ in units of $10^8\Msun$, $a$ is the orbital separation, $r_G$ is the Schwarzschild radius of the primary SMBH, $q$ is the SMBHB mass ratio, and $M_{star}/\Mbh$ is the total stellar mass enclosed within the binary, here taken to be 5.  For $M_8=0.3$ \citep{Kondratko08}, $a=130\,$pc and $0.1<q<1$, we find $P_{orb}\sim(9-10)\times10^{6}\,$yr.  Outflow rotation may also be induced due to geodetic precession of the primary \cite{Begelman80, Roos88} or tidal precession in an inclined secondary \cite{Katz97}.  These effects are expected to become important when the binary is close (sub-parsec) and may be shorter than kiloyears depending upon the orbital parameters \citep{LC07}.

We can compare this $P_{orb}$ with plausible timescales inferred from rising bubbles that shape the outer edge of the S-shaped arm.  We can assume that the edge of the arms indicates leading edge of rotating outflow, and that when a rising outflow reaches the edge it decelerates to approximately terminal velocity; if a bubble rises $\sim0.5$\arcsec\ as the outflow sweeps through $\sim30\degr$ with  $P_{orb}\sim9.5\times10^6\,$yr, consistent with edge of the NE arm, its mean rising velocity is $v_{rise}\sim160\kms$.  Given numerous uncertainties, $v_{rise}$ is not significantly faster than the expected terminal velocity $v_{term}\sim110\kms$ calculated according to \cite{Braithwaite10} (assuming local Keplerian velocity $v_{kep}\sim\sigma_{gas}$, where $\sigma_{gas}$ is inferred from \citealt{Fischer13}, and the radius of the bubble is half its distance to the nucleus).

A simpler alternative to SMBHB precession is precession of the disk of a single SMBH.  Such precession would be due to the \cite{BP75} effect, where Lense-Thirring precession causes a misaligned disk to warp and align with the SMBH spin.  From \cite{LZ05}, expected period ranges from $P_{BP}\sim7.8\times10^{5}\,$yr (assuming disk viscosity $\alpha=0.1$, $M_8=0.31$, dimensionless black hole spin $a=1$, and accretion rate $\Mdot=0.1\,\Msun\rm{yr}^{-1}$) to $P_{BP}\sim1.2\times10^{7}\,$yr \citep[assuming $a=0.1$, $\Mdot=0.04\,\Msun\rm{yr}^{-1}$; \Mdot\ values from ][]{Kondratko08}.  A shorter period than our derived  SMBHB $P_{orb}$ implies an even faster rising velocity, which is less likely given the speed of sound.  Also, the critical radius for the precessing disk is $\sim0.02\,$pc \citep[derived according to][]{LZ05}, which is a factor of $\sim10$ smaller than the inner megamaser disk size of 0.17\,pc \citep{Kondratko08}.  Since there appears to be good alignment between the axis of the megamaser disk and the ionization cones \citep{Kondratko08}, there is no evidence for misalignment between the extended and inner disks.  Given these problems, a single-SMBH scenario would therefore disfavor the idea that the S-shape is predominantly due to disk precession.  SMBHB precession may therefore be more likely than single disk precession, if the S-shaped structure is not entirely due to environmental asymmetry.

Deeper high-resolution radio observations could help determine the origin of the S-shaped structure.  Line-emitting filaments that surrounding cavities which do not appear to be associated with radio emission could be formed by a low-energy population of electrons from a earlier stage of the outflow.  If so, then deeper VLA observations at comparable resolution to the pre-existing data ($\theta_{FWHM}\lesssim0.3\arcsec$) should be capable of detecting them.  Alternately, VLBA observations at 326 MHz have a largest angular scale $\theta_{LAS}\sim0.3\arcsec$ and could therefore detect plasma structure at these scales.  Detection of such extended structure could support the precessing outflow model of S-shape formation here.

\section{Summary and Conclusions}\label{sec-con}

As part of the continuing {\it CHEERS} project, we investigate the spatially resolved relative contributions of photoionization and shocks in NGC 3393, a nearby Seyfert 2 AGN.  Using narrow filter {\it Hubble} WFC3 images, we directly compare the relative spatial distributions of [\ion{O}{3}], [\ion{S}{2}] and H$\alpha$ against deep PSF-deconvolved {\it Chandra} images with $\sim0.2\arcsec$ ($\sim50\;$pc) resolution, as well as older {\it VLA} 8.4 GHz images of the sub-kpc radio jet, where the jet has long been suspected \citep[e.g.][]{Cooke00,Koss15} to play an important role in the formation of S-shaped structures emitting [\ion{O}{3}]  and H$\alpha$ lines, as well as ultraviolet.  These observations show that the circumnuclear ISM of NGC 3393 is a {\it complex multi-phase medium}, with several specific implications:

\smallskip
\noindent$\bullet$\,1) The ionization structure is highly stratified with respect to outflow-driven bubbles in the bicone, and varies dramatically on scales of $\sim10$ pc.

\smallskip
On $\sim50$-pc-scales ($\sim0.2\arcsec$) [\ion{O}{3}]/H$\alpha$, [\ion{S}{2}]/H$\alpha$ and $0.3-8\;$keV X-rays trace very different physical regions.  A kpc-scale [\ion{S}{2}]/H$\alpha$ cocoon envelopes the ionization cones, and low-ionization features spanning $\sim10$s of pc are associated with the radio jet boundaries, consistent with jet-gas kinematic interaction and supporting a local role for shocks, given typical [\ion{S}{2}]/H$\alpha>0.4$ on all scales of this ENLR cocoon.  These shocks may play a role in local \ion{O}{3} excitation and X-ray production via Bremsstrahlung \citep[see, e.g.][]{DS95,Contini12}.  We also find a small, $\sim$100-pc-scale N-S region at the nucleus where [\ion{S}{2}]/H$\alpha$ is weak relative to its surroundings ($\simless0.7$ vs. $\sim1.0$) and likely associated with very strong ionization from the AGN, either through photoionization or strong jet-gas interactions.  The knot of strong X-ray emission associated with the SW radio lobe suggests that the `hotspot' described by \cite{Koss15} is a site of such strong interactions.

\smallskip
\noindent$\bullet$\,2) High-energy emission from the mid-ultraviolet and X-rays is dominated by the S-shaped arms, with few photons from the nucleus.
\smallskip

Although much of the X-ray emission is associated with the ENLR, as per \cite{Bianchi06}, \cite{Levenson06} and \cite{Koss15}, we find that the bulk of this emission ($\sim50\%$) arises from narrow sub-arcsecond filaments associated with the brightest emission line features.

\smallskip
\noindent$\bullet$\,3) Few X-ray and mid-ultraviolet photons are observed from the nucleus.
\smallskip

In our deconvolved X-ray images, we see only a modest ($<10\%$) contribution from the nuclear point source in the $0.3-8\;$keV band, consistent with previous X-ray spectroscopy establishing the AGN as Compton-thick \citep{Maiolino98, Guainazzi05, Burlon11, Koss15}.   From the MUV, we infer that the IUE emission observed by \citep{DPW88} is not only extended, but is also predominantly cospatial with the S-shaped arms.

\smallskip
\noindent$\bullet$\,4) Multiple observational indicators support a role for shocks from AGN outflows for some parts of the ENLR in close contact with the radio plasma.
\smallskip

This role of shocks is consistent with line ratio mapping, H$\alpha$ compression, STIS kinematics from \cite{Fischer13}, and UV imaging.  X-rays (such as in the knot coincident with the SW radio lobe) are key to this interpretation since the optical lines are incomplete indicators.  \cite{Wang11c} and \cite{Paggi12} observe excess \ion{Ne}{9} emission relative to \ion{O}{7} at the leading edges of AGN outflows in NGC 4151 and Mrk 573, consistent with emission from a thermal plasma, so we might expect shocks contributing to the ENLR of NGC 3393 to display similar features when spatially resolved.  In Paper III (Maksym et al, {\it in prep}), we will investigate the resolved X-ray imaging spectroscopy in detail.

\smallskip
\noindent$\bullet$\,5) Cavities formed by filamentary X-ray and optical line emission appear to extend beyond the radio plasma, possibly indicating an undetected outflow component.
\smallskip

On larger scales, [\ion{O}{3}]/H$\alpha$, X-rays and radio from NGC 3393 preferentially occupy cavities of low [\ion{S}{2}]/H$\alpha$ values.  The excellent match between the faintest 8.4 GHz radio contours and [\ion{S}{2}]/H$\alpha$ cavity boundaries in some regions also supports shock formation, but raises the question of why the {\it whole} cavity does not demonstrate radio emission.  If jet expansion is the root cause of cavity formation, then we would expect deeper radio observations with the {\it VLA} (or with {\it VLBI}) to reveal more extended structure within the ENLR.  Otherwise, we must invoke a more complicated picture for cavity formation.

\smallskip
\noindent$\bullet$\,6) The origin of the NUV emission is complex, and likely signifies large spatial variations in temperature, ionization parameter, and scattered light contributions on sub-kpc scales.  
\smallskip

NUV decomposition similar to \cite{MM09} shows that although the nuclear stellar component is well-modeled by an old ($\sim$few Gyr) population, the extended emission is not well-suited to a single $f$ ratio between the NUV continuum and [\ion{O}{3}].  The complexity is consistent with a bubbles of plasma photoionized from one side, with possible contributions from shocks.  Ultraviolet emission line imaging or spatially resolved spectroscopy is necessary to investigate the origins of these effects.

At optical wavebands, IFU observations (such as with {\it MUSE} and {\it JWST}) would help address the relative roles of photoionization and kinematics in the feedback which we see in NGC 3393.  Full kinematic fitting with a wider range of emission lines is necessary to determine the amount of energy deposited into the ENLR via jet interaction, and would greatly improve upon previous work by \cite{Cooke00}, \cite{Contini12}, and \cite{Fischer13} using sparser spatial sampling.  Spectrally resolving lines indicative of density (such as the [\ion{S}{2}] doublet) would be particularly useful, as the assumed densities from \cite{Cooke00} may not be generally applicable.  Direct measurement of [\ion{N}{2}] would be superior to the assumptions used here, and full spectroscopic measurement would help address the origin of a possible excess in continuum observations of the S-shaped arms.  Given the obvious structure on scales of $\sim$few {\it Hubble} pixels, however, additional  {\it Hubble} narrow-line imaging of \ion{He}{2}, [\ion{O}{2}] $\lambda$3727 and [\ion{O}{3}] $\lambda$4363 is necessary to trace the ionizing flux, collisional ionization, and local gas temperatures.  And very deep {\it Chandra} observations will allow high-fidelity PSF-deconvolved imaging of diagnostic emission lines such as \ion{O}{7}, \ion{O}{8} and \ion{Ne}{9}, which will allow us to trace the direct impact of shocks at an angular resolution directly comparable to the complexity seen in {\it HST} imaging.



%% Included in this acknowledgments section are examples of the
%% AASTeX hypertext markup commands. Use \url without the optional [HREF]
%% argument when you want to print the url directly in the text. Otherwise,
%% use either \url or \anchor, with the HREF as the first argument and the
%% text to be printed in the second.

\acknowledgments

We thank the referee for numerous helpful comments which have significantly improved the quality of the manuscript.
WPM acknowledges support from {\it Chandra} grants GO4-15107X, GO5-16099X, and GO2-13127X, and {\it Hubble} grant GO-13741.002-A.
WPM thanks Dan Milisavljevic and Travis Fischer for helpful discussions.
STSDAS and PyRAF are products of the Space Telescope Science Institute, which is operated by AURA for NASA.  


%% in parentheses, after the keyword, but they will not be verified.

{\it Facilities:} \facility{CXO}, \facility{HST}, \facility{VLA}.




%\bibliographystyle{apj}  

%\bibliography{apj-jour,pete_tidal,biblio_mel_marc}

\bibliographystyle{apj}  

\bibliography{apj-jour,pete_tidal,biblio_mel_marc,pete_agn}

%% Tables may also be prepared as separate files. See the accompanying
%% sample file table.tex for an example of an external table file.
%% To include an external file in your main document, use the \input
%% command. Uncomment the line below to include table.tex in this
%% sample file. (Note that you will need to comment out the \documentclass,
%% \begin{document}, and \end{document} commands from table.tex if you want
%% to include it in this document.)


%% The following command ends your manuscript. LaTeX will ignore any text
%% End of file `sample.tex'.

\end{document}